%% file: main-paper.tex
\documentclass[sigconf]{acmart}

\usepackage{soul}
\usepackage{fontawesome5}
\usepackage{algorithm2e}
\usepackage{hyperref}
\usepackage{balance}
\newcommand{\tlcomment}[1]{\noindent{\\\textcolor{black}{\textbf{\#\#\# TL:} \textsf{#1} \#\#\#\\}}}

\newcommand{\zzhighlight}[1]{{\textcolor{black}{#1}}}
\newcommand{\tlhighlight}[1]{{\textcolor{black}{#1}}}

\newcommand{\crhighlight}[1]{{\textcolor{black}{#1}}}

\newcommand{\cmt}[1]{\ignorespaces}

\SetKwInput{KwInput}{Input}
\SetKwInput{KwOutput}{Output}
\SetKwInput{KwName}{Name}

\usepackage{mathtools}
\DeclarePairedDelimiter\floor{\lfloor}{\rfloor}

\AtBeginDocument{%
  \providecommand\BibTeX{{%
    \normalfont B\kern-0.5em{\scshape i\kern-0.25em b}\kern-0.8em\TeX}}}

\copyrightyear{2023}
\acmYear{2023}
\setcopyright{acmlicensed}\acmConference[UIST '23]{The 36th Annual ACM Symposium on User Interface Software and Technology}{October 29-November 1, 2023}{San Francisco, CA, USA}
\acmBooktitle{The 36th Annual ACM Symposium on User Interface Software and Technology (UIST '23), October 29-November 1, 2023, San Francisco, CA, USA}
\acmPrice{15.00}
\acmDOI{10.1145/3586183.3606776}
\acmISBN{979-8-4007-0132-0/23/10}

\begin{document}

\title{PEANUT: A Human-AI Collaborative Tool for Annotating Audio-Visual Data}


\author{Zheng Zhang}
\authornote{Both authors contributed equally to this work.}
\affiliation{%
  \institution{University of Notre Dame}
  \city{Notre Dame}
  \state{Indiana}
  \country{USA}}
\email{zzhang37@nd.edu}

\author{Zheng Ning}
\authornotemark[1]
\affiliation{%
  \institution{University of Notre Dame}
  \city{Notre Dame}
  \state{Indiana}
  \country{USA}}
\email{zning@nd.edu}

\author{Chenliang Xu}
\affiliation{%
  \institution{University of Rochester}
  \city{Rochester}
  \state{New York}
  \country{USA}}
\email{chenliang.xu@rochester.edu}

\author{Yapeng Tian}
\affiliation{%
  \institution{The University of Texas at Dallas}
  \city{Richardson}
  \state{TX}
  \country{USA}}
\email{yapeng.tian@utdallas.edu}

\author{Toby Jia-Jun Li}
\affiliation{%
  \institution{University of Notre Dame}
  \city{Notre Dame}
  \state{Indiana}
  \country{USA}}
\email{toby.j.li@nd.edu}


\begin{abstract}

% Audio-visual learning is an emerging machine learning domain that seeks to enhance the computer's multi-modal perception leveraging the correlation between the auditory and visual modalities. Despite their many useful downstream tasks such as video retrieval, AR/VR, and accessibility, the performance and the wide adoption of existing audio-visual models have been impeded by the availability of high-quality datasets. Annotating audio-visual datasets is laborious, expensive, and time-consuming. To address this challenge, we design and develop \textsc{Peanut}, a more efficient audio-visual annotation tool. \textsc{Peanut}'s novel human-AI collaborative pipeline separates the multi-modal task into two single-modal tasks and utilizes state-of-the-art object detection and sound-tagging models to reduce the effort and the cognitive load for human annotators to process each frame, as well as lower the number of manually-annotated frames needed. \textsc{Peanut} uses active learning to optimize its models for higher accuracy in the current domain in real-time as users use the tool. The design of \textsc{Peanut} also presents features that address the ``overreliance'' problem reported in prior work on human-AI collaborative data annotation. A within-subject user study with 15 participants found that \textsc{Peanut} can significantly accelerate the audio-visual data annotation process and improve the annotation accuracy. \looseness=-1

Audio-visual learning seeks to enhance the computer's multi-modal perception leveraging the correlation between the auditory and visual modalities. Despite their many useful downstream tasks, such as video retrieval, AR/VR, and accessibility, the performance and adoption of existing audio-visual models have been impeded by the availability of high-quality datasets. Annotating audio-visual datasets is laborious, expensive, and time-consuming. To address this challenge, we designed and developed an efficient audio-visual annotation tool called \textsc{Peanut}. \textsc{Peanut}'s human-AI collaborative pipeline separates the multi-modal task into two single-modal tasks, and utilizes state-of-the-art object detection and sound-tagging models to reduce the annotators' effort to process each frame and the number of manually-annotated frames needed. A within-subject user study with 20 participants found that \textsc{Peanut} can significantly accelerate the audio-visual data annotation process while maintaining high annotation accuracy.

\end{abstract}




\begin{teaserfigure}
  \includegraphics[width=\linewidth]{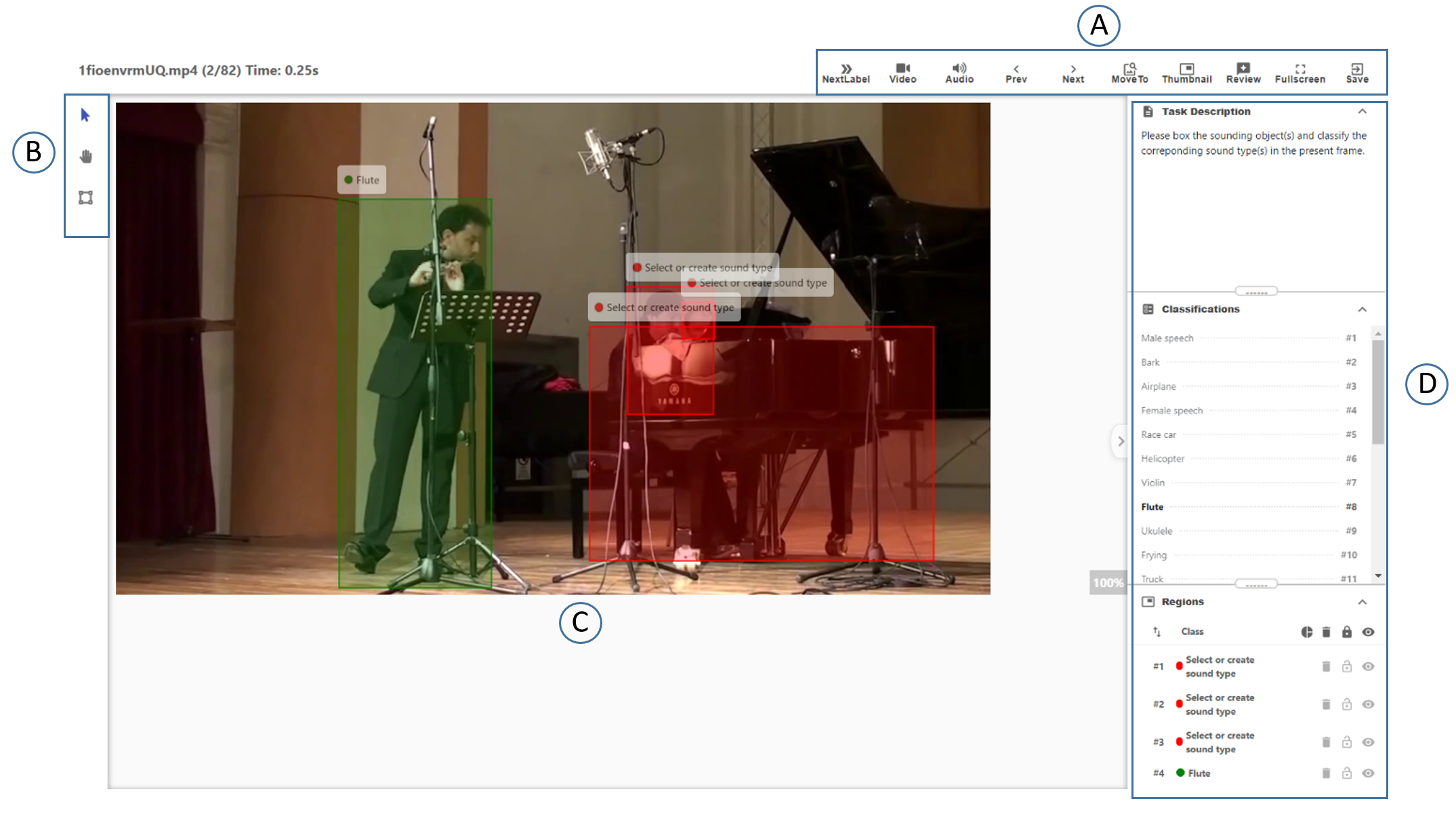}
  \caption{The interface of \textsc{Peanut} for audio-visual data annotation. \zzhighlight{A and B: toolbars with editing and annotation assistance functions; C: the annotation workspace; D: the information panel showing meta-information about the annotations on the current frame\looseness=-1} }
  \label{fig:system_screenshot}
\end{teaserfigure}

\maketitle

\input{1-Intro}
\input{2-Related}
\input{3-System}

\input{4-Study}

\input{5-Discussion}

\begin{acks}
This work was supported in part by an AnalytiXIN Faculty Fellowship, an NVIDIA Academic Hardware Grant, a Google Cloud Research Credit Award, a Google Research Scholar Award, and the NSF Grant 2211428. Yapeng Tian was supported by a gift from Cisco systems. Any opinions, findings or recommendations expressed here are those of the authors and do not necessarily reflect views of the sponsors. We would like to thank Dakuo Wang, Yuwen Lu, and Simret Araya Gebreegziabher for useful discussions.
\end{acks}

\balance
\bibliographystyle{ACM-Reference-Format}
\bibliography{main-paper}

\end{document}

%% file: 1-Intro.tex
\section{Introduction}

Most of our real-world perceptual experiences are specified by multiple cooperating human senses with multi-sensory integration~\cite{murray2011neural}. For example, we can perceive spoken language words and sentiments from lip movements, facial expressions, and speech sounds of the other speakers.  
To mimic human perception capability, researchers in Artificial Intelligence (AI) community have begun to explore audio-visual machine learning (ML) approaches~\cite{aytar2016soundnet,arandjelovic2017look,owens2018audio,hu2019deep,hu2020discriminative}. As an emerging research field, audio-visual learning has attracted a lot of attention from both the academic community and the industry because of its potential to solve many challenging problems in real-world applications such as video retrieval~\cite{zeng2018audio,liu2019use}, AR/VR~\cite{morgado2018self,richard2021audio}, and accessibility~\cite{pavel2020rescribe,wang2021toward}. 

\tlhighlight{A fundamental task in audio-visual learning is \textit{sounding object localization}, which identifies and localizes sounds to visual objects in videos. This task associates audio data with the corresponding visual data, which represents the critical step for many downstream audio-visual tasks such as audio-visual scene-aware dialogs~\cite{alamri2019audio}, video sound separation~\cite{gao2019co,Tian_2021_CVPR}, audio spatialization~\cite{Gao_2019_CVPR,morgado2018self}, audio-visual captioning~\cite{tian2018attempt,rahman2019watch}, and multimodal embodied AI~\cite{chen2020learning,gao2021objectfolder}.} 

% \tlhighlight{However, the progress in audio-visual learning tasks such as sounding object localization is impeded by the lack of high-quality datasets. Since there are various audible and inaudible visual objects in video frames and even a single sound source can make different sounds with different intensities, the sounding object localization task is hungry for large-scale and high-quality training data to capture the diverse audio-visual correspondences and mitigate data variations.}\looseness=-1

\tlhighlight{However, the performance and wide adoption of audio-visual learning have been impeded by the availability of high-quality datasets. For example, commonly used ``gold-standard'' datasets such as AVE~\cite{tian2018audio} is only weakly supervised for this task (i.e., events are annotated for video segments, but object-level ground-truth labels are not available for frames). Therefore, previous works~\cite{aytar2016soundnet,arandjelovic2017look,owens2018audio,hu2019deep,hu2020discriminative} focus on weakly-supervised or self-supervised methods. These methods not only result in lower model accuracy~\cite{senocak2018learning,tian2018audio}, but also introduce biases in the models. In these methods, existing common label noises will be propagated into learned models, leading to compromised results with ethical issues~\cite{tian2020unified,wyatte2019biasing} (e.g., \crhighlight{introducing biases and stereotypes presented in datasets into annotation results}). Since there are various audible and inaudible visual objects in video frames and even a single sound source can make different sounds with different intensities, the sounding object localization task is hungry for large-scale and high-quality training data to capture the diverse audio-visual correspondences and mitigate data variations.}\looseness=-1

% Due to the lack of high-quality annotated data, previous works~\cite{aytar2016soundnet,arandjelovic2017look,owens2018audio,hu2019deep,hu2020discriminative} focus on weakly-supervised or self-supervised methods. These methods not only result in lower model accuracy~\cite{senocak2018learning,tian2018audio}, but also introduce biases in the models. In these methods, existing common label noises will be propagated into learned models, leading to compromised results with ethical issues~\cite{tian2020unified,wyatte2019biasing} (\emph{e.g.,} localizing women's faces as sound sources of men's speech). To mitigate the problem, high-quality audio-visual annotated datasets are desired.

% A particularly useful audio-visual machine learning task that can benefit from high-quality datasets is \textit{sounding object localization}. It aims to identify and localize visible sounding objects in videos. The task is a fundamental problem in audio-visual learning and can help solve a wide range of downstream tasks,  Since there are various audible and inaudible visual objects in video frames and even a single sound source can make different sounds with different intensities, the sounding object localization task is hungry for large-scale and high-quality training data to capture the diverse audio-visual correspondences and mitigate data variations.
% \hl{summarize what it is, useful applications, and why it can benefit from more data}. 

The lack of high-quality datasets is a direct consequence of the large effort required to create such datasets. Annotating audio-visual data is laborious, expensive, and time-consuming. \tlhighlight{With the current annotation tools of for audio-visual data (e.g., VIA~\cite{dutta_2019_via}), annotators need to watch each frame of the video, listen to the corresponding sound, identify the sounding object, draw a bounding box, and indicate the type of sound.} Considering that even a short video would require the annotation of hundreds, if not thousands, of frames, this process is highly repetitive and tedious. \looseness=-1

% The bottleneck imposed by the lack of annotated data is not unique to audiovisual learning, but is also common in many machine learning (ML) tasks in different application domains~\cite{brew2010interaction,halevy2009unreasonable,askhtorab_aiassisted_2021}. The accuracy, quantity, and domain-specificity of data play a crucial role in ensuring the quality and performance of AI systems~\cite{sambasivan_2021_everyone}. 

\tlhighlight{Some prior intelligent tools (e.g.,~\cite{evensen-etal-2020-ruler, askhtorab_aiassisted_2021, demirkus2014robust, vajda_2015_semiautomatic, desmond_increasing_2021, 10.1145/3411764.3445165} have been introduced to provide AI-enabled assistance to users in the data annotation process with promising outcomes to improve annotation efficiency using strategies such as  batching~\cite{askhtorab_aiassisted_2021}, rule synthesis~\cite{evensen-etal-2020-ruler}, and active learning~\cite{labelsleuth2022}. However, these existing tools are limited to the annotation of data in \textit{one} modality (e.g., text categorization~\cite{askhtorab_aiassisted_2021, desmond_increasing_2021}, head post recognition~\cite{demirkus2014robust}, handwriting recognition~\cite{vajda_2015_semiautomatic}, image labeling~\cite{10.1145/3411764.3445165, berg2019ilastik}, and video segmentation~\cite{Qiao2022HumanintheLoopVS}) while audio-visual data annotation requires the user and its AI assistance to process data from \textit{two} modalities and explicitly connect them together.}

In this paper, we present \textsc{Peanut}\footnote{The name \textsc{Peanut} is an acronym for \textbf{P}latform for \textbf{E}fficient \textbf{A}nnotation with \textbf{N}o \textbf{U}nnecessary \textbf{T}edium.}, a new data annotation tool for improving the efficiency of audio-visual data annotation. \zzhighlight{To tackle the unique challenge in \textit{multi-modal} data annotation, \textsc{Peanut} encapsulates a novel human-AI collaborative active learning pipeline where the user validates, revises, and connects the output from multiple single-modal models through a mixed-initiative interface~\cite{horivitz1999principles}.} 

\zzhighlight{Instead of using a fixed ML model to pre-label data, \textsc{Peanut} uses an active learning architecture ~\cite{cohn1996active, settles_activelearning_2012} that} allows these partial-automation models to incrementally learn from the user's annotations in real-time to improve model performance, {learn about visual and auditory data in new video topics, and adapt to the specific domain of the current video}. Several design features are presented to ensure the user's sense of agency and control~\cite{dakuo_autods_2021, wang_designing_2021} and alleviate users' overreliance on AI~\cite{askhtorab_aiassisted_2021}, both are notable issues found in human-AI collaboration of data works from previous studies. A within-subjects study with 20 participants showed that \textsc{Peanut} can significantly accelerate the annotation of audio-visual data (annotate almost 3 times the number of frames compared to the baseline condition) while also achieving high data accuracy.

% \zzcomment{Write the audio grounding process: sound identification, localization and tagging}

In summary, this paper presents the following three main contributions:
\begin{itemize}
    \item \tlhighlight{A set of interaction mechanisms for incorporating outcomes of \textit{single-modal} ML models into a new human-AI collaborative annotation workflow of \textit{multi-modal} audio-visual data while improving model performance with user annotations in real time using an active learning approach.}
    
    \item \textsc{Peanut}, a human-AI collaborative annotation tool that implements these strategies to \tlhighlight{reduce user efforts and improve efficiency in annotating audio-visual data for sounding object localization.}
    
    \item A within-subjects user study with 20 participants \tlhighlight{with diverse annotation and ML expertises} showing that \textsc{Peanut} can improve efficiency in the annotation of audio-visual data while also achieving high data accuracy.
\end{itemize}

%% file: 2-Related.tex
\section{Background and Related Work}

\subsection{Audio-Visual Learning}

Audio-visual learning aims to build a multi-sensory perception system that learns from perceived auditory and visual scenes. Mimicking human perception capacity, it can enable a variety of novel applications in many fields, such as multimedia~\cite{zeng2018audio,liu2019use,hamroun2021multimodal}, affective computing~\cite{noroozi2017audio,ma2019audio}, accessibility~\cite{pavel2020rescribe,wang2021toward}, and AR/VR~\cite{pavel2020rescribe,wang2021toward}. Utilizing and learning from both auditory and visual modalities has attracted significant attention in the AI community. 

We have seen great progress in the development of new audiovisual learning problems and applications, such as representation learning~\cite{aytar2016soundnet,arandjelovic2017look,owens2018audio},
audio-visual sound separation~\cite{ephrat2018looking,gao2018learning,owens2018audio,zhao2018sound}, sounding object localization~\cite{owens2018audio,senocak2018learning,tian2018audio,hu2020discriminative}, audio-visual event localization~\cite{tian2018audio,lin2019dual,wu2019DAM}, audio-visual captioning~\cite{tian2018attempt,rahman2019watch}, and multimodal embodied AI~\cite{chen2020learning,gao2021objectfolder}. 

% Among these tasks, one core problem in audio-visual learning is sounding object visual localization, which aims to identify visible sound sources in our daily life. Pioneering works utilize mutual information~\cite{hershey2000audio} and canonical correlation analysis (CCA)~\cite{kidron2005pixels} to localize sounding visual regions. Very recently, deep audio-visual networks are developed to spatially locate sound sources based on cross-modal embedding similarity~\cite{arandjelovic2018objects, owens2018audio, hu2019deep}, audio-guided visual attention~\cite{senocak2018learning, tian2018audio}, audio-visual class activation mapping~\cite{qian2020multiple}, class-aware object localization map, and sounding object visual grounding~\cite{Tian_2021_CVPR}. These approaches typically take advantage of natural synchronization between audio and visual content and are learned in self-supervised and weakly-supervised manners. Since ground-truth locations of sounding objects are usually not available during training, they tend to make inaccurate predictions. Consequently, certain ethical concerns would be raised. For example, their learned models without using precise labels might incorrectly localize women faces given men's speech.  To solve the problems, training videos with high-quality human annotated sounding object annotations are desired.

\subsubsection{Sounding Object Localization}
\label{sec:sounding_object_localization}
Among audio-visual learning tasks, one important task is the localization of sounding objects. For example, in a symphony concert scenario, the model should identify which instrument is making a particular sound and where the instrument is located in the video. Early work in this area utilized mutual information~\cite{hershey2000audio} and canonical correlation analysis (CCA)~\cite{kidron2005pixels} to localize the sounding visual regions. Recently, deep audio-visual networks have been developed to spatially locate sound sources based on cross-modal embedding similarity~\cite{arandjelovic2018objects, owens2018audio, hu2019deep}, audio-guided visual attention~\cite{senocak2018learning, tian2018audio}, audio-visual class activation mapping~\cite{qian2020multiple}, class-aware object localization map~\cite{hu2020discriminative}, and sounding object visual grounding~\cite{Tian_2021_CVPR}. 

These approaches take advantage of the natural synchronization between audio and visual contents and are trained using self-supervised or weakly-supervised methods (i.e., using no or limited annotated training data). Due to the lack of ground truth annotation for training, they tend to make inaccurate predictions. A relevant audio-visual ML task is active speaker detection which focuses on detecting the active speaker from speech, due to the narrower task domain (human speech only) and the availability of datasets such as~\cite{kim21k_interspeech}, the state-of-the-art active speaker detection models generally perform better than the domain-general sounding object localization ones~\cite{alcazar2021maas,kim21k_interspeech,hu2020discriminative,hu2020discriminative}. Our work seeks to address this problem by making it easier to annotate large audio-visual datasets.

% It is also hard for the model to train when audio and vision are out of sync or there are objects of the same category making sounds together (e.g. several violins playing at the same time). 

\subsection{Data Annotation for Machine Learning}
High-quality data is the foundation of most ML models. The lack of high-quality data has been a long-time bottleneck for many ML tasks~\cite{brew2010interaction,halevy2009unreasonable}. While the undervaluing of data work compared to the lionized work of building novel models and algorithms is common in AI development~\cite{sambasivan_2021_everyone}, ``data excellence'' played a crucial role in the quality of AI systems~\cite{sambasivan_2021_everyone}.\looseness=-1

Despite that many unsupervised, self-supervised, and weakly-supervised models have been found useful in many task domains~\cite{doersch2015unsupervised,zhou2018brief}, supervised learning (i.e., models learning from annotated example input-output pairs) still shows important advantages in model performances and robustness. However, collecting annotated ground truth data is usually a costly and time-consuming process that requires extensive human labor. 

There are two types of data annotation---(1) Explicit data annotation, where human annotators use their perceptive and cognitive abilities to categorize and label data for the purpose of creating annotated datasets~\cite{askhtorab_aiassisted_2021}. This is often a repetitive and tedious process especially because datasets need to be large in order to be effective; (2) Implicit data annotation, where users of a computing system generate useful datasets as a side product of interacting with the system (e.g., users of a recommender system generate useful data when interacting with recommended items). While this approach does not incur additional user efforts, it requires the system to be deployed at a large scale in order to collect sufficient data, which requires significant effort and is not always feasible in all task domains and at all stages of the project. Implicit annotation also faces the ``cold start'' problem~\cite{schein2002methods}---it still needs a dataset for training the initial model to provide acceptable performance at the beginning before implicitly-annotated data from user interactions come in.

\textsc{Peanut} is designed to reduce the human effort required in \textit{explicit} data annotation, making the process more efficient while maintaining data accuracy. Meanwhile, \textsc{Peanut} also uses \textit{implicit} data annotation strategies to improve the performance of its own object detection model in real-time while the user is in the process of explicit interactive data annotation (detail in Section~\ref{sec:active_learning}).

\subsubsection{Assistance for Explicit Data Annotation}
Explicit data annotation is traditionally a fully manual process---a human annotator examines the input data and determines the output result that the ML model should produce using their human knowledge and cognitive abilities~\cite{cassidy_2017_tools} (e.g., the commonly used VIA tool~\cite{dutta_2019_via} for manual audio-visual annotation). Recently, several interactive tools have been developed to assist human annotators with the process~\cite{zhang2022onelabeler,evensen-etal-2020-ruler,ratner2017snorkel,shnarch2022label,rietz2021cody}. Notably, \textsc{Ruler}~\cite{evensen-etal-2020-ruler} is an interactive system that synthesizes labeling rules while the human annotator manually assigns labels in textual data (known as the \textit{data programming} process~\cite{ratner2017snorkel}). Like \textsc{Peanut}, \textsc{Ruler} uses a partial-automation approach where an intelligent system helps the human annotator by automating some parts, but not the full end-to-end annotation task. The two systems also share the ``explicit+implicit'' approach as \textsc{Ruler} learns to synthesize new labeling rules while the user manually labels the data. Desmond et al. introduced an AI labeling assistant that uses a semi-supervised learning algorithm to predict the most probable labels for each example in the labeling intents of user natural language inquiries~\cite{desmond_increasing_2021}. \tlhighlight{\textsc{Peanut}'s interfaces for users to verify labels predicted by a model and correct model-generated bounding boxes are similar to prior work in improving object detection models in computer vision~\cite{kaul2022label,liu2020faster}.}

\zzhighlight{Some ML-enabled annotation assistance tools use off-the-shelf models to pre-label data as suggestions for users. For example, CVAT\footnote{\url{https://github.com/opencv/cvat}} uses a deep learning model to pre-label the images. Similarly, Ilastik~\cite{berg2019ilastik} provides pre-labels to support semi-automatic image segmentation using edge detection and watershed models. Although pre-labeling is effective for accelerating the annotation, it risks annotating data with model biases, especially when the data is in a new domain previously unseen in the model's training process~\cite{Xiao2018AddressingTB}. The approach used in \textsc{Peanut} has significant differences from these pre-labeling approaches. Instead of performing pre-labeling independent of human annotation, \textsc{Peanut} grounds its annotation suggestions on human-labeled key frames in real-time to balance model performance and human effort for achieving high-quality annotations and ensuring user control of the annotation process at the same time. The key frames are determined in real-time according to video contents and intermediate annotation results (Section~\ref{sec:algo_key_frame}). This approach also allows \textsc{Peanut} to learn new topics from the user's few shot annotations.}\looseness=-1

\tlhighlight{The problem domain in \textsc{Peanut} is also more complex than the domain in existing AI-enabled annotation support tools for single-modal data (e.g., images~\cite{berg2019ilastik,10.1145/3411764.3445165}, text~\cite{rietz2021cody,labelsleuth2022,desmond_increasing_2021, gao2023collabcoder}). \textsc{Peanut} works in a task domain with data in multiple modalities (audio and visual). The task explicitly addresses the explicit and implicit relations between data in different frames.}\looseness=-1

Another type of assistance for explicit data annotation uses the strategy of \textit{batching} (e.g.,~\cite{askhtorab_aiassisted_2021,vajda_2015_semiautomatic,demirkus2014robust}). The system first puts ``similar'' data into batches using unsupervised clustering models or pre-trained models (e.g., semantic similarity for NLP tasks) and then asks the human annotator to annotate data by batch. The underlying assumption is that it would be easier and faster to annotate similar data together than to annotate them individually because they are likely to be assigned with the same or similar labels. The batching strategy has been shown to be effective in accelerating the data annotation process~\cite{askhtorab_aiassisted_2021}. A potential concern with batching is users' overreliance on AI---the human annotator might assign the same label to a batch without carefully examining each data point because ``the AI model thought that they were all similar''~\cite{askhtorab_aiassisted_2021}. \textsc{Peanut} also  uses batching---but it was not achieved using an unsupervised clustering model. Instead, \textsc{Peanut} leverages the characteristics of videos so that adjacent frames are often similar to each other. The auditory and visual models used in \textsc{Peanut} detect changes in the scene or sudden movements, which are used to batch frames so that the human annotator only annotates key frames. A guided workflow for human annotation (Section~\ref{sec:automatic_annotation}), video playback, and thumbnail preview features (Section~\ref{sec:annotation_review}) in \textsc{Peanut} alleviate the overreliance issue. \looseness=-1

\subsubsection{Implicit Data Annotation}
Systems that use implicit data annotation strategy include (1) those that collect data from their interactions with users for the purpose of a \textit{different} data task, such as reCAPTCHA~\cite{von2008recaptcha} that collects user-annotated data for training computer vision models through its interactive process of distinguishing human users from bots for authentication purposes and the Foldit game~\cite{cooper2010predicting} that collects user-annotated protein structures through an online game; (2) those that collect data from their interactions with users for the purpose of improving the \textit{same} interaction, such as recommender systems that learn about user personal preferences as the user interacts with the recommended items~\cite{rashid_getting_2002} and intelligent agents that learn about tasks while helping users with task automation~\cite{li_sugilite:_2017, li_pumice:_2019}.

\textsc{Peanut}'s use of implicit data annotation falls into the latter category. As discussed in Section~\ref{sec:active_learning}, the human annotation result for each keyframe is used to fine-tune the visual-sound grounding model, which reduces human effort in annotating the rest of the frames. This strategy is also an example of \textit{active learning}~\cite{settles_activelearning_2012,cohn1996active}, where the system chooses which data the visual-sound grounding model should learn from by querying the user through \textsc{Peanut}'s selection of keyframes (Section~\ref{sec:algo_key_frame}).

\subsection{Human-AI Collaboration in Data Science}
\textsc{Peanut} belongs to a fast-growing list of AI-powered interactive tools that assist and augment human capabilities in different subtasks in the data science workflow~\cite{dakuo_autods_2021,wang_human-ai_2019,heer2019agency, ning2023empirical}. Besides data annotation, human-AI collaborative tools have also been developed for data wrangling (i.e., cleaning and formatting data to make it suitable for analysis~\cite{kandel_enterprise_2012}) (e.g.,~\cite{guo_proactive_2011}), exploratory data analysis and sensemaking (e.g.,~\cite{wongsuphasawat_voyager_2017}), selection of ML models (e.g.,~\cite{he2021automl}), generating new data features (e.g.,~\cite{galhotra_automated_2019}), testing and debugging ML models (e.g.,~\cite{wu2021polyjuice}), and fine-tuning parameters in ML models (e.g.,~\cite{liu2020admm}).\looseness=-1

The design of \textsc{Peanut} is informed by empirical studies on how data science workers work with data~\cite{muller_how_2019, muller_2021_ground-truth} and data workers' perceptions and mental models of human-AI collaborative data science tools~\cite{wang_human-ai_2019}. For example, studies~\cite{muller_how_2019,muller_2021_ground-truth} reported that the difficulty with finding reliable labels for ground truth is a common problem that data science practitioners encounter. In industry settings, external domain experts often need to be hired~\cite{muller_2021_ground-truth}. Spreadsheet is commonly used as a tool for labeling---while specialized tools such as CrowdFlower\footnote{https://appen.com/} have also been used, they are used to facilitate group collaboration on data annotation ~\cite{muller_2021_ground-truth} with no intelligent automation feature that reduces the workload of the annotation.\looseness=-1

% \hl{Discuss the implications}.

More broadly, facilitating effective collaboration between human and intelligent systems has been a long-standing topic since the origin of HCI research in the seminal paper on man-computer symbiosis~\cite{licklider1960man} where computers can ``do the routinizable work that must be done to prepare the way for insights'' meanwhile human users can leverage their domain expertise to make decisions that computers cannot. The design of \textsc{Peanut} follows this pattern where the system marks potential object candidates and identifies keyframes for users to annotate using single-modal partial-automation models while the user ''connects the final dots'' with their annotations that finish the end-to-end multi-modal process using human perceptive and cognitive capabilities that ML models do not yet possess. 

Later work such as the principles in mixed-initiative interactions~\cite{horivitz1999principles} identified strategies such as considering uncertainties in user intents, assessing the added-value of automation, providing mechanisms for refining automation results, and maintaining user working memory of interactions. More recently, guidelines in human-AI interaction~\cite{amershi_2019_guidelines} have been proposed to address challenges that came with the popularity of ``black-box-like'' data-driven AI models in interaction systems (as opposed to the ``traditional'' planning-based techniques). These guidelines, principles, and theoretical frameworks have been widely used in the design of human-AI collaborative systems in domains like healthcare~\cite{cai_human-centered_2019}, creativity support~\cite{louie_noivce_2020}, and error repairs in chatbots~\cite{li_sovite:_2020}. Two key human-AI collaboration challenges we specifically address in the design of \textsc{Peanut} are to accommodate the \textit{imperfection} of AI models and to enable the continuous learning of partial-automation models, which we discuss in Section~\ref{sec:design_challenges}.

%% file: 3-System.tex
\section{The \textsc{Peanut} System}

% TL: reorganize --> 
\subsection{Task: Sounding Object Localization}

\zzhighlight{\textsc{Peanut} focuses on annotating data for the sounding object localization task. As defined in~\cite{hu2021class}, given a video, sounding object localization aims to semantically correlate each sound to the visual regions containing the sounding source and recognize the category of the sound in each frame. Therefore, in the annotation task, the annotator should first identify all the sounds in the current frame, and for each sound, provide a semantic label (e.g., church bell, dog bark) and associate it to its sounding object (represented as a bounding box) for each and every frame in a video. A frame may contain multiple sounding objects at the same time as well as silent objects that are physically capable of making sounds.} \looseness=-1

% A video may contain multiple sounding objects at the same time as well as silent objects that are physically capable of making sounds. The sounding object localization task has recently gained increasing attention from the computer vision community as it is critical for enabling machine to proceed towards human understandings of complex scenes by leveraging the temporal and cross-modal correspondences between visual and auditory cues \cite{li2021space}.

\subsection{Design Goals}
\label{sec:design_challenges}
Informed by results from prior studies on data annotation~\cite{askhtorab_aiassisted_2021,mitra_comparing_2015,ratner2017snorkel}, \zzhighlight{status quo of relevant ML models,} and our experience with data annotation and human-AI collaborative tools, \crhighlight{we summarized the following four design goals in our design of an AI-assisted audio-visual data annotation tool to address potential human and AI challenges in sounding object localization task}:
% \subsubsection{Challenges}
\paragraph{DG1: Improving annotation efficiency without compromising accuracy with imperfect AI models} \crhighlight{The ultimate goal of the annotation is to collect data to train an end-to-end ML model for the sounding object localization task. Although there are some end-to-end models for this task ~\zzhighlight{\cite{owens2018audio,senocak2018learning,tian2018audio,hu2020discriminative}}, their performance is limited due to the lack of annotated data, which our work seeks to address. While some off-the-shelf models (e.g., object detectors, audio tagging models) can contribute to the annotation process of audio-visual data, they are often limited in several ways: (1) they often process input data in only one modality; (2) they have limited accuracy (as they are not specifically trained for our domain); and (3) they only assist with a part of the annotation process. In contrast, our annotation task requires: (1) processing multi-modal audio-visual data; (2) achieving high accuracy in annotation; and (3) providing an end-to-end annotation from raw videos to annotated sounding objects in each frame of the video.}

\crhighlight{However, the limitations of AI models do not preclude their potential to function as assistant to human annotators. Prior works \cite{berg2019ilastik, choi2022ai, kehl2021artificial} have demonstrated that using AI models to automate parts of data annotation tasks could significantly improve annotation efficiency. Nonetheless, previous research ~\cite{mitra_comparing_2015} also suggested that without effective human intervention strategy, imperfect AI could result in lower quality in annotations despite the higher efficiency. Therefore, our main motivation is: how can we make annotating audio-visual data more efficient by introducing AI assistance that reduces human effort and cognitive load? Meantime, it is also crucial that improved efficiency does not come at the cost of compromising accuracy. As discussed in~\cite{sambasivan_2021_everyone}, data quality issues can cause significant compounding negative effects on downstream tasks, resulting in ``data cascades'' that harm users and communities.}

\paragraph{DG2: Supporting users' agency and mitigating their overreliance on AI models}

\zzhighlight{Prior work \cite{langer1989minding} find that users tend to give premature cognitive commitment to automation if the assisted task is routine, repetitive, and demanding. In particular, Ashktorab et al. \cite{askhtorab_aiassisted_2021} showed that AI-assisted data labeling tasks conform to these characteristics in which users are inclined to overrely on AI: despite the inaccuracies of the models, human users sometimes perceive \textit{higher} qualities and \textit{higher} capabilities of the models than what they actually are when they observe their high performance in common situations. However, when uncommon situations arise, human users may overrely on AI automation \crhighlight{which could threat} human discernment \crhighlight{and agency} in annotation~\cite{askhtorab_aiassisted_2021}. This poses a challenge in our task given that the frame-wise annotation is highly repetitive and demanding, which may make users less wary of the cross-modal mismatches and inaccurate annotations by the AI.} Therefore, our system should make it easy for users to identify potential errors in AI results and ensure that users fulfill their duties of reviewing and validating AI results in the annotation workflow. \crhighlight{Furthermore, our system should support user agency in the annotation process, providing the flexibility to control the degree of human involvement based on their perception of AI's accuracy change over the annotation process.}\looseness=-1

% \subsubsection{Goals}
% \paragraph{\zzhighlight{DG3: Improving annotation efficiency without compromising accuracy}} \zzhighlight{The primary goal of AI-assisted annotation tools is to help annotators label data more accurately and faster. However, prior work~\cite{mitra_comparing_2015} suggests that, if no effective human intervention strategy is enabled, imperfect AI could result in lower quality in annotations despite the higher efficiency. Therefore, our main motivation is: } how can we make annotating audio-visual data more efficient by introducing AI assistance that reduces human effort and cognitive load? Meantime, it is also crucial that improved efficiency does not come at the cost of compromising accuracy. As discussed in~\cite{sambasivan_2021_everyone}, data quality issues can cause significant compounding negative effects on downstream tasks, resulting in ``data cascades'' that harm users and communities.

\paragraph{DG3: Minimizing the learning barrier}
Data annotation tasks are often conducted by people without significant AI/ML expertise---many users of data annotation tools are either domain experts (e.g., physicians and radiologists who annotate medical imaging data \zzhighlight{\cite{neves2014survey}}) or laypeople who only annotate data either as a one-off task or only occasionally (e.g., Mechanic Turk workers \zzhighlight{\cite{mitra_comparing_2015}}). Therefore, it would be prohibitive if the tool has a high learning barrier or requires extensive expertise from the users. Ideally, the system should not require users to learn new skills beyond what they would already need if they labeled the data manually.

\paragraph{\zzhighlight{DG4: Supporting annotation for diverse video topics}}

\zzhighlight{Audio-visual scenarios are intrinsically diverse in terms of the object, sound, and event type involved in videos \cite{arandjelovic2017look}. Therefore, our annotation tool should be able to provide users with effective AI assistance regardless of the topics in the videos being labeled. To meet this need, models used in the pipeline should be generalizable and not limited to specific domains. While some of these models might be pre-trained to bootstrap the system's performance in cold-start situations for common topics, they should also be able to learn about new object and audio types quickly from the user's few-shot example data in real time.}

% Lastly, it is important that the AI assistance should not rely on any domain-specific knowledge, as the tool should support video annotation in any domain. Pre-trained models used in the tool should be trained on large domain-general datasets and has the capability to adapt to new domains when needed.

\begin{figure*}[t!]
  \centering
  \vspace{-2mm}
  \includegraphics[width=\linewidth]{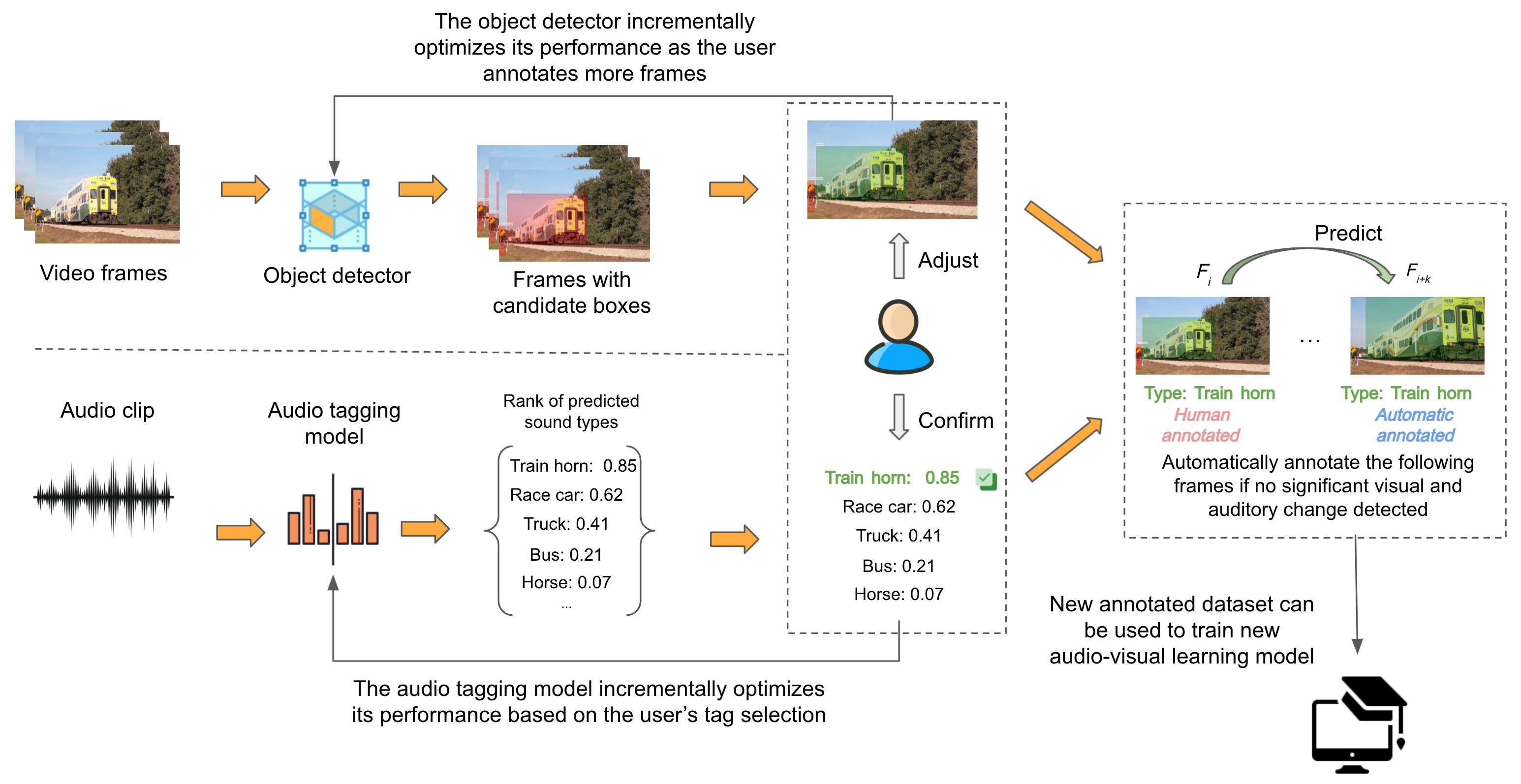}
  \caption{The system architecture of \textsc{Peanut}. The video frames and the audio clips are processed separately by single-modal partial-automation models before they are presented to the human annotator for the end-to-end result that leverages the annotator's multi-modal perceptive and cognitive capabilities. Both the object detector and the audio tagging model incrementally improve their own performances using an active learning approach.}
  \label{fig:architecture}

\end{figure*}
% \tlcomment{The ``imperfection'' of AI models: 1. single modal; 2. limited accuracy; 3. not end-to-end vs. The requirement of good data annotation: 1. multi-modal; 2. high accuracy; 3. end-to-end === strategies to help users work with imperfect AI models}
% \tlcomment{Over-reliance -- expect users to do (1)(2)(3), otherwise the accuracy will suffer}
% \tlcomment{No new learning barrier or additional effort -- data annotators often not-trained e.g., gig worker}
% \tlcomment{Need to be domain-general -- work with data from all domains}
% ------------
% \tlcomment{DG: faster}
% \tlcomment{DG: less tedious/effort required}

% TL: need to SEPARATE design goals from approaches -- design goals should not be BIASED towards a certain approach e.g., "enable the active learning..." is not a goal, but an approach
\subsection{System Design}

% \cmt {
%     To help lay users annotate audio-visual data both efficiently and accurately, we developed \textsc{Peanut}. The design of \textsc{Peanut} was guided by the following goals:
%     \begin{itemize}
%         \item \textbf{DG1}: Realize human-AI collaborative annotation where AI can automate audio-visual labeling of most video frames under the guidance of a few human annotations
%         \item \textbf{DG2}: Enable annotators to efficiently review and modify the labeling resulting from the human-AI collaboration
%         \item \textbf{DG3}: Reduce the cognitive effort of the annotator to identify and locate the sounding objects
%         \item \textbf{DG4}: Enable active learning of the AI "annotator" so that the accuracy of automatic labeling can implicitly improve from a stream of human-annotated data
%     \end{itemize}
% }

To address the aforementioned design challenges and design goals, we designed and implemented \textsc{Peanut}, a human-AI collaborative audio-visual annotation tool that seeks to make annotation more efficient using novel interaction strategies, features, and algorithms. Figure~\ref{fig:system_screenshot} shows the main annotation workspace of \textsc{Peanut}, which consists of three components: (1) a canvas that displays the video frame and its current annotation state (C); (2) top and side toolbars that allow the annotator to perform different operations (A, B); (3) an information panel that summarizes the meta-information about the current object types and displays the operation history (D). The system architecture of \textsc{Peanut} is shown in Figure~\ref{fig:architecture}.

\cmt{
    Following these four design goals, we implemented \textsc{Peanut} as shown in Figure \ref{fig:system_screenshot}. The workspace of \textsc{Peanut} consists of three components: (1) A canvas that bears the human and AI labeling (C); (2) Top and side toolbars that allow the annotator to perform different operations (A, B); (3) An information panel that summarizes the meta-information (D).
}

\subsubsection{Human annotation} For each frame, \textsc{Peanut} lowers the user's effort and cognitive load to annotate it by: (1) facilitating the user's access to both global and local audio-visual contexts; (2) inferring bounding boxes for potential candidates of sounding objects; and (3) predicting the audio tags for sounds. 

When the user moves to a frame that corresponds to the \textit{i}-th millisecond (ms) of the video, \textsc{Peanut} will auto-play a one-second-long audio clip that contains the soundtrack from \textit{i-500} to \textit{i+500} ms, which gives the annotator a sense of the local audio context. The annotator can replay the local soundtrack by clicking the ``Audio'' button. In addition, \textsc{Peanut} allows the annotator to watch the entire video at any time during the annotation process, which helps the annotator disambiguate between the identities of local sound sources with the information of the global context. 

\textsc{Peanut} helps the annotator locate and tag the sounding object. When using \textsc{Peanut} to annotate a frame, instead of manually recognizing the audio type and drawing the bounding box for its sounding object, the user, in most cases, selects a sound type from a list of predictions made by an audio-tagging model and matches it to one of the visual objects detected by an object detector (red boxes shown in Figure~\ref{fig:system_screenshot}). This process can reduce the annotator's cognitive load for locating and labeling the sounding objects \crhighlight{(DG1)} \crhighlight{and demand no domain knowledge from annotators (DG3)}. If none of the predicted sound types or visual objects is correct, the annotator can manually enter a sound type and draw a new bounding box by clicking the \faIcon{expand} button.

\begin{figure*}[t!]
  \centering
  \includegraphics[width=\linewidth]{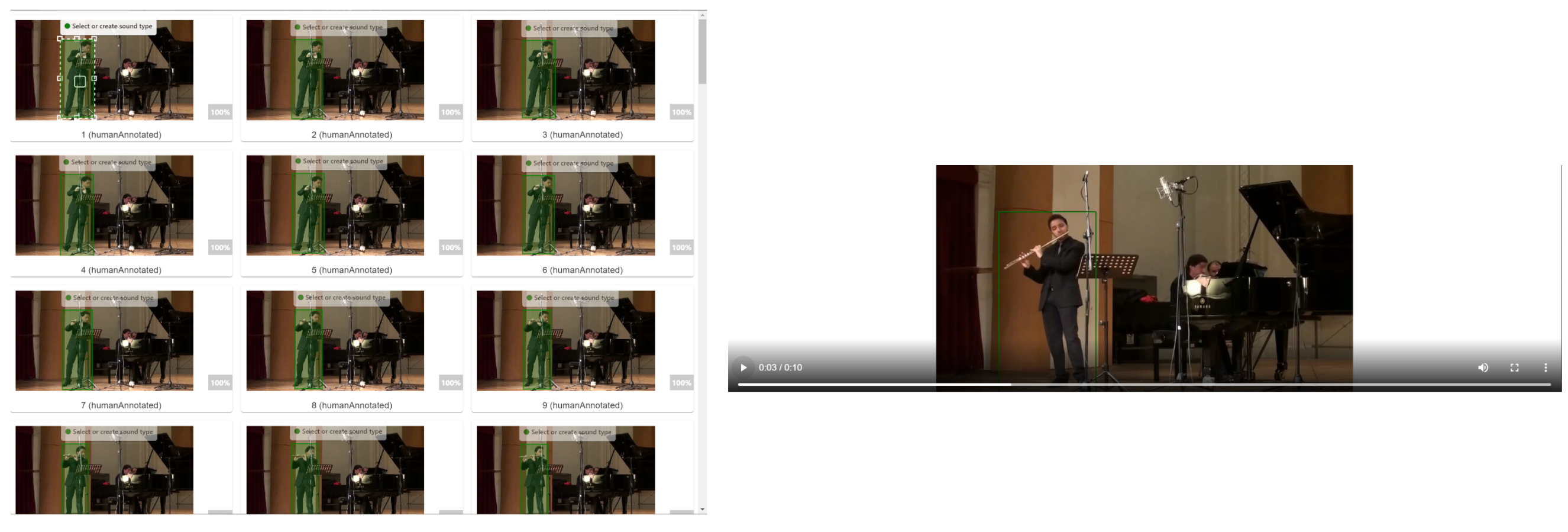}
  \caption{\textsc{Peanut} provides two interfaces for users to review the annotation result: the frame-by-frame thumbnail review and the annotated video playback preview.}
  \label{fig:system_review_screenshot}
  \vspace{-3mm}
\end{figure*}

\subsubsection{Automatic annotation}
\label{sec:automatic_annotation}
Besides reducing the efforts required for the user to annotate each frame, \textsc{Peanut} allows the user to annotate fewer frames. Instead of asking the annotator to label every frame in the video, \textsc{Peanut} uses two complementary strategies to automatically infer the annotation result of the remaining frames based on the human annotation of ``key frames''. The combination of two strategies provides the annotator with the flexibility to adjust the granularity of automatic annotation.

For the first strategy, \textsc{Peanut} dynamically navigates the annotator to the next keyframe that requires human annotation, from which \textsc{Peanut} can infer the annotation results of the frames between two human-annotated ones. The annotator can go to the next recommended keyframe by clicking the \faIcon{angle-double-right} button. Under the hood, \textsc{Peanut} uses an \textit{audio-visual-sensitive binary search} algorithm (see Section~\ref{sec:recomendFrame}) to identify the next keyframe that needs human annotation. Note that the recommended key frames may not be in sequential order. For example, after an annotator annotates the 10th and 20th frame consecutively, \textsc{Peanut} may roll back to the 15th frame if there was a significant visual or auditory change between the 10th frame and the 20th frame according to the human annotation results. When the annotator finishes the annotation of two adjacent recommended keyframes, if both are determined to be continuous in both auditory and visual spaces, \textsc{Peanut} will automatically interpolate the annotation results of the in-between frames by tracking the movement of known sounding objects, as explained in Section~\ref{sec:recomendFrame}.

In the second strategy, the annotator can choose to automate the annotation of each frame by clicking on the \faIcon{angle-right} button. \textsc{Peanut} will preempt the annotation of the immediate next frame for the annotator to review and confirm. In this way, the annotator can closely inspect the automatic annotation result to ensure its correctness. The second strategy is especially useful when the auditory or visual context changes quickly in the video.

\crhighlight{ It's important to note that PEANUT provides users with the choice to select their preferred strategy at any moment during the video annotation, thus allowing dynamic control over the level of human engagement in the annotation process (DG2). }

\subsubsection{Annotation result review}
\label{sec:annotation_review}
\textsc{Peanut} uses two interfaces for the annotator to review the annotation result: \textit{frame-by-frame thumbnail} and \textit{annotated video playback preview} (Figure \ref{fig:system_review_screenshot}). The \textit{frame-by-frame thumbnail} interface displays the annotation result of each frame in a grid view, enabling the annotator to quickly detect frames with inappropriate annotations. The \textit{annotated video playback preview} interface allows the annotator to examine how the entire video looks after the annotation process is finished. The sound types are also displayed in semi-transparent white boxes above the bounding boxes. The annotator can also go to a specific frame to review and modify the annotation result by clicking the \textit{Move To} button. \crhighlight{These two features are designed to assist humans in spotting potential inaccuracies in AI annotations by employing supplementary review strategies, in order to mitigate  possible overreliance (DG2).}

% \hl{talk about how this help alleviate the overreliance problem}

\subsubsection{Active learning}
\label{sec:active_learning}

% \tlcomment{generalize to different topics -- how active learning can help}

\zzhighlight{Human-in-the-loop audio-visual data annotation may face two predominant challenges. First, the pre-trained object detector and sound tagging model could only be able to tackle certain objects or sound types and provide little support in those unfamiliar data. Plus, the detection accuracy of object and sound type may be contingent to event scenarios, while audio-visual scenarios are often highly diverse. It is possible that the model has trouble in recognizing a learned type from an unseen scenario. To address these challenges, as shown in Figure~\ref{fig:architecture}, \textsc{Peanut} adopts an active learning strategy to optimize the visual sound grounding network model, object detector and audio tagging model incrementally in real time as the users annotate more data. Because video frames and sounding objects in the same video are usually similar to each other, this active learning strategy allows the model to learn from ground truth data that likely closely resemble input data that it will process in the future, effectively adapting the model to the domain of the video. In this way, when encountering data type or event scenario unknown to AI model, human annotators can provide the model with a few ground truth annotations to enable it to classify data in the current specific scenario. This strategy also eliminates the need of granting a model with a generalized ability to handle a wide range of diverse scenarios in a single training \crhighlight{(DG4)}, which is formidable for current audio-visual models. } \looseness=-1

\begin{algorithm}
\caption{The audio-visual-sensitive binary search algorithm}\label{alg:nextFrame}
\KwName{\textbf{AudioVisualSensitiveBinarySearch}}
\KwInput{$LeftBoundFrame, CurrentHumanAnnotatedFrame,$\\$ GlobalRightBoundStack$}
\KwOutput{$NextHumanAnnotatedFrame$}

$left \gets LeftBoundFrame $\;
$right \gets CurrentHumanAnnotatedFrame $\;
$grs \gets GlobalRightBoundFrameStack$\;
$predObjects \gets PredictSoundingObjectFromPriorAnnotation(left, right) $\;

{
    \uIf{$right$ ==0 \textbf{or} ($predObjects == SelectedObjects[right]$ \textbf{and} $grs.length == 0$)} {
        \Return FarthestFrameNeedHumanAnnotation()\;
    }\uElseIf{$predObjects == SelectedObjects[right]$ \textbf{and} $grs.length > 0$)} {
        $cur \gets right$\; 
        $right \gets grs.pop()$\;
        \While{right != Null}{
            $predObjects \gets PredictSoundingObjectFromPriorAnnotation(cur, right) $\;
            \uIf{pred\_objects == SelectedObjects[right]}{
                PopulateFrameAnnotation($cur$, $right$)\;
                $cur \gets right$\; 
                $right \gets grs.pop()$\;
            } 
            \Else{
                $next \gets \floor{\frac{cur+right}{2}}$\;
                \Return $next$\;
            }
        }
        \Return FarthestFrameNeedHumanAnnotation()\;
    }
    \Else{
        $grs.push(right)$\;
        $next \gets \floor{\frac{left+right}{2}}$\;
        \Return $next$\;
    }
}
\label{alg:search}
\end{algorithm}

% \begin{algorithm}
%     \caption{Predict the sounding objects based on prior human annotation} \label{alg:predict}
%     \KwName{\textbf{PredictSoundingObjectFromPriorAnnotation}}
%     \KwInput{$RefFrame, TargetFrame$}
%     \KwOutput{$PredObjects$}
%     $targetObjects \gets DetectedObjects[TargetFrame] $\;
%     $PredObjects \gets \textit{[ ]} $\;
%     \For{$refObject$ \textbf{in} $DetectedObjects[RefFrame]$} 
%     {
%         $candidateObject \gets GetObjectwithClosestBox(refObject, targetObjects) $\;
%         \uIf{CheckObjectMatch(refObject, candidateObject) == true}{
%             $PredObjects.append(candidateObject)$\;
%         }
%         \Else {
%             $refObjectCopy \gets refObject.copy()$\;
%             $PredObjects.append(refObjectCopy)$\;
%         }
%     }
% \Return $PredObjects$\;
% \end{algorithm}

% \begin{algorithm}
%     \caption{Examine the auditory and visual consistency between two frames} \label{alg:change}
%     \KwName{\textbf{DetectAudioVisualChange}}
%     \KwInput{$left, right$}
%     \KwOutput{$isChanged$}
%     % $offset \gets right-left$\;
%     %  \For{$i \gets left+1$ \KwTo $right$}{
%         \If{DetectedObjects[left] != DetectedObjects[right]} {
%             \Return $true$\;
%         }
     
%         $left\_tags \gets AudioTags[left][0:3] $\;
%         $right\_tags \gets AudioTags[right][0:3]$\;
%         \If{haveCommonTags(left\_tags, right\_tags) == false} {
%             \Return $true$\;
%         }
%     % }
%     \Return  $false$\;
% \end{algorithm}

    \begin{algorithm}
        \caption{Automatically interpolate the annotation results for the frames between two human-annotated frames} \label{alg:populate}
        \KwName{\textbf{PopulateFrameAnnotation}}
        \KwInput{$left, right$}
        \KwOutput{$None$}
        \For{$i \gets left+1$ \KwTo $right$}{
            $predAnnotation \gets PredictSoundingObjectFromPriorAnnotation(left, i) $\;
            \If{AduioTags[i].contains(videoTag)} {
                $Annotate(i, predAnnotation)$\;
            }
        }
    \end{algorithm}
    
    \begin{algorithm}
        \caption{Return the farthest frame that needs human annotation } \label{alg:farthest}
        \KwName{\textbf{FarthestFrameNeedHumanAnnotation}}
        \KwInput{$None$}
        \KwOutput{$next$}
        
        $frameIndex \gets getFarthestHumanAnnotatedFrame()$\;
        \For{$i \gets 1$ \KwTo k}{
            \If{DetectAudioVisualChange(frameIndex, frameIndex+i) == false} {
                $continue$
            }
            \Else {
                \Return $frameIndex+i$\;
            }
        }
    \end{algorithm}

\begin{figure*}[htp]
  \centering
  \includegraphics[width=\linewidth]{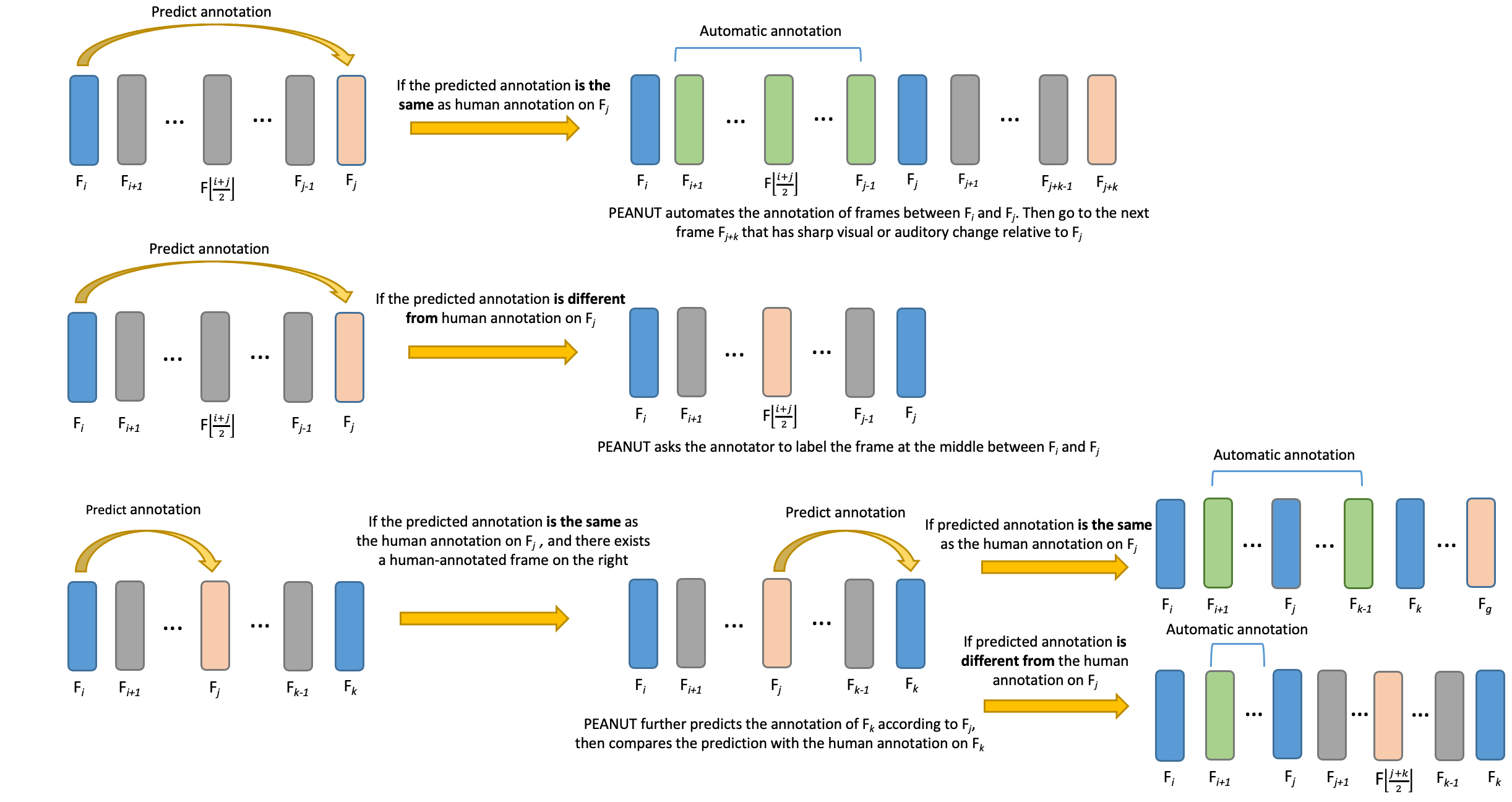}
  \caption{The illustration of the visual-audio sensitive binary search algorithm (Algorithm \ref{alg:nextFrame}). Blue rectangles represent the frames that have already been  annotated by a human annotator, grey rectangles represent the frames that have not been annotated, pink rectangles represent the frames that the human annotator is working on, and green rectangles represent the frames automatically annotated by \textsc{Peanut}.}
 
  \label{fig:algorithm_illustration}
  \vspace{-3mm}
\end{figure*}

  %\textsc{Peanut} first predicts the annotation for the present frame $(F_{\textit{j}})$ based on the nearest human-annotated frame to the left ($F_{i}$). Then \textsc{Peanut} uses the similarity between the predicted annotation and the actual human annotation on $F_{j}$ to decide the next frame for human labeling. If the similarity is high and no human-annotated frame exists to the right, \textsc{Peanut} will bring the annotator to the next frame that has a significant visual or auditory change relative to $F_j$ (or $F_{j+k}$ if no change is detected within $k$ steps); If the similarity is low, \textsc{Peanut} will move to the middle point between $F_i$ and $F_j$; If the similarity is high but a human-annotated frame ($F_k$) exists to the right, \textsc{Peanut} will use the same strategy to predict the annotation on $F_k$, and decide the next step according to the predicted annotation result's similarity with the human annotation result on $F_k$. 

\subsection{Algorithmic Methods}

\subsubsection{Recommending the next frame for human annotation}

\label{sec:recomendFrame}

We developed an \textit{audio-visual sensitive binary search} algorithm for \textsc{Peanut} to decide the next ``keyframe'' that needs human annotation (illustrated in Figure \ref{fig:algorithm_illustration}). The details of the algorithm are shown in Algorithm \ref{alg:nextFrame}. The index of the next key frame is decided by the similarity of the annotation between the left bound frame (\textit{lb}), the current human annotation frame (\textit{cur}), and a stack of right bound frames (\textit{rbs}). The left-bound frame refers to the preceding human-annotated frame that is the closest to the current frame on the timeline. On the contrary, the right-bound frames are those after the current frame, while the annotation similarity between the current frame and the right-bound frames needs to be confirmed. The calculation of the annotation similarity uses a "\textit{predict-select-compare}" strategy. Given the current frame, \textsc{Peanut} first predicts the sounding objects in the current frame by inheriting the human annotation of \textit{lb}. Then \textsc{Peanut} compares the user-selected sounding objects and the predicted sounding objects. If these two sets of objects are not the same, this indicates that there is a visual discontinuity between \textit{lb} and \textit{cur} that is not captured by \textsc{Peanut}. To locate the first frame of this ``discontinuity'' efficiently, \textsc{Peanut} adopts binary search and asks the human annotator to label the frame $\floor{\frac{lb+cur}{2}}$, meanwhile \textsc{Peanut} pushes \textit{cur} into \textit{rbs}. On the other hand, if the predicted sounding objects are the same as the user-selected one, \textsc{Peanut} will further examine whether a right-bound frame exists. If \textit{rbs} is not empty, \textsc{Peanut} will calculate the similarity between \textit{cur} and \textit{tf}, where \textit{tf} is the frame at the top of \textit{rbs}. \textsc{Peanut} will remove \textit{tf} from \textit{rbs} if the similarity holds and automatically interpolate the annotation results for the intermediate frames in between (Algorithm \ref{alg:populate}). After that, \textsc{Peanut} continues to compare \textit{cur} and the new frame at the top of \textit{rbs} until \textit{rbs} becomes empty. If the similarity does not hold, \textsc{Peanut} will ask the human annotator to annotate the frame $\floor{\frac{cur+tf}{2}}$. 

Otherwise, if the predicted sounding objects are the same as the user-selected ones and no right bound frame exists, \textsc{Peanut} will move to the first frame that follows the farthest human-annotated frame \textit{hf} and has a significant auditory or visual change compared to \textit{hf} (Algorithm \ref{alg:farthest}). If there is no significant auditory or visual change in the next $k$ frames (we used $k=10$ in our implementation of \textsc{Peanut}), \textsc{Peanut} will ask the human annotator to annotate the frame $hf+k$.

\subsubsection{Detecting visual and auditory changes}
\label{sec:algo_key_frame}

When a frame has a significant visual or auditory change compared to the last human-annotated frame denoted \textit{src}, \textsc{Peanut} needs a human annotator to annotate it. A frame denoted \textit{target} is considered to have a significant visual change relative to \textit{src} if (1) the number of detected objects varies between these two frames or (2) a bounding box in \textit{src} does not have a correspondence in \textit{target}, where the correspondence establishes if, given a bounding box \textit{i} on left, there is a bounding box in \textit{target} that overlaps \textit{i} and the overlapping area satisfies the Condition~\ref{cond:1}, where overlapping parameters $\alpha$ is 0.8 and $\beta$ is 0.05. The overlapping threshold decreases as the time difference between the target and source frames increases.

  \begin{equation}
    \label{cond:1}
    \begin{aligned}
        & OverlapArea(B_{highest}, A_i)  > \\ & [\alpha - (Index_{target}-Index_{src}) \times \beta] \times Area(A_i)
    \end{aligned}
    \end{equation}

The detection of significant auditory changes is based on an audio tagging model. \textsc{Peanut} uses a state-of-the-art pre-trained Audio Neural Networks~\cite{Kong2020PANNsLP} to predict the audio tags for each frame. An audio tag describes the possible type (e.g., train horn, race car, truck, as shown in Figure~\ref{fig:architecture}) of the sound corresponding to that frame.

\cmt{
    \subsubsection{Predicting the sounding objects for unannotated frame}
    \label{sec:predictNext}
    
    Given a human-annotated frame denoted as \textit{Src}, \textsc{Peanut} can automatically annotate the following frame \textit{Target}. For each bounding box $A_i$ selected in \textit{Src}, \textsc{Peanut} scores each candidate bounding box $B_j$ in \textit{Target} by adding up the distance of their center coordinates and the width and height difference. \textsc{Peanut} will inherit the tag and annotate the box of the highest score as a sound source in \textit{Target} if the overlapping area between $B_{highest}$ and $A_i$ satisfies:
    
    \begin{equation}
    \label{cond:1}
    \begin{aligned}
        & OverlapArea(B_{highest}, A_i)  > \\ & [\alpha - (Index_{target}-Index_{src}) \times \beta] \times Area(A_i)
    \end{aligned}
    \end{equation}
        
    \noindent where overlapping thresholds $\alpha$ is 0.8 and $\beta$ is 0.05. These thresholds decrease as the time difference between the target and the source frame grows; such thresholds accommodate the accumulated spatial offset of a sound source across a series of frames.  Otherwise, \textsc{Peanut} will copy $B_i$ to \textit{Target} and treat it as the sound source.
}

\subsubsection{Tackling complex audio-visual scenarios}

\crhighlight{Audio-visual data often inherently possess various complexities. For example, there might be situations where multiple sound-producing objects are active at the same time, making it challenging to discern the specific source of a sound. Besides, some sounds may start at different times and overlap with each other. Additionally, the video may lack a visual indicator of the sounding object, which could complicate the correlation between two modalities. To tackle those complexities, our algorithm focuses on identifying discrepancies or ambiguity in the auditory or visual modality, and asks humans to annotate keyframes. For example, in cases of simultaneous sounds from multiple objects, \textsc{Peanut} requests human assistance when the audio-tagging model's confidence score is low, indicating auditory uncertainty. Also, the detection of sound changes prompts annotation at the frames with new sound sources. When sounding objects are not visible, \textsc{Peanut} invokes human intervention at the initial sound frame, predicting the same annotation for subsequent frames until significant changes occur (sound cessation or the introduction of a new sound). The ``jumpback'' (Algorithm \ref{alg:search}) solicits human input at the midpoint when successive key frames have inconsistent annotations due to sounding object changing with no visual cue. }

\crhighlight{Our approach echos prior research \cite{gebreegziabher2023patat, zhang2023visar, bobes2021improving, jarrahi2018artificial} in human-AI collaboration for the completion of complex tasks where machine learning models primarily target the automation of repetitive and mundane tasks, resorting to human assistance when the uncertainty score surpasses a threshold.  }

\subsection{Implementation}
\subsubsection{Web app}

The front-end web application of \textsc{Peanut} is implemented in React based on \textit{react-image-annotate}\footnote{\url{https://github.com/UniversalDataTool/react-image-annotate}}, an open-source framework for the development of image annotation tools and hosted using Python's built-in HTTP server. The back-end server is developed using the Flask framework with a MongoDB database that stores user annotations and log data.

\subsubsection{Object detector}
\label{sec:object_detector}
\textsc{Peanut} uses the off-the-shelf Detectron2 \cite{wu2019detectron2} library for implementing its object detector. The object detector is built using the Faster R-CNN~\cite{ren2015faster} neural network architecture with R101-FPN feature pyramid networks~\cite{lin2017feature} and is pre-trained on the domain-general MS-COCO dataset~\cite{lin2014microsoft}. The object detector does not need to be retrained when using \textsc{Peanut} on videos from a new topic domain. From the input of a video frame in bitmap format, the object detector can identify objects of 80 different types (e.g., ``train'', ``violin'', ``dog'', etc.) and return the corresponding coordinates of the bounding box and the type of each object. \looseness=-1

\subsubsection{Active visual sound grounding}
The Visual Sound Grounding (VSG) network model is trained to identify objects that make sounds among the candidate objects in each video frame. The network is built on top of~\cite{Tian_2021_CVPR}. The network is first pre-trained on the AVE dataset~\cite{tian2018audio}. Afterward, it is iteratively and incrementally fine-tuned with newly annotated data during the active learning stage. The network takes the feature of the current audio clip and objects that are proposed by the object detector (Section~\ref{sec:object_detector}) as inputs and predicts the likelihood that each object is associated with the sound, which allows the model to identify and remove bounding boxes that correspond to likely-silent objects from the potential candidates of bounding boxes.

 \looseness=-1

%% file: 4-Study.tex
\section{User Evaluation}
To evaluate \textsc{Peanut}, we conducted a user study\footnote{The study protocol has been reviewed and approved by the IRB at our institution.} with 20 users to compare the efficiency and accuracy of audio-visual data annotation using \textsc{Peanut} with those using a baseline system without intelligent features. The results of the study suggest that \textsc{Peanut} can help users annotate data for the sounding object localization task at a faster speed than in the baseline condition, while also improving the annotation accuracy at the same time. The user study also validated the usability of \textsc{Peanut} and provided insight into user reflections on their experiences using \textsc{Peanut}. 

% We conducted a user evaluation to assess the efficiency and usability of \hl{SystemName}, and user's annotation accuracy. \hl{\{\textit{assess model performance improvement or human-in-the-loop object detector improvement?}\}} 

% Specifically, our evaluation study aims (1) to see whether participants are able to annotate raw videos efficiently and correctly using \hl{SystemName}, (2) to gain insights into the advantages, disadvantages, and usability of \hl{SystemName}.

% In a within-subjects experiment, we compared the full features of \hl{SystemName} to a baseline version of the interface. The quantitative and subjective results were strongly in favor of the efficiency, usability and annotation accuracy supplied by \hl{StoryName} over the baseline version.

% \subsection{Baseline interface}

% Because there is no existing audio-visual data annotation tool, we created a baseline interface to compare with the \hl{SystemName}. The baseline interface has the same UI layout as the \hl{StoryName}, while it does not feature automatic object detection and sounding object recommendation. Therefore, users need to manually box and classify the sounding object(s) at each frame. The design of baseline interface is intended to ensure that the speed, accuracy, and usability difference between two conditions are mainly attributed to the presence of advanced features implemented in \hl{StoryName}.  

% \subsection{Dataset}

\subsection{Participants}

For this study, we recruited 20 participants through university mailing lists. 7 were undergraduate students, 8 were Master's students, 4 were doctoral students, and 1 was a high school student. Each participant was compensated with \$15 USD for their time.

Our participants had varied levels of prior data annotation experiences and ML backgrounds. 8 participants had no prior experience with data annotation and 12 had annotated data at least once. 5 participants had no ML background, 5 participants had at least a beginner level of ML knowledge (have taken introductory ML courses or had basic ML knowledge), 5 had an intermediate level of ML expertise (have taken advanced ML courses or with equivalent expertise), and 5 identified themselves as experts in ML (experienced researchers or practitioners of ML).

\subsection{Study Design}

Each user study session lasted around 70 minutes and was conducted remotely on Zoom due to the impact of the COVID-19 global pandemic. Before the beginning of each session, the participant signed the consent form and completed a demographic questionnaire. Participants accessed \textsc{Peanut} using the browser on their own computers and shared their screens with the experimenter. After receiving a 5-minute tutorial on how to use \textsc{Peanut}'s interface, each participant completed the annotation tasks under two conditions in random order (see Section~\ref{sec:condition}) and completed a post-study questionnaire on their perceived usability and usefulness of \textsc{Peanut}. The study session ended with a 10-minute semi-structured interview with the participant in which they reflected on their experience interacting with \textsc{Peanut}. All user study sessions were video recorded. 

\subsubsection{Dataset}
\label{sec:dataset}
% \begin{itemize}
%     \item AVE dataset
%     \item Frame sampling
%     \item Sampling of audio clips
%     \item Expert annotation
%     \item Data split for different condition groups
% \end{itemize}
In this study, we trained and evaluated \textsc{Peanut} using the Audio-Visual Event (AVE) dataset~\cite{tian2018audio}, which is a widely used benchmark dataset for the audio-visual localization task. The full AVE dataset contains 4,143 video clips from a wide range of topics and domains (e.g., ``Church Bell'', ``Male speech'', ``Dog Bark'') in 28 event categories. Each video clip in the AVE dataset is about 10 seconds long. We re-sampled each video at 8 FPS (a common practice in audio-visual data annotation so there are fewer frames to annotate in each video). For the user study, we used a sample of 30 video clips from the AVE dataset. \zzhighlight{The sampled dataset contains 10 different event categories such as music play, car race, male/female speech to investigate the effectiveness of \textsc{Peanut} on videos with a variety of topics.} Each of these 30 video clips was manually annotated by two experts as ground truth data for evaluating the accuracy of user annotation. Two authors, who were experts in audio-visual learning, annotated the ground truth data independently using the baseline Full Manual version of \textsc{Peanut}. We used cIoU (see Section \ref{sec:accuracy}) to measure the inter-annotator agreement. The average cIoU score between the annotation results by two experts is 0.96, suggesting a very high agreement between the two expert annotators.

\subsubsection{Conditions}
\label{sec:condition}
The study used a within-subject design, where each participant performed tasks under two conditions in random order. In the experiment condition (\textsc{Peanut}), the participant used the fully functional \textsc{Peanut} tool to label videos from a split of our sample dataset in 25 minutes. In the control condition (Full Manual), the participant used a baseline version of \textsc{Peanut} with all ``intelligent features'' (object detector, active visual sound grounding, and annotation interpolation) turned off to label the videos from the other split of our sample dataset in 25 minutes. The control condition reflected the essential practices of the current video or image data annotation tools \cite{labelme2016, xtract, dutta_2019_via}. The split of the videos between the two conditions and the order of the videos in each condition were randomized in each study session. In Table~\ref{tab:performance}, we also include the accuracy score of a ``Fully Automated'' model using the pre-trained VSG network as a baseline for annotation accuracy.

\subsubsection{Procedure}

In the study, participants were asked to annotate audio-visual data for sounding object localization using \textsc{Peanut} and the baseline tool. We randomized the order to control for learning effects. The study procedure consisted of three parts: a 30-minute session with the first interface, a 30-minute session with the second interface, and a 10-minute session for the post-study interview and a post-study questionnaire. In each 30-minute session, the experimenter started with a 5-minute tutorial teaching participants how to use the interface in the condition. Subsequently, participants had 25 minutes to annotate as many video frames as possible using the tool provided. After completing both sessions, each participant filled out a post-study questionnaire. The study session ended with a 10-minute semi-structured interview.

\begin{table}[t!]
    \def\arraystretch{1.2}
    \centering
    \begin{tabular}{|c|c|c|c|}
    \hline
      & \textbf{Average SoC $\downarrow$} &\textbf{ \# of Frames $\uparrow$} & \textbf{cIoU $\uparrow$} \\
     \hline
        Full Automated & N/A & N/A & 0.33\\
        Full Manual & 7.73 & 169.45 & 0.72\\
        \textsc{Peanut} & 5.12 & 488.85 & 0.93\\ \hline
    \end{tabular}
    \caption{Statistics of participants' performance in control (Full Manual) and experiment (\textsc{Peanut}) conditions. \zzhighlight{SoC means second of completion per frame, \# of frames means the total number of frames that a participant annotates in the condition, cIoU means consensus intersection over union, which is a well-accepted measure for the accuracy of bounding box annotation.} The cIoU accuracy of the fully automated model is also provided as a reference.}
    \label{tab:performance}
\end{table}

\subsection{Results on Annotation Performance}
We expect \textsc{Peanut} to accelerate the annotation task for participants in two ways: (H1) the user can annotate each frame faster because of the model-suggested bounding boxes of detected visual objects and the predicted audio tags; (H2) the user needs to manually annotate fewer frames due to the visual-audio-sensitive binary search process.

As shown in Table~\ref{tab:performance}, we report three statistics. The average \zzhighlight{second of completion (SoC)} validates H1, and the \# of Frames validates the combined effect of H1 and H2. cIoU measures
the impact of using \textsc{Peanut} on the accuracy of the annotated data. \zzhighlight{When reporting those statistics, we also report the standard deviation among all users in the target condition.} We will explain each statistics and its results in this section.\looseness=-1

\subsubsection{Time to completion}

We calculated the average seconds of completion (Average SoC) on human-annotated frames. SoC measures the average time participants spent annotating each video frame (not including those automatically annotated by the model). The difference between the two conditions in SoC demonstrates the effectiveness of \textsc{Peanut} in reducing the effort and cognitive load in the annotation of individual frames, such as displaying the potential candidates of objects and removing silent objects. As shown in Table \ref{tab:performance}, the average SoC is 5.12 per frame \zzhighlight{($SD = 1.87$)} in the experiment condition and 7.73 per frame \zzhighlight{($SD = 2.23$)} in the control condition. The paired t-test showed that there is a significant difference between the average SoC under the two conditions ($p<0.01$), indicating that participants can annotate a frame faster with \textsc{Peanut} than with the baseline interface.

\subsubsection{The number of annotated frames}

We calculated the average number of frames that a participant annotated in a 25-minute session (\# of Frames). Note that the count includes the frames automatically annotated by \textsc{Peanut} in the experiment condition. In addition to capturing the effect of \textsc{Peanut} features reflected in average SoC, the difference in the average number of frames between two conditions also reflects the effectiveness of automatic annotation in \textsc{Peanut}---instead of annotating all the video frames as in the baseline condition, participants only need to annotate key frames identified by \textsc{Peanut} and verify automatic annotation in the experiment condition. As shown in Table \ref{tab:performance}, the average number of annotated frames is 488.85 \zzhighlight{($SD = 167.93$)} in the experiment condition, and 169.45 \zzhighlight{($SD = 68.26$)} in the control condition. The paired t-test showed that there is a significant difference between the average number of frames annotated under the two conditions ($p<0.001$), indicating that participants can annotate more frames with \textsc{Peanut} than with the baseline interface in a 25-minute session. \looseness=-1

% We found no significant difference in the average number of annotated frames among participants with different levels of ML expertise when using \textsc{Peanut}. Interestingly, users who had no prior annotation experience, on average, annotated more frames using \textsc{Peanut} (\zzhighlight{$AVG = 630.4, SD = 187.78$)} than those who had previously annotated data \zzhighlight{($AVG = 394.5, SD = 95.46$) ($p<0.05$)}. A possible explanation of this phenomenon is that experienced data annotations would be more careful with examining and validating model-generated results. We will leave the follow-up investigation of this phenomenon for future work.

\subsubsection{Annotation accuracy}
\label{sec:accuracy}

We used \textit{consensus intersection over union} (cIoU)~\cite{senocak2018learning} to assess the participants' annotation accuracy in each condition. cIoU is a common metric for quantitatively evaluating the accuracy of bounding box annotations. Given a video frame, cIoU assigns scores to each pixel according to the consensus of multiple expert annotations. In specific, the ground-truth bounding boxes annotated by experts are first converted into binary maps $\{\textbf{b}_j\}^{N}_{j=1}$, where N is the number of expert annotators. Then, we calculate a representative score map $\textbf{g}$ from $\{\textbf{b}_j\}$ considering the consensus of experts:  

\begin{equation}
    \textbf{g} = min(\sum_{j=1}^{N} \frac{\textbf{b}_j}{\#consensus}, 1)
\end{equation}

\noindent where $\#consensus$ is a parameter indicating the minimum number of expert annotations to reach agreement. Since we have two expert annotators, we set $\#consensus$=1 by the majority rule in our study. Given this weighted score map $\textbf{g}$ and participant's annotation \textbf{$\alpha$}, we define cIoU as:

\begin{equation}
    cIoU = \frac{\sum_{i \in A} \textbf{g}_i}{\sum_i \textbf{g}_i + \sum_{i \in A-G} 1}
\end{equation}

\noindent where \textit{i} indicates the pixel index of the map, $A = \{i | \alpha_i = 1 \}$ means the set of pixels that the participant annotates, $ G = \{i | g_i > 0\}$ means the set of pixels in the weighted ground truth annotation.\looseness=-1

We calculate the cIoU score for each annotated frame of each participant. The average cIoU score for all annotated frames (all by human) in the control condition is 0.72 \zzhighlight{($SD = 0.14$)}, and is 0.93 \zzhighlight{($SD = 0.09$)} for all annotated frames (by human or by the system) in the experiment condition. A paired sample t-test showed a significant difference between the average cIoU under the two conditions ($p<0.001$). For comparison, the cIoU of a fully-automated VSG (off-the-shelf pre-trained without any human annotation) on the same sample of videos is 0.33 \zzhighlight{($SD = 0.08$)}. The paired t-test shows that the cIoU for the fully-automated model is significantly lower ($p<0.001$) than both human-annotated conditions (Full Manual and \textsc{Peanut}). The result indicates that the use of \textsc{Peanut} can achieve a high annotation accuracy (and even higher than the full-manual control condition in our experiment).

We suspect two possible reasons for the observed improvement in accuracy in the \textsc{Peanut} condition compared to the Full Manual condition: (1) The use of object detectors may have improved consistency in selecting the bounding boxes. The annotation of some sounding objects can be inherently ambiguous---for example, when a sounding object is a person playing the violin, should the annotator draw a box for the person (that includes the violin) or only the violin? The labeling of situations like this can be inconsistent in the Full Manual condition (e.g., the violin is selected in some frames, while the person is selected in other frames). In this example situation, the objector detector would consistently go with the person because it prefers larger boxes when choosing from two overlapped ones (which is often consistent with the common best practice of sounding object localization), reducing the bounding box inconsistency in data annotation. (2) Drawing accurate bounding boxes is a challenging task on its own. The bounding boxes drawn manually by users often do not align as accurately with the edges of the object as the model-detected boxes.  Moreover, because data annotation with the baseline interface is a highly repetitive and tedious process, the precision of participants' manually created bounding boxes could fluctuate due to fatigue (e.g., not modifying a box when an object has moved ``a little bit'', but still aligns mostly with the box). We plan to further investigate the factors that contribute to the improvement of accuracy as a future work direction.\looseness=-1

\subsubsection{The impact of user expertise}

% \tlhighlight{We found no significant difference in annotation accuracy between participants who had no prior annotation experience and participants who had previously annotated data. We found no significant difference in annotation accuracy among participants with different levels of ML expertise, either.}

\zzhighlight{Comparing the annotation experience and efficiency, users without prior annotation experience annotated faster than those with prior annotation experience in the \textsc{Peanut} condition. The average numbers of frames that participants with and without data annotation experience annotated in the experiment condition were 394.5 ($SD = 95.46$) and 630.4 ($SD = 187.78$) respectively. An unpaired t-test shows that participants without data annotation experience annotated significantly more frames than those with annotation experience ($p=0.043<0.05$). Participants without annotation experience also on average spent less time on each frame ($AVG = 4.35s, SD = 0.75$) than those with annotation experience ($AVG = 6.26s, SD = 0.58$) ($p=0.031<0.05$). There was no significant difference between their average number of frames in the Full Manual condition. The efficiency improvement from using \textsc{Peanut} (comparing \# of Frames between \textsc{Peanut} and Full Manual conditions) is significant for both groups.}

\zzhighlight{In terms of ML expertise, there was no significant difference in all three metrics (Average SoC, number of Frames, cIoU) among groups of participants with different levels of ML expertise in a one-way ANOVA test. We found no significant difference in annotation accuracy between participants who had no prior annotation experience and those who had prior annotation experience, either.}

% the average of total frames finished by participants with no ML background, beginner, intermediate and expert ML level were 732.2, 812.4, 405.4, and 683.2 respectively. The one-way ANOVA test did not find statistical significant difference among them. As to SoC, these four groups on average took 6.67s, 7.79s, 8.89s, and 5.34s respectively, and no statistically significant difference was found among them. We neither found statistically significant difference for cIoU among different ML expertise levels.}

\zzhighlight{For the observed difference in efficiency between groups with and without prior annotation experiences, a possible explanation is that users with prior data annotation experience may be more skeptical to automated annotation and thus spent more time double-checking the results. We plan to investigate this phenomenon more closely in future studies.}\looseness=-1

% \hl{There are two possible reasons accounting for the accuracy increase}: First, since we used a manually optimized object detector, the AI-recommended bounding boxes usually cover whole sounding objects which are consistent with our experts’ annotations. However, certain participants sometimes might draw smaller sounding regions due to the ambiguity of labeling boxes. Second, the precision of one participant's annotation could fluctuate in the study due to the fatigue or their implicitly varied standard for drawing an appropriate bounding box.

% We also analyzed the sources of annotation errors raised in the experiment condition by reviewing the study recordings. We found that most of errors still arose from the participants' overreliance on \textsc{Peanut}. For example, participants sometimes ignored the wrong \textsc{Peanut} annotation as they kept clicking \textit{Next} button fast. Besides, when clicking \textit{Next Label} button, some participants did not check the automatic annotation of the intermediate frames properly though \textsc{Peanut} made mistake. Nevertheless, many participants were able to realize the limitation of the automatic annotation via the frame-by-frame thumbnail and annotation playback preview at the end of each video annotation session, and manually modified the incorrect annotations by themselves.

\begin{figure*}[t!]
  \centering
  \includegraphics[width=\linewidth]{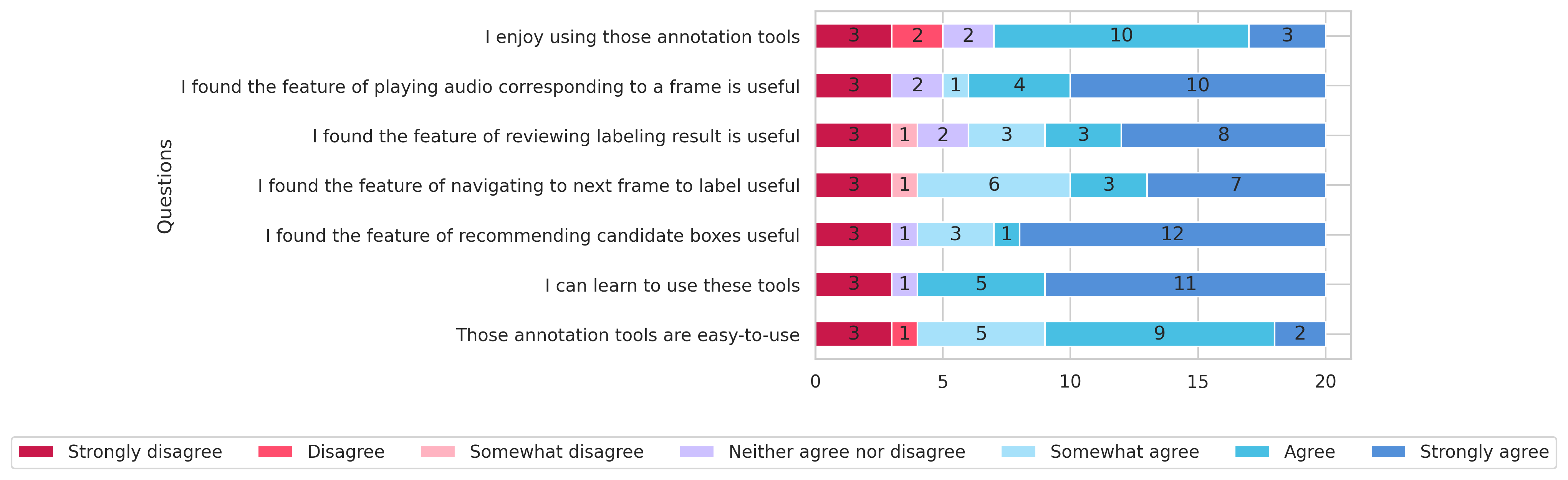}
  \caption{Results of the post-study questionnaire}
  \label{fig:questionnaire}
\end{figure*}

\subsection{Results on User Behaviors and Experiences}

\subsubsection{Usage statistics}
We analyzed the interaction logs and screen recordings of participants in the experiment (\textsc{Peanut}) condition to understand how participants interacted with \textsc{Peanut}'s AI assistance, especially on the ratio between manual vs. automated annotation and how often they edit the automated annotation results. As we see in Table~\ref{tab:performance}, each participant, on average, annotated 488.85 frames in a session. Among them, an average of 37.2 (7.6\%) frames are manually annotated by the participant, 421.7 (86.2\%) are automatically annotated without any user modification, 29.9 (6.1\%) are first automatically annotated by the model and then modified by the user. In each session, a participant on average resized/moved the model-predicted bounding boxes of visual objects in 12.6 frames (2.6\%), created new bounding boxes for visual objects in 14.2 frames (2.9\%), and edited the model-predicted audio tags in 3.1 frames (0.6\%).\looseness=-1

\zzhighlight{Among the annotations made in the \textsc{Peanut} condition, the average cIoU of the 13.8\% human annotated frames is 0.81 ($SD = 0.12$) and average cIoU of the 86.2\% automated annotated frames is 0.96 ($SD = 0.07$). This result indicates that with a small portion of user annotation on mostly ``key frames'' identified by our algorithm, the model can get very accurate at automated annotations (compared to the average cIoU of 0.33 in the Full Automated condition), indicating the effectiveness of \textsc{Peanut}'s active learning pipeline.}

\subsubsection{Post-study questionnaire}

In a post-study questionnaire, we asked each participant to rate seven statements about the usability, usefulness, and user experience of \textsc{Peanut} on a 7-point Likert scale from ``strongly agree'' to ``strongly disagree''. The results are shown in Figure \ref{fig:questionnaire}. Specifically, \textsc{Peanut} scored on average 4.9 ($SD = 1.92$) on ``\textit{\textsc{Peanut} is easy-to-use}'', 5.7 ($SD = 2.10$) on ``\textit{I can learn to use \textsc{Peanut} easily}'', 5.6 ($SD = 2.19$) on ``\textit{I found the feature of recommending candidate boxes is useful}'', 5.15 ($SD = 2.08$) on ``\textit{I found the feature of navigating to next frame to label is useful}'', 5.15 ($SD = 2.16$) on ``\textit{I found the features of reviewing annotation result are useful}'', 5.5 ($SD = 2.16$) ``\textit{I found the feature of playing audio corresponding to a frame is useful}'', and 4.8 ($SD = 2.11$) on ``\textit{I enjoy using \textsc{Peanut}}''. The results indicate that our participants generally found \textsc{Peanut} easy to use, and the design features are useful for their annotations.

When a participant rated a statement lower than ``agree'', the experimenter would take a note and ask the participant about the specific issues they had encountered and solicit their suggestions for addressing these issues in the interview. We will report on these findings in Section \ref{sec:interview} below.

\subsubsection{User experiences, challenges, and feedback}
\label{sec:interview}

In the interview, we discussed with the participants about their post-study questionnaire responses, the difficulties they encountered when using \textsc{Peanut}, the different user experiences of annotating data in two conditions, and their suggestions for the design of \textsc{Peanut}. The leading author first coded all the interview transcripts independently and discussed the codebook with another author to reach a consensus. The unified codebook was then used by the other author to code all the interview transcripts independently again to validate the result. The overall inter-rater reliability is 0.72.

Most of the reported difficulties originated from the default positions and sizes of the canvas and the annotation box. Due to an implementation bug, the video frames would be shown in a smaller window in the upper left corner and users ``\textit{had to drag and resize them every time}'' (P13). The panel for predicted bounding boxes may also appear outside the frame sometimes, so the user has to drag it back to the current view. Both of these implementation issues can be easily fixed in new versions of \textsc{Peanut}. Besides, after carefully watching the video multiple times, participants still sometimes had trouble recognizing the sound type or locating the source due to ``\textit{the background noise, ambiguity of the sound, and existence of multiple objects of a similar type}'' (P15). Lastly, a few participants found the difference between the \textit{Next} button and \textit{Next Label} confusing, especially in videos that had rapid visual or auditory changes, for which \textsc{Peanut} tends to conservatively recommend the immediate next frame when participants clicked the \textit{Next Label} button, which could confuse the user (P13).

When asked to describe the pros and cons of the AI-assisted \textsc{Peanut} system compared to the baseline, the 20 participants mentioned that \textsc{Peanut} significantly accelerated their annotation process and reduced their workload. Participants thought ``\textit{\textsc{Peanut} saved their time from doing repetitive and tedious manual labeling, especially when frames varied little once a time}'' (P6). With the assistance of \textsc{Peanut}, they only need to ``\textit{focus on a handful of key frames}'' (P10) or ``\textit{keep clicking the Next button to oversee the automatic annotation frame-by-frame}'' (P12). In addition, some participants thought that the systems recommended more accurate bounding boxes than what they can manually create, thus they did not need to ``\textit{struggle with creating precise boxes manually}'' (P13). Compared to the baseline tool, \textsc{Peanut} ``\textit{makes the annotation process much less exhausting}'' (P13).  On the weakness of \textsc{Peanut}, participants thought that with the \textit{Next Label} button they may easily overtrust the system from the beginning, and ``\textit{merely realized the wrong or missing annotation when reviewing back at the end}'' (P4).\looseness=-1

\crhighlight{With regard to human agency, participants thought the system ``\textit{allow them to jump in and take control back at any time when they feel the AI model start going wrong}'' (P16). P9 suggested that ``\textit{the system should provide more information or clear warning to them so that they can better notice the issue and decide when to get in}''. Moreover, as to their reliance on AI models, some participants said they ``\textit{tend to trust AI more when the sounding object was obvious in the video and AI correctly labelled them for consecutive frames}'' (P10). In contrast, they ``\textit{would choose to do it on their own when sounding object changes a lot}'' (P5). }

For the design of \textsc{Peanut}. P4 suggested enabling keyboard shortcuts for frequently used operations such as drawing, deletion, and frame switch. Furthermore, some participants thought "\textit{the algorithm behind the \textit{Next Label} button should be more robust}" (P15) especially ``\textit{when the scene contains complex objects or has a large variation between the frames}'' (P12).\looseness=-1

%% file: 5-Discussion.tex
\section{Discussion and Implications}
In this section, we discuss several novel interaction strategies that we used in the design of \textsc{Peanut}, the lessons we learned in the user study,  and their implications for human-AI collaboration.

\subsection{Connecting and Unifying Different Data Modalities}
\zzhighlight{The design of \textsc{Peanut} introduces a strategy for human-AI collaboration in multi-modal tasks that demand high perceptive and cognitive overhead to process and associate input from different modalities.} \textsc{Peanut} uses two single-modal models that can provide partial automation. The human user then contributes by (1) validating the partial automation result; (2) associating the partial automation result of one model from one input modality with that of a different model from a different input modality to achieve the multi-modal end goal. Many participants found this partial automation approach effective for reducing cognitive load and fatigue when annotating a large number of similar frames. For example, P8 said ``\textit{I almost fell asleep when using the first interface (baseline). In comparison, the second one (\textsc{Peanut}) allowed me to focus on those (frames) worth my effort so that I did not need to do a tedious, repeated annotation for every picture}''. \looseness=-1

This new interaction strategy represents the adoption and development of classic theories in multi-modal interfaces~\cite{oviatt_ten_1999, oviatt_2000_perceptual} in a new domain. Instead of an interface that processes user input data in multiple (usually complementary) modalities and tries to better understand the input from one modality using the input from another modality (mutual disambiguation~\cite{oviatt_mutual_1999}), \textsc{Peanut} works in the reversed direction, utilizing humans' perceptive and cognitive capabilities of understanding inputs from one modality using inputs from another modality in order to ground auditory data to visual data. \crhighlight{The strategy discussed in this paper and design implications from the study could translate to other multi-modal data annotation tasks like natural language visual grounding~\cite{anayurt2019} visual question answering~\cite{antol2015vqa}. Besides,} we expect this interaction strategy to be useful in other application domains beyond data annotation, such as helping users with visual or hearing impairments better understand videos as discussed in Section~\ref{sec:future_work}.

% \tlcomment{talk about the connection of this method to classic multi-modal paper, and the applications}

% \tlcomment{implications on combining audio and visual modalities in data annotation -- how purely visual algorithms (e.g., object detection) and purely auditory algorithms (what we used for detecting changes in audio?) can assist in annotations in complementary modalities}

% \tlcomment{Represents a new paradigm of partial automation -- partial automation by modality}

\subsection{Minimizing the Overhead Cost of AI in Human-AI Collaboration}
When we design the user workflow of \textsc{Peanut}, an important goal is that the use of \textsc{Peanut} should not introduce additional burdens or learning barriers compared to what the user would experience if they manually annotated the same data. The user does not need to learn a new skill or make any configuration of \textsc{Peanut} before starting to use it. Anything the user does in \textsc{Peanut} is either a sub-process of what they would need to do in manual annotation (e.g., choosing the sounding object from several candidate boxes instead of identifying the sounding object in the video and drawing a box for it) or the same process but less repetitive (e.g., annotating for only the key frames instead of all the frames). This is different from many other human-AI collaboration scenarios, where the use of AI incurs significant overhead, requiring justifying the use of AI by assessing whether the benefit exceeds the cost.

To achieve this goal, we used a strategy of keeping the user's original workflow as much as possible. For example, \textsc{Peanut} can identify the next key frame that the user can label (Section~\ref{sec:automatic_annotation}). This intelligent feature does not require any additional input from the user, as it relies on only the user's annotation result for the current frame and the current state of the frame sequence. The user's interaction with the system also remains the same as in a manual labeling tool---they click on the ``Next'' button to work on a different frame after finishing annotating the current one with the only difference being that the ``next'' frame is no longer necessarily the $(n+1)^{th}$ frame after the $n^{th}$ frame. The user retains the ability to freely move along the sequence of frames, edit previously annotated frames, or annotate frames out of the ``recommended'' order as they wish. \textsc{Peanut} will automatically track the next recommended frame to be annotated without requiring user intervention, allowing users to retain control and agency in the human-AI collaborative data annotation process. The participants confirmed the effectiveness of human-AI collaboration in the interview. For example, P5 said that ``\textit{I felt it is smooth to work with the algorithm behind \textsc{Peanut}. It takes care of many simple annotations that had to be done by myself in the previous fully manual version}''.

Another common type of ``cost'' in human-AI collaboration that we address in \textsc{Peanut} is the degraded accuracy due to users' overreliance on AI, which has been identified as a key issue in AI-assisted data annotation~\cite{askhtorab_aiassisted_2021}. When the audio-visual sound grounding model identifies candidate objects and removes likely silent objects from the candidates, we still expect the user to maintain their attention and edit the results if needed.  To address this challenge, \textsc{Peanut} provides video playback and thumbnail preview features (Section~\ref{sec:annotation_review}) that allow the user to quickly validate the result and identify annotation issues if there are any. In the user study, these features indeed helped participants locate incorrect automatic annotations and consequently calibrate their trust in AI automation. For example, P14 said that ``\textit{I totally trusted the system to label the frames at the beginning. However, as I looked at the review of the annotation result, I realized the system could make mistakes and I have to go back to improve the annotation}''.

\cmt {
    \tlcomment{Goals: no additional burden/learning barrier for the user -- the user doesn't need to do anything extra or learn any new tasks compared with manual annotation}
    
    \tlcomment{Fulfill a secondary agent initiative (locating key frames with changes) through the manipulation of human annotation order using a binary search approach}
    
    \tlcomment{Perhaps discuss the overreliance issue in human-AI collaboration in labeling~\cite{askhtorab_aiassisted_2021} -- how designs, e.g., video review address the overreliance issue}
}

\subsection{The Role of Partial Automation in Pursuit of Full Automation}
The traditional workflow of data annotation for ML models represents an approach that goes from \textit{full manual} efforts directly to \textit{full automation}---human users go through a fully manual process to create a dataset, which is then used to train a fully-automated ML model. In contrast, \textsc{Peanut}'s approach highlights the role of incrementally-trained partial-automation models that can bridge the two ends in pursuit of full automation.

In \textsc{Peanut}'s model, as the user is annotating the data in a manual process, partial-automation models are incrementally trained with the user's incomplete annotation of each frame. Thank to the characteristic of video data that frames in the same video are usually similar to each other, domain-general partial-automation models can quickly adapt to the specific domain of the video in a \textit{few-shot} fashion. The active learning process guided by the key frame selections (Section~\ref{sec:algo_key_frame}) in \textsc{Peanut} incrementally improves the performance of partial-automation models as the user annotates more frames, reducing human efforts to reach the data size and data quality required for training an end-to-end fully-automated model. During the interview, P4 commented: ``\textit{One feature I can imagine is that, with my annotation on a few frames at the beginning of a video, the system can learn to label the following frames or even other videos with similar content and do the remaining annotation on behalf of me}''. \looseness=-1

We expect that such an approach can be useful for a variety of other human-in-the-loop ML applications. An example is the human/ML hybrid sensing approach such as Zensor~\cite{laput_2015_zensors} where sensors can switch between crowd intelligence and ML to adapt to environmental changes. However, unlike Zensor which toggles between either full automation or full manual for the primary sensing task and uses human annotation results as a validation method, \textsc{Peanut}'s approach allows for more flexible partial-automation states in between, taking advantage of the partially annotated data to accelerate the annotation before a fully automated model is ready.

\subsection{Mitigating Biases in AI-Assisted Data Annotation}

\zzhighlight{While the issue of biases is less prominent in task domain of \textsc{Peanut} since unlike many other domains vulnerable to subjective bias (e.g., hate speech detection \cite{Mollas2020ETHOSAM}), the result of sounding object localization does come with an objective truth, mitigating the biases of AI and human annotators in this task is still an important consideration. Presumably, the addition of AI assistance in data annotation could introduce or amplify two kinds of biases in annotation.} 

\zzhighlight{First, AI models in \textsc{Peanut} may introduce intrinsic biases to annotation results~\cite{Dar2021AFT, Leavy2020MitigatingGB, Blanzeisky2021AlgorithmicFI}. These biases originated from the dataset on which these models are pre-trained on. To address them, \textsc{Peanut} identifies key frames with previously unseen objects and proactively requests human annotation in these key frames. On those key frames, the role of AI is to \textit{assist} with human annotation by suggesting label candidates for user to use rather than attempting to fully automating their annotations using the pre-trained model. The design allows users to take control and accountability of annotations on keyframes and offsets AI biases with human judgement.}

\zzhighlight{In addition to introducing biases themselves, AI models may also amplify human errors when performing automated annotation based on previous human annotations \cite{Ahmed2021AttenuationOH, Chen2021UnderstandingAM, Bhargava2019ExposingAC}. For example, when there are multiple guitars in a key frame in the scene, the human annotator may have difficulty in identifying which guitar is currently making the sound  and select a wrong guitar as the sound source. In this case, the object detector in \textsc{Peanut}  will propagate this wrong annotation to automated annotations in subsequent frames. The playback reviews (Section~\ref{sec:annotation_review}) could be useful to mitigate this issue by providing a global context that facilitates users to identify errors in continuous scenes.}

\zzhighlight{We also expect to introduce other bias-reducing strategies in the future version of \textsc{Peanut}. For example, we may implement assistance functions to support human decision-making of the sound source at key frames when the human annotator is unsure. For example, \textsc{Peanut}  may leverage the model-inferred depth and direction information of the sound to indicate the likelihood of each visual region in the frame containing the sound source object. \looseness=-1}

\section{Limitations \& Future work}
\label{sec:future_work}

\subsection{Incorporating Speech and Natural Language Models}
The current version of \textsc{Peanut} only supports the use of auditory and visual models, but not natural language understanding (NLU) models that process the content of speech in videos. Speech is one of the most ubiquitous sound sources in videos. Human speech usually contains important semantic information relevant to the surrounding audio and visual scenes in physical environments. Moreover, unlike other sounds (\emph{e.g., traffic, rain, dog barking}), speech can be transcribed into text using automatic speech recognition~\cite{hinton2012deep}. Benefitting from advances in NLU~\cite{schmitz2012open,berant2014semantic,chen2015event,song2018graph}, we can automatically extract structured and abstracted semantic content from transcribed text. This could provide opportunities for us to incorporate NLU models with \textsc{Peanut} to facilitate more powerful multimodal data collection to support multidisciplinary research across audio, visual, and language.

\subsection{Expanding to Audio-Visual Tasks beyond Sounding Object Localization}
 We implemented and evaluated the current version of \textsc{Peanut} in the context of the sounding object localization task. With its support for user interaction with both audio and visual modalities, \textsc{Peanut} can be easily expanded to a range of multi-modal audio-visual tasks, such as audio-visual event localization~\cite{tian2018audio,wu2019DAM}, audio-visual video parsing~\cite{tian2020unified,wu2021exploring}, and audio-visual video captioning~\cite{tian2018attempt,rahman2019watch}. 

Audio-visual event localization aims to temporally localize audio-visual events\footnote{Audio-visual events are synchronized video segments in which the sound sources are visible and their sounds are audible.} and recognize the categories of events. Toward more unified multi-sensory perception, audio-visual video parsing aims to recognize event categories bind to sensory modalities and find temporal boundaries of when such an event starts and ends. To train models for addressing these two tasks, two datasets: AVE~\cite{tian2018audio} and LLP~\cite{tian2020unified} have been collected, respectively. Due to the lack of efficient annotation tools, only second-level temporal boundaries with the corresponding categories are fully manually annotated in these datasets. We believe that more precise frame-level annotations will enable more accurate audio-visual ML models and facilitate the development of future research.

In addition to facilitating existing tasks, our system has the potential to help researchers investigate new problems. For example, by collecting temporal boundary, object box, and category annotations, we can formulate a new space-time audio-visual parsing task that aims to perform spatio-temporal multi-modal analysis over videos to predict temporal boundaries of audio, visual, and audio-visual events, their associated semantic categories, and spatially localized sounding objects.   

\subsection{Release and Deployment}
We plan to release the \textsc{Peanut} tool for public use. We are also currently planning a large-scale deployment to complete the annotation of all 4,143 video clips in the AVE dataset~\cite{tian2018audio} for sounding object localization. \cmt{The deployment will use Amazon Mechanical Turk workers as human annotators, which can help us understand \textsc{Peanut}'s performance with a big group of users without any ML background and annotation experience} The annotation result on the full AVE dataset will allow us to train a new \textit{supervised} sounding object localization model with the dataset, compare its performance with the current state-of-art model, and illustrate \textsc{Peanut}'s effectiveness in improving the performance of ML models by enabling the creation of better-quality annotated datasets.
    
    % \tlcomment{use implications of this system on tools to help users with visual/hearing impairment to understand videos}

\section{Conclusion}
In this paper, we presented \textsc{Peanut}, a human-AI collaborative audio-visual annotation tool for improving the data annotation efficiency of the sounding object localization task. A controlled user study of \textsc{Peanut} demonstrated that a human-AI collaborative approach with several new mixed-initiative partial-automation strategies can enable human annotators to perform the data annotation task faster while maintaining high accuracy. Our findings provide design implications for AI assistance in data annotation as well as human-AI collaboration tools for working with multi-modal data. \looseness=-1

%% file: main-paper.bbl

\begin{thebibliography}{126}

%%% the Web address.

\ifx \showCODEN    \undefined \def \showCODEN     #1{\unskip}     \fi
\ifx \showDOI      \undefined \def \showDOI       #1{#1}\fi
\ifx \showISBNx    \undefined \def \showISBNx     #1{\unskip}     \fi
\ifx \showISBNxiii \undefined \def \showISBNxiii  #1{\unskip}     \fi
\ifx \showISSN     \undefined \def \showISSN      #1{\unskip}     \fi
\ifx \showLCCN     \undefined \def \showLCCN      #1{\unskip}     \fi
\ifx \shownote     \undefined \def \shownote      #1{#1}          \fi
\ifx \showarticletitle \undefined \def \showarticletitle #1{#1}   \fi
\ifx \showURL      \undefined \def \showURL       {\relax}        \fi
% invisible to TeX
\providecommand\bibfield[2]{#2}
\providecommand\bibinfo[2]{#2}
\providecommand\natexlab[1]{#1}
\providecommand\showeprint[2][]{arXiv:#2}

\bibitem[\protect\citeauthoryear{Ahmed, Athyaab, and Muqtadeer}{Ahmed
  et~al\mbox{.}}{2021}]%
        {Ahmed2021AttenuationOH}
\bibfield{author}{\bibinfo{person}{Saad~Bin Ahmed}, \bibinfo{person}{Saif~Ali
  Athyaab}, {and} \bibinfo{person}{Shaik~Abdul Muqtadeer}.}
  \bibinfo{year}{2021}\natexlab{}.
\newblock \showarticletitle{Attenuation of Human Bias in Artificial
  Intelligence: An Exploratory Approach}.
\newblock \bibinfo{journal}{\emph{2021 6th International Conference on
  Inventive Computation Technologies (ICICT)}} (\bibinfo{year}{2021}),
  \bibinfo{pages}{557--563}.
\newblock


\bibitem[\protect\citeauthoryear{Alamri, Cartillier, Das, Wang, Cherian, Essa,
  Batra, Marks, Hori, Anderson, et~al\mbox{.}}{Alamri et~al\mbox{.}}{2019}]%
        {alamri2019audio}
\bibfield{author}{\bibinfo{person}{Huda Alamri}, \bibinfo{person}{Vincent
  Cartillier}, \bibinfo{person}{Abhishek Das}, \bibinfo{person}{Jue Wang},
  \bibinfo{person}{Anoop Cherian}, \bibinfo{person}{Irfan Essa},
  \bibinfo{person}{Dhruv Batra}, \bibinfo{person}{Tim~K Marks},
  \bibinfo{person}{Chiori Hori}, \bibinfo{person}{Peter Anderson},
  {et~al\mbox{.}}} \bibinfo{year}{2019}\natexlab{}.
\newblock \showarticletitle{Audio visual scene-aware dialog}. In
  \bibinfo{booktitle}{\emph{Proceedings of the IEEE/CVF Conference on Computer
  Vision and Pattern Recognition}}. \bibinfo{pages}{7558--7567}.
\newblock


\bibitem[\protect\citeauthoryear{Alc{\'a}zar, Caba, Thabet, and
  Ghanem}{Alc{\'a}zar et~al\mbox{.}}{2021}]%
        {alcazar2021maas}
\bibfield{author}{\bibinfo{person}{Juan~Le{\'o}n Alc{\'a}zar},
  \bibinfo{person}{Fabian Caba}, \bibinfo{person}{Ali~K Thabet}, {and}
  \bibinfo{person}{Bernard Ghanem}.} \bibinfo{year}{2021}\natexlab{}.
\newblock \showarticletitle{Maas: Multi-modal assignation for active speaker
  detection}. In \bibinfo{booktitle}{\emph{Proceedings of the IEEE/CVF
  International Conference on Computer Vision}}. \bibinfo{pages}{265--274}.
\newblock


\bibitem[\protect\citeauthoryear{Amershi, Weld, Vorvoreanu, Fourney, Nushi,
  Collisson, Suh, Iqbal, Bennett, Inkpen, Teevan, Kikin-Gil, and
  Horvitz}{Amershi et~al\mbox{.}}{2019}]%
        {amershi_2019_guidelines}
\bibfield{author}{\bibinfo{person}{Saleema Amershi}, \bibinfo{person}{Dan
  Weld}, \bibinfo{person}{Mihaela Vorvoreanu}, \bibinfo{person}{Adam Fourney},
  \bibinfo{person}{Besmira Nushi}, \bibinfo{person}{Penny Collisson},
  \bibinfo{person}{Jina Suh}, \bibinfo{person}{Shamsi Iqbal},
  \bibinfo{person}{Paul~N. Bennett}, \bibinfo{person}{Kori Inkpen},
  \bibinfo{person}{Jaime Teevan}, \bibinfo{person}{Ruth Kikin-Gil}, {and}
  \bibinfo{person}{Eric Horvitz}.} \bibinfo{year}{2019}\natexlab{}.
\newblock \showarticletitle{Guidelines for Human-AI Interaction}. In
  \bibinfo{booktitle}{\emph{Proceedings of the 2019 CHI Conference on Human
  Factors in Computing Systems}} (Glasgow, Scotland Uk)
  \emph{(\bibinfo{series}{CHI '19})}. \bibinfo{publisher}{Association for
  Computing Machinery}, \bibinfo{address}{New York, NY, USA},
  \bibinfo{pages}{1–13}.
\newblock
\showISBNx{9781450359702}
\urldef\tempurl%
\url{https://doi.org/10.1145/3290605.3300233}
\showDOI{\tempurl}


\bibitem[\protect\citeauthoryear{Anayurt, Ozyegin, Cetin, Aktas, and
  Kalkan}{Anayurt et~al\mbox{.}}{2019}]%
        {anayurt2019}
\bibfield{author}{\bibinfo{person}{Hazan Anayurt}, \bibinfo{person}{Sezai~Artun
  Ozyegin}, \bibinfo{person}{Ulfet Cetin}, \bibinfo{person}{Utku Aktas}, {and}
  \bibinfo{person}{Sinan Kalkan}.} \bibinfo{year}{2019}\natexlab{}.
\newblock \showarticletitle{Searching for Ambiguous Objects in Videos using
  Relational Referring Expressions}. In \bibinfo{booktitle}{\emph{Proceedings
  of the British Machine Vision Conference (BMVC)}}.
\newblock


\bibitem[\protect\citeauthoryear{Antol, Agrawal, Lu, Mitchell, Batra, Zitnick,
  and Parikh}{Antol et~al\mbox{.}}{2015}]%
        {antol2015vqa}
\bibfield{author}{\bibinfo{person}{Stanislaw Antol}, \bibinfo{person}{Aishwarya
  Agrawal}, \bibinfo{person}{Jiasen Lu}, \bibinfo{person}{Margaret Mitchell},
  \bibinfo{person}{Dhruv Batra}, \bibinfo{person}{C~Lawrence Zitnick}, {and}
  \bibinfo{person}{Devi Parikh}.} \bibinfo{year}{2015}\natexlab{}.
\newblock \showarticletitle{VQA: Visual question answering}. In
  \bibinfo{booktitle}{\emph{Proceedings of the IEEE international conference on
  computer vision}}. \bibinfo{pages}{2425--2433}.
\newblock


\bibitem[\protect\citeauthoryear{Arandjelovic and Zisserman}{Arandjelovic and
  Zisserman}{2017}]%
        {arandjelovic2017look}
\bibfield{author}{\bibinfo{person}{Relja Arandjelovic} {and}
  \bibinfo{person}{Andrew Zisserman}.} \bibinfo{year}{2017}\natexlab{}.
\newblock \showarticletitle{Look, listen and learn}. In
  \bibinfo{booktitle}{\emph{Proceedings of the IEEE International Conference on
  Computer Vision}}. \bibinfo{pages}{609--617}.
\newblock


\bibitem[\protect\citeauthoryear{Arandjelovic and Zisserman}{Arandjelovic and
  Zisserman}{2018}]%
        {arandjelovic2018objects}
\bibfield{author}{\bibinfo{person}{Relja Arandjelovic} {and}
  \bibinfo{person}{Andrew Zisserman}.} \bibinfo{year}{2018}\natexlab{}.
\newblock \showarticletitle{Objects that sound}. In
  \bibinfo{booktitle}{\emph{ECCV}}.
\newblock


\bibitem[\protect\citeauthoryear{Ashktorab, Desmond, Andres, Muller, Joshi,
  Brachman, Sharma, Brimijoin, Pan, Wolf, Duesterwald, Dugan, Geyer, and
  Reimer}{Ashktorab et~al\mbox{.}}{2021}]%
        {askhtorab_aiassisted_2021}
\bibfield{author}{\bibinfo{person}{Zahra Ashktorab}, \bibinfo{person}{Michael
  Desmond}, \bibinfo{person}{Josh Andres}, \bibinfo{person}{Michael Muller},
  \bibinfo{person}{Narendra~Nath Joshi}, \bibinfo{person}{Michelle Brachman},
  \bibinfo{person}{Aabhas Sharma}, \bibinfo{person}{Kristina Brimijoin},
  \bibinfo{person}{Qian Pan}, \bibinfo{person}{Christine~T. Wolf},
  \bibinfo{person}{Evelyn Duesterwald}, \bibinfo{person}{Casey Dugan},
  \bibinfo{person}{Werner Geyer}, {and} \bibinfo{person}{Darrell Reimer}.}
  \bibinfo{year}{2021}\natexlab{}.
\newblock \showarticletitle{AI-Assisted Human Labeling: Batching for Efficiency
  without Overreliance}.
\newblock \bibinfo{journal}{\emph{Proc. ACM Hum.-Comput. Interact.}}
  \bibinfo{volume}{5}, \bibinfo{number}{CSCW1}, Article \bibinfo{articleno}{89}
  (\bibinfo{date}{April} \bibinfo{year}{2021}), \bibinfo{numpages}{27}~pages.
\newblock
\urldef\tempurl%
\url{https://doi.org/10.1145/3449163}
\showDOI{\tempurl}


\bibitem[\protect\citeauthoryear{Aytar, Vondrick, and Torralba}{Aytar
  et~al\mbox{.}}{2016}]%
        {aytar2016soundnet}
\bibfield{author}{\bibinfo{person}{Yusuf Aytar}, \bibinfo{person}{Carl
  Vondrick}, {and} \bibinfo{person}{Antonio Torralba}.}
  \bibinfo{year}{2016}\natexlab{}.
\newblock \showarticletitle{Soundnet: Learning sound representations from
  unlabeled video}.
\newblock \bibinfo{journal}{\emph{Advances in neural information processing
  systems}}  \bibinfo{volume}{29} (\bibinfo{year}{2016}),
  \bibinfo{pages}{892--900}.
\newblock


\bibitem[\protect\citeauthoryear{Berant and Liang}{Berant and Liang}{2014}]%
        {berant2014semantic}
\bibfield{author}{\bibinfo{person}{Jonathan Berant} {and}
  \bibinfo{person}{Percy Liang}.} \bibinfo{year}{2014}\natexlab{}.
\newblock \showarticletitle{Semantic parsing via paraphrasing}. In
  \bibinfo{booktitle}{\emph{Proceedings of the 52nd Annual Meeting of the
  Association for Computational Linguistics (Volume 1: Long Papers)}}.
  \bibinfo{pages}{1415--1425}.
\newblock


\bibitem[\protect\citeauthoryear{Berg, Kutra, Kroeger, Straehle, Kausler,
  Haubold, Schiegg, Ales, Beier, Rudy, et~al\mbox{.}}{Berg
  et~al\mbox{.}}{2019}]%
        {berg2019ilastik}
\bibfield{author}{\bibinfo{person}{Stuart Berg}, \bibinfo{person}{Dominik
  Kutra}, \bibinfo{person}{Thorben Kroeger}, \bibinfo{person}{Christoph~N
  Straehle}, \bibinfo{person}{Bernhard~X Kausler}, \bibinfo{person}{Carsten
  Haubold}, \bibinfo{person}{Martin Schiegg}, \bibinfo{person}{Janez Ales},
  \bibinfo{person}{Thorsten Beier}, \bibinfo{person}{Markus Rudy},
  {et~al\mbox{.}}} \bibinfo{year}{2019}\natexlab{}.
\newblock \showarticletitle{Ilastik: interactive machine learning for (bio)
  image analysis}.
\newblock \bibinfo{journal}{\emph{Nature methods}} \bibinfo{volume}{16},
  \bibinfo{number}{12} (\bibinfo{year}{2019}), \bibinfo{pages}{1226--1232}.
\newblock


\bibitem[\protect\citeauthoryear{Bhargava and Forsyth}{Bhargava and
  Forsyth}{2019}]%
        {Bhargava2019ExposingAC}
\bibfield{author}{\bibinfo{person}{Shruti Bhargava} {and}
  \bibinfo{person}{David Forsyth}.} \bibinfo{year}{2019}\natexlab{}.
\newblock \showarticletitle{Exposing and Correcting the Gender Bias in Image
  Captioning Datasets and Models}.
\newblock \bibinfo{journal}{\emph{ArXiv}}  \bibinfo{volume}{abs/1912.00578}
  (\bibinfo{year}{2019}).
\newblock


\bibitem[\protect\citeauthoryear{Blanzeisky and Cunningham}{Blanzeisky and
  Cunningham}{2021}]%
        {Blanzeisky2021AlgorithmicFI}
\bibfield{author}{\bibinfo{person}{William Blanzeisky} {and}
  \bibinfo{person}{Padraig Cunningham}.} \bibinfo{year}{2021}\natexlab{}.
\newblock \showarticletitle{Algorithmic Factors Influencing Bias in Machine
  Learning}. In \bibinfo{booktitle}{\emph{PKDD/ECML Workshops}}.
\newblock


\bibitem[\protect\citeauthoryear{Bobes-Bascar{\'a}n, Mosqueira-Rey, and
  Alonso-R{\'\i}os}{Bobes-Bascar{\'a}n et~al\mbox{.}}{2021}]%
        {bobes2021improving}
\bibfield{author}{\bibinfo{person}{Jos{\'e} Bobes-Bascar{\'a}n},
  \bibinfo{person}{Eduardo Mosqueira-Rey}, {and} \bibinfo{person}{David
  Alonso-R{\'\i}os}.} \bibinfo{year}{2021}\natexlab{}.
\newblock \showarticletitle{Improving medical data annotation including humans
  in the machine learning loop}.
\newblock \bibinfo{journal}{\emph{Engineering Proceedings}}
  \bibinfo{volume}{7}, \bibinfo{number}{1} (\bibinfo{year}{2021}),
  \bibinfo{pages}{39}.
\newblock


\bibitem[\protect\citeauthoryear{Brew, Greene, and Cunningham}{Brew
  et~al\mbox{.}}{2010}]%
        {brew2010interaction}
\bibfield{author}{\bibinfo{person}{Anthony Brew}, \bibinfo{person}{Derek
  Greene}, {and} \bibinfo{person}{P{\'a}draig Cunningham}.}
  \bibinfo{year}{2010}\natexlab{}.
\newblock \showarticletitle{The interaction between supervised learning and
  crowdsourcing}. In \bibinfo{booktitle}{\emph{NIPS workshop on computational
  social science and the wisdom of crowds}}.
\newblock


\bibitem[\protect\citeauthoryear{Cai, Reif, Hegde, Hipp, Kim, Smilkov,
  Wattenberg, Viegas, Corrado, Stumpe, and Terry}{Cai et~al\mbox{.}}{2019}]%
        {cai_human-centered_2019}
\bibfield{author}{\bibinfo{person}{Carrie~J. Cai}, \bibinfo{person}{Emily
  Reif}, \bibinfo{person}{Narayan Hegde}, \bibinfo{person}{Jason Hipp},
  \bibinfo{person}{Been Kim}, \bibinfo{person}{Daniel Smilkov},
  \bibinfo{person}{Martin Wattenberg}, \bibinfo{person}{Fernanda Viegas},
  \bibinfo{person}{Greg~S. Corrado}, \bibinfo{person}{Martin~C. Stumpe}, {and}
  \bibinfo{person}{Michael Terry}.} \bibinfo{year}{2019}\natexlab{}.
\newblock \showarticletitle{Human-Centered Tools for Coping with Imperfect
  Algorithms During Medical Decision-Making}. In
  \bibinfo{booktitle}{\emph{Proceedings of the 2019 CHI Conference on Human
  Factors in Computing Systems}} (Glasgow, Scotland Uk)
  \emph{(\bibinfo{series}{CHI '19})}. \bibinfo{publisher}{Association for
  Computing Machinery}, \bibinfo{address}{New York, NY, USA},
  \bibinfo{pages}{1–14}.
\newblock
\showISBNx{9781450359702}
\urldef\tempurl%
\url{https://doi.org/10.1145/3290605.3300234}
\showDOI{\tempurl}


\bibitem[\protect\citeauthoryear{Cassidy and Schmidt}{Cassidy and
  Schmidt}{2017}]%
        {cassidy_2017_tools}
\bibfield{author}{\bibinfo{person}{Steve Cassidy} {and} \bibinfo{person}{Thomas
  Schmidt}.} \bibinfo{year}{2017}\natexlab{}.
\newblock \bibinfo{booktitle}{\emph{Tools for multimodal annotation}}.
\newblock \bibinfo{publisher}{Springer, Springer Nature},
  \bibinfo{address}{United States}, \bibinfo{pages}{209--227}.
\newblock
\showISBNx{9789402408799}
\urldef\tempurl%
\url{https://doi.org/10.1007/978-94-024-0881-2_7}
\showDOI{\tempurl}


\bibitem[\protect\citeauthoryear{Chang, Lee, and Igarashi}{Chang
  et~al\mbox{.}}{2021}]%
        {10.1145/3411764.3445165}
\bibfield{author}{\bibinfo{person}{Chia-Ming Chang},
  \bibinfo{person}{Chia-Hsien Lee}, {and} \bibinfo{person}{Takeo Igarashi}.}
  \bibinfo{year}{2021}\natexlab{}.
\newblock \showarticletitle{Spatial Labeling: Leveraging Spatial Layout for
  Improving Label Quality in Non-Expert Image Annotation}. In
  \bibinfo{booktitle}{\emph{Proceedings of the 2021 CHI Conference on Human
  Factors in Computing Systems}} (Yokohama, Japan) \emph{(\bibinfo{series}{CHI
  '21})}. \bibinfo{publisher}{Association for Computing Machinery},
  \bibinfo{address}{New York, NY, USA}, Article \bibinfo{articleno}{306},
  \bibinfo{numpages}{12}~pages.
\newblock
\showISBNx{9781450380966}
\urldef\tempurl%
\url{https://doi.org/10.1145/3411764.3445165}
\showDOI{\tempurl}


\bibitem[\protect\citeauthoryear{Chen, Majumder, Al-Halah, Gao, Ramakrishnan,
  and Grauman}{Chen et~al\mbox{.}}{2020}]%
        {chen2020learning}
\bibfield{author}{\bibinfo{person}{Changan Chen}, \bibinfo{person}{Sagnik
  Majumder}, \bibinfo{person}{Ziad Al-Halah}, \bibinfo{person}{Ruohan Gao},
  \bibinfo{person}{Santhosh~Kumar Ramakrishnan}, {and} \bibinfo{person}{Kristen
  Grauman}.} \bibinfo{year}{2020}\natexlab{}.
\newblock \showarticletitle{Learning to set waypoints for audio-visual
  navigation}.
\newblock \bibinfo{journal}{\emph{arXiv preprint arXiv:2008.09622}}
  (\bibinfo{year}{2020}).
\newblock


\bibitem[\protect\citeauthoryear{Chen and Joo}{Chen and Joo}{2021}]%
        {Chen2021UnderstandingAM}
\bibfield{author}{\bibinfo{person}{Yunliang Chen} {and}
  \bibinfo{person}{Jungseock Joo}.} \bibinfo{year}{2021}\natexlab{}.
\newblock \showarticletitle{Understanding and Mitigating Annotation Bias in
  Facial Expression Recognition}.
\newblock \bibinfo{journal}{\emph{2021 IEEE/CVF International Conference on
  Computer Vision (ICCV)}} (\bibinfo{year}{2021}),
  \bibinfo{pages}{14960--14971}.
\newblock


\bibitem[\protect\citeauthoryear{Chen, Xu, Liu, Zeng, and Zhao}{Chen
  et~al\mbox{.}}{2015}]%
        {chen2015event}
\bibfield{author}{\bibinfo{person}{Yubo Chen}, \bibinfo{person}{Liheng Xu},
  \bibinfo{person}{Kang Liu}, \bibinfo{person}{Daojian Zeng}, {and}
  \bibinfo{person}{Jun Zhao}.} \bibinfo{year}{2015}\natexlab{}.
\newblock \showarticletitle{Event extraction via dynamic multi-pooling
  convolutional neural networks}. In \bibinfo{booktitle}{\emph{Proceedings of
  the 53rd Annual Meeting of the Association for Computational Linguistics and
  the 7th International Joint Conference on Natural Language Processing (Volume
  1: Long Papers)}}. \bibinfo{pages}{167--176}.
\newblock


\bibitem[\protect\citeauthoryear{Choi, Garcia, Raman, Enzmann, and Brown}{Choi
  et~al\mbox{.}}{2022}]%
        {choi2022ai}
\bibfield{author}{\bibinfo{person}{Youngwon Choi}, \bibinfo{person}{Marlena
  Garcia}, \bibinfo{person}{Steven~S Raman}, \bibinfo{person}{Dieter~R
  Enzmann}, {and} \bibinfo{person}{Matthew~S Brown}.}
  \bibinfo{year}{2022}\natexlab{}.
\newblock \showarticletitle{AI-human interactive pipeline with feedback to
  accelerate medical image annotation}. In \bibinfo{booktitle}{\emph{Medical
  Imaging 2022: Computer-Aided Diagnosis}}, Vol.~\bibinfo{volume}{12033}. SPIE,
  \bibinfo{pages}{741--747}.
\newblock


\bibitem[\protect\citeauthoryear{Cohn, Ghahramani, and Jordan}{Cohn
  et~al\mbox{.}}{1996}]%
        {cohn1996active}
\bibfield{author}{\bibinfo{person}{David~A Cohn}, \bibinfo{person}{Zoubin
  Ghahramani}, {and} \bibinfo{person}{Michael~I Jordan}.}
  \bibinfo{year}{1996}\natexlab{}.
\newblock \showarticletitle{Active learning with statistical models}.
\newblock \bibinfo{journal}{\emph{Journal of artificial intelligence research}}
   \bibinfo{volume}{4} (\bibinfo{year}{1996}), \bibinfo{pages}{129--145}.
\newblock


\bibitem[\protect\citeauthoryear{Cooper, Khatib, Treuille, Barbero, Lee,
  Beenen, Leaver-Fay, Baker, Popovi{\'c}, et~al\mbox{.}}{Cooper
  et~al\mbox{.}}{2010}]%
        {cooper2010predicting}
\bibfield{author}{\bibinfo{person}{Seth Cooper}, \bibinfo{person}{Firas
  Khatib}, \bibinfo{person}{Adrien Treuille}, \bibinfo{person}{Janos Barbero},
  \bibinfo{person}{Jeehyung Lee}, \bibinfo{person}{Michael Beenen},
  \bibinfo{person}{Andrew Leaver-Fay}, \bibinfo{person}{David Baker},
  \bibinfo{person}{Zoran Popovi{\'c}}, {et~al\mbox{.}}}
  \bibinfo{year}{2010}\natexlab{}.
\newblock \showarticletitle{Predicting protein structures with a multiplayer
  online game}.
\newblock \bibinfo{journal}{\emph{Nature}} \bibinfo{volume}{466},
  \bibinfo{number}{7307} (\bibinfo{year}{2010}), \bibinfo{pages}{756--760}.
\newblock


\bibitem[\protect\citeauthoryear{Dar, Muthukumar, and Baraniuk}{Dar
  et~al\mbox{.}}{2021}]%
        {Dar2021AFT}
\bibfield{author}{\bibinfo{person}{Yehuda Dar}, \bibinfo{person}{Vidya
  Muthukumar}, {and} \bibinfo{person}{Richard Baraniuk}.}
  \bibinfo{year}{2021}\natexlab{}.
\newblock \showarticletitle{A Farewell to the Bias-Variance Tradeoff? An
  Overview of the Theory of Overparameterized Machine Learning}.
\newblock \bibinfo{journal}{\emph{ArXiv}}  \bibinfo{volume}{abs/2109.02355}
  (\bibinfo{year}{2021}).
\newblock


\bibitem[\protect\citeauthoryear{Demirkus, Clark, and Arbel}{Demirkus
  et~al\mbox{.}}{2014}]%
        {demirkus2014robust}
\bibfield{author}{\bibinfo{person}{Meltem Demirkus}, \bibinfo{person}{James~J
  Clark}, {and} \bibinfo{person}{Tal Arbel}.} \bibinfo{year}{2014}\natexlab{}.
\newblock \showarticletitle{Robust semi-automatic head pose labeling for
  real-world face video sequences}.
\newblock \bibinfo{journal}{\emph{Multimedia Tools and Applications}}
  \bibinfo{volume}{70}, \bibinfo{number}{1} (\bibinfo{year}{2014}),
  \bibinfo{pages}{495--523}.
\newblock


\bibitem[\protect\citeauthoryear{Desmond, Muller, Ashktorab, Dugan,
  Duesterwald, Brimijoin, Finegan-Dollak, Brachman, Sharma, Joshi, and
  Pan}{Desmond et~al\mbox{.}}{2021}]%
        {desmond_increasing_2021}
\bibfield{author}{\bibinfo{person}{Michael Desmond}, \bibinfo{person}{Michael
  Muller}, \bibinfo{person}{Zahra Ashktorab}, \bibinfo{person}{Casey Dugan},
  \bibinfo{person}{Evelyn Duesterwald}, \bibinfo{person}{Kristina Brimijoin},
  \bibinfo{person}{Catherine Finegan-Dollak}, \bibinfo{person}{Michelle
  Brachman}, \bibinfo{person}{Aabhas Sharma}, \bibinfo{person}{Narendra~Nath
  Joshi}, {and} \bibinfo{person}{Qian Pan}.} \bibinfo{year}{2021}\natexlab{}.
\newblock \bibinfo{booktitle}{\emph{Increasing the Speed and Accuracy of Data
  Labeling Through an AI Assisted Interface}}.
\newblock \bibinfo{publisher}{Association for Computing Machinery},
  \bibinfo{address}{New York, NY, USA}, \bibinfo{pages}{392–401}.
\newblock
\showISBNx{9781450380171}
\urldef\tempurl%
\url{https://doi.org/10.1145/3397481.3450698}
\showURL{%
\tempurl}


\bibitem[\protect\citeauthoryear{Doersch, Gupta, and Efros}{Doersch
  et~al\mbox{.}}{2015}]%
        {doersch2015unsupervised}
\bibfield{author}{\bibinfo{person}{Carl Doersch}, \bibinfo{person}{Abhinav
  Gupta}, {and} \bibinfo{person}{Alexei~A Efros}.}
  \bibinfo{year}{2015}\natexlab{}.
\newblock \showarticletitle{Unsupervised visual representation learning by
  context prediction}. In \bibinfo{booktitle}{\emph{Proceedings of the IEEE
  international conference on computer vision}}. \bibinfo{pages}{1422--1430}.
\newblock


\bibitem[\protect\citeauthoryear{Dutta and Zisserman}{Dutta and
  Zisserman}{2019}]%
        {dutta_2019_via}
\bibfield{author}{\bibinfo{person}{Abhishek Dutta} {and}
  \bibinfo{person}{Andrew Zisserman}.} \bibinfo{year}{2019}\natexlab{}.
\newblock \showarticletitle{The VIA Annotation Software for Images, Audio and
  Video}. In \bibinfo{booktitle}{\emph{Proceedings of the 27th ACM
  International Conference on Multimedia}} (Nice, France)
  \emph{(\bibinfo{series}{MM '19})}. \bibinfo{publisher}{Association for
  Computing Machinery}, \bibinfo{address}{New York, NY, USA},
  \bibinfo{pages}{2276–2279}.
\newblock
\showISBNx{9781450368896}
\urldef\tempurl%
\url{https://doi.org/10.1145/3343031.3350535}
\showDOI{\tempurl}


\bibitem[\protect\citeauthoryear{Ephrat, Mosseri, Lang, Dekel, Wilson,
  Hassidim, Freeman, and Rubinstein}{Ephrat et~al\mbox{.}}{2018}]%
        {ephrat2018looking}
\bibfield{author}{\bibinfo{person}{Ariel Ephrat}, \bibinfo{person}{Inbar
  Mosseri}, \bibinfo{person}{Oran Lang}, \bibinfo{person}{Tali Dekel},
  \bibinfo{person}{Kevin Wilson}, \bibinfo{person}{Avinatan Hassidim},
  \bibinfo{person}{William~T Freeman}, {and} \bibinfo{person}{Michael
  Rubinstein}.} \bibinfo{year}{2018}\natexlab{}.
\newblock \showarticletitle{Looking to listen at the cocktail party: A
  speaker-independent audio-visual model for speech separation}.
\newblock \bibinfo{journal}{\emph{TOG}} (\bibinfo{year}{2018}).
\newblock


\bibitem[\protect\citeauthoryear{Evensen, Ge, and Demiralp}{Evensen
  et~al\mbox{.}}{2020}]%
        {evensen-etal-2020-ruler}
\bibfield{author}{\bibinfo{person}{Sara Evensen}, \bibinfo{person}{Chang Ge},
  {and} \bibinfo{person}{Cagatay Demiralp}.} \bibinfo{year}{2020}\natexlab{}.
\newblock \showarticletitle{Ruler: Data Programming by Demonstration for
  Document Labeling}. In \bibinfo{booktitle}{\emph{Findings of the Association
  for Computational Linguistics: EMNLP 2020}}. \bibinfo{publisher}{Association
  for Computational Linguistics}, \bibinfo{address}{Online},
  \bibinfo{pages}{1996--2005}.
\newblock
\urldef\tempurl%
\url{https://doi.org/10.18653/v1/2020.findings-emnlp.181}
\showDOI{\tempurl}


\bibitem[\protect\citeauthoryear{Galhotra, Khurana, Hassanzadeh, Srinivas,
  Samulowitz, and Qi}{Galhotra et~al\mbox{.}}{2019}]%
        {galhotra_automated_2019}
\bibfield{author}{\bibinfo{person}{Sainyam Galhotra}, \bibinfo{person}{Udayan
  Khurana}, \bibinfo{person}{Oktie Hassanzadeh}, \bibinfo{person}{Kavitha
  Srinivas}, \bibinfo{person}{Horst Samulowitz}, {and} \bibinfo{person}{Miao
  Qi}.} \bibinfo{year}{2019}\natexlab{}.
\newblock \showarticletitle{Automated Feature Enhancement for Predictive
  Modeling using External Knowledge}. In \bibinfo{booktitle}{\emph{2019
  International Conference on Data Mining Workshops (ICDMW)}}.
  \bibinfo{pages}{1094--1097}.
\newblock
\urldef\tempurl%
\url{https://doi.org/10.1109/ICDMW.2019.00161}
\showDOI{\tempurl}


\bibitem[\protect\citeauthoryear{Gao, Guo, Lim, Zhan, Zhang, Li, and
  Perrault}{Gao et~al\mbox{.}}{2023}]%
        {gao2023collabcoder}
\bibfield{author}{\bibinfo{person}{Jie Gao}, \bibinfo{person}{Yuchen Guo},
  \bibinfo{person}{Gionnieve Lim}, \bibinfo{person}{Tianqin Zhan},
  \bibinfo{person}{Zheng Zhang}, \bibinfo{person}{Toby Jia-Jun Li}, {and}
  \bibinfo{person}{Simon~Tangi Perrault}.} \bibinfo{year}{2023}\natexlab{}.
\newblock \showarticletitle{CollabCoder: A GPT-Powered Workflow for
  Collaborative Qualitative Analysis}.
\newblock \bibinfo{journal}{\emph{arXiv preprint arXiv:2304.07366}}
  (\bibinfo{year}{2023}).
\newblock


\bibitem[\protect\citeauthoryear{Gao, Chang, Mall, Fei-Fei, and Wu}{Gao
  et~al\mbox{.}}{2021}]%
        {gao2021objectfolder}
\bibfield{author}{\bibinfo{person}{Ruohan Gao}, \bibinfo{person}{Yen-Yu Chang},
  \bibinfo{person}{Shivani Mall}, \bibinfo{person}{Li Fei-Fei}, {and}
  \bibinfo{person}{Jiajun Wu}.} \bibinfo{year}{2021}\natexlab{}.
\newblock \showarticletitle{ObjectFolder: A Dataset of Objects with Implicit
  Visual, Auditory, and Tactile Representations}.
\newblock \bibinfo{journal}{\emph{arXiv preprint arXiv:2109.07991}}
  (\bibinfo{year}{2021}).
\newblock


\bibitem[\protect\citeauthoryear{Gao, Feris, and Grauman}{Gao
  et~al\mbox{.}}{2018}]%
        {gao2018learning}
\bibfield{author}{\bibinfo{person}{Ruohan Gao}, \bibinfo{person}{Rogerio
  Feris}, {and} \bibinfo{person}{Kristen Grauman}.}
  \bibinfo{year}{2018}\natexlab{}.
\newblock \showarticletitle{Learning to separate object sounds by watching
  unlabeled video}. In \bibinfo{booktitle}{\emph{Proceedings of the European
  Conference on Computer Vision (ECCV)}}. \bibinfo{pages}{35--53}.
\newblock


\bibitem[\protect\citeauthoryear{Gao and Grauman}{Gao and Grauman}{2019a}]%
        {Gao_2019_CVPR}
\bibfield{author}{\bibinfo{person}{Ruohan Gao} {and} \bibinfo{person}{Kristen
  Grauman}.} \bibinfo{year}{2019}\natexlab{a}.
\newblock \showarticletitle{2.5D Visual Sound}. In
  \bibinfo{booktitle}{\emph{Proceedings of the IEEE/CVF Conference on Computer
  Vision and Pattern Recognition (CVPR)}}.
\newblock


\bibitem[\protect\citeauthoryear{Gao and Grauman}{Gao and Grauman}{2019b}]%
        {gao2019co}
\bibfield{author}{\bibinfo{person}{Ruohan Gao} {and} \bibinfo{person}{Kristen
  Grauman}.} \bibinfo{year}{2019}\natexlab{b}.
\newblock \showarticletitle{Co-separating sounds of visual objects}. In
  \bibinfo{booktitle}{\emph{Proceedings of the IEEE/CVF International
  Conference on Computer Vision}}. \bibinfo{pages}{3879--3888}.
\newblock


\bibitem[\protect\citeauthoryear{Gebreegziabher, Zhang, Tang, Meng, Glassman,
  and Li}{Gebreegziabher et~al\mbox{.}}{2023}]%
        {gebreegziabher2023patat}
\bibfield{author}{\bibinfo{person}{Simret~Araya Gebreegziabher},
  \bibinfo{person}{Zheng Zhang}, \bibinfo{person}{Xiaohang Tang},
  \bibinfo{person}{Yihao Meng}, \bibinfo{person}{Elena~L Glassman}, {and}
  \bibinfo{person}{Toby Jia-Jun Li}.} \bibinfo{year}{2023}\natexlab{}.
\newblock \showarticletitle{Patat: Human-ai collaborative qualitative coding
  with explainable interactive rule synthesis}. In
  \bibinfo{booktitle}{\emph{Proceedings of the 2023 CHI Conference on Human
  Factors in Computing Systems}}. \bibinfo{pages}{1--19}.
\newblock


\bibitem[\protect\citeauthoryear{Guo, Kandel, Hellerstein, and Heer}{Guo
  et~al\mbox{.}}{2011}]%
        {guo_proactive_2011}
\bibfield{author}{\bibinfo{person}{Philip~J. Guo}, \bibinfo{person}{Sean
  Kandel}, \bibinfo{person}{Joseph~M. Hellerstein}, {and}
  \bibinfo{person}{Jeffrey Heer}.} \bibinfo{year}{2011}\natexlab{}.
\newblock \showarticletitle{Proactive Wrangling: Mixed-Initiative End-User
  Programming of Data Transformation Scripts}. In
  \bibinfo{booktitle}{\emph{Proceedings of the 24th Annual ACM Symposium on
  User Interface Software and Technology}} (Santa Barbara, California, USA)
  \emph{(\bibinfo{series}{UIST '11})}. \bibinfo{publisher}{Association for
  Computing Machinery}, \bibinfo{address}{New York, NY, USA},
  \bibinfo{pages}{65–74}.
\newblock
\showISBNx{9781450307161}
\urldef\tempurl%
\url{https://doi.org/10.1145/2047196.2047205}
\showDOI{\tempurl}


\bibitem[\protect\citeauthoryear{Halevy, Norvig, and Pereira}{Halevy
  et~al\mbox{.}}{2009}]%
        {halevy2009unreasonable}
\bibfield{author}{\bibinfo{person}{Alon Halevy}, \bibinfo{person}{Peter
  Norvig}, {and} \bibinfo{person}{Fernando Pereira}.}
  \bibinfo{year}{2009}\natexlab{}.
\newblock \showarticletitle{The unreasonable effectiveness of data}.
\newblock \bibinfo{journal}{\emph{IEEE Intelligent Systems}}
  \bibinfo{volume}{24}, \bibinfo{number}{2} (\bibinfo{year}{2009}),
  \bibinfo{pages}{8--12}.
\newblock


\bibitem[\protect\citeauthoryear{Hamroun, Tamine, and Crespin}{Hamroun
  et~al\mbox{.}}{2021}]%
        {hamroun2021multimodal}
\bibfield{author}{\bibinfo{person}{Mohamed Hamroun}, \bibinfo{person}{Karim
  Tamine}, {and} \bibinfo{person}{Beno{\^\i}t Crespin}.}
  \bibinfo{year}{2021}\natexlab{}.
\newblock \showarticletitle{Multimodal Video Indexing (MVI): A New Method Based
  on Machine Learning and Semi-Automatic Annotation on Large Video
  Collections}.
\newblock \bibinfo{journal}{\emph{International Journal of Image and Graphics}}
  (\bibinfo{year}{2021}), \bibinfo{pages}{2250022}.
\newblock


\bibitem[\protect\citeauthoryear{He, Zhao, and Chu}{He et~al\mbox{.}}{2021}]%
        {he2021automl}
\bibfield{author}{\bibinfo{person}{Xin He}, \bibinfo{person}{Kaiyong Zhao},
  {and} \bibinfo{person}{Xiaowen Chu}.} \bibinfo{year}{2021}\natexlab{}.
\newblock \showarticletitle{AutoML: A Survey of the State-of-the-Art}.
\newblock \bibinfo{journal}{\emph{Knowledge-Based Systems}}
  \bibinfo{volume}{212} (\bibinfo{year}{2021}), \bibinfo{pages}{106622}.
\newblock


\bibitem[\protect\citeauthoryear{Heer}{Heer}{2019}]%
        {heer2019agency}
\bibfield{author}{\bibinfo{person}{Jeffrey Heer}.}
  \bibinfo{year}{2019}\natexlab{}.
\newblock \showarticletitle{Agency plus automation: Designing artificial
  intelligence into interactive systems}.
\newblock \bibinfo{journal}{\emph{Proceedings of the National Academy of
  Sciences}} \bibinfo{volume}{116}, \bibinfo{number}{6} (\bibinfo{year}{2019}),
  \bibinfo{pages}{1844--1850}.
\newblock


\bibitem[\protect\citeauthoryear{Hershey and Movellan}{Hershey and
  Movellan}{2000}]%
        {hershey2000audio}
\bibfield{author}{\bibinfo{person}{John~R Hershey} {and}
  \bibinfo{person}{Javier~R Movellan}.} \bibinfo{year}{2000}\natexlab{}.
\newblock \showarticletitle{Audio vision: Using audio-visual synchrony to
  locate sounds}. In \bibinfo{booktitle}{\emph{Advances in neural information
  processing systems}}. \bibinfo{pages}{813--819}.
\newblock


\bibitem[\protect\citeauthoryear{Hinton, Deng, Yu, Dahl, Mohamed, Jaitly,
  Senior, Vanhoucke, Nguyen, Sainath, et~al\mbox{.}}{Hinton
  et~al\mbox{.}}{2012}]%
        {hinton2012deep}
\bibfield{author}{\bibinfo{person}{Geoffrey Hinton}, \bibinfo{person}{Li Deng},
  \bibinfo{person}{Dong Yu}, \bibinfo{person}{George~E Dahl},
  \bibinfo{person}{Abdel-rahman Mohamed}, \bibinfo{person}{Navdeep Jaitly},
  \bibinfo{person}{Andrew Senior}, \bibinfo{person}{Vincent Vanhoucke},
  \bibinfo{person}{Patrick Nguyen}, \bibinfo{person}{Tara~N Sainath},
  {et~al\mbox{.}}} \bibinfo{year}{2012}\natexlab{}.
\newblock \showarticletitle{Deep neural networks for acoustic modeling in
  speech recognition: The shared views of four research groups}.
\newblock \bibinfo{journal}{\emph{IEEE Signal processing magazine}}
  \bibinfo{volume}{29}, \bibinfo{number}{6} (\bibinfo{year}{2012}),
  \bibinfo{pages}{82--97}.
\newblock


\bibitem[\protect\citeauthoryear{Horvitz}{Horvitz}{1999}]%
        {horivitz1999principles}
\bibfield{author}{\bibinfo{person}{Eric Horvitz}.}
  \bibinfo{year}{1999}\natexlab{}.
\newblock \showarticletitle{Principles of Mixed-Initiative User Interfaces}. In
  \bibinfo{booktitle}{\emph{Proceedings of the SIGCHI Conference on Human
  Factors in Computing Systems}} (Pittsburgh, Pennsylvania, USA)
  \emph{(\bibinfo{series}{CHI ’99})}. \bibinfo{publisher}{ACM},
  \bibinfo{address}{New York, NY, USA}, \bibinfo{pages}{159–166}.
\newblock
\showISBNx{0201485591}
\urldef\tempurl%
\url{https://doi.org/10.1145/302979.303030}
\showDOI{\tempurl}


\bibitem[\protect\citeauthoryear{Hu, Nie, and Li}{Hu et~al\mbox{.}}{2019}]%
        {hu2019deep}
\bibfield{author}{\bibinfo{person}{Di Hu}, \bibinfo{person}{Feiping Nie}, {and}
  \bibinfo{person}{Xuelong Li}.} \bibinfo{year}{2019}\natexlab{}.
\newblock \showarticletitle{Deep Multimodal Clustering for Unsupervised
  Audiovisual Learning}. In \bibinfo{booktitle}{\emph{CVPR}}.
\newblock


\bibitem[\protect\citeauthoryear{Hu, Qian, Jiang, Tan, Wen, Ding, Lin, and
  Dou}{Hu et~al\mbox{.}}{2020}]%
        {hu2020discriminative}
\bibfield{author}{\bibinfo{person}{Di Hu}, \bibinfo{person}{Rui Qian},
  \bibinfo{person}{Minyue Jiang}, \bibinfo{person}{Xiao Tan},
  \bibinfo{person}{Shilei Wen}, \bibinfo{person}{Errui Ding},
  \bibinfo{person}{Weiyao Lin}, {and} \bibinfo{person}{Dejing Dou}.}
  \bibinfo{year}{2020}\natexlab{}.
\newblock \showarticletitle{Discriminative Sounding Objects Localization via
  Self-supervised Audiovisual Matching}. In \bibinfo{booktitle}{\emph{Advances
  in Neural Information Processing Systems}},
  \bibfield{editor}{\bibinfo{person}{H.~Larochelle},
  \bibinfo{person}{M.~Ranzato}, \bibinfo{person}{R.~Hadsell},
  \bibinfo{person}{M.F. Balcan}, {and} \bibinfo{person}{H.~Lin}} (Eds.),
  Vol.~\bibinfo{volume}{33}. \bibinfo{publisher}{Curran Associates, Inc.},
  \bibinfo{pages}{10077--10087}.
\newblock
\urldef\tempurl%
\url{https://proceedings.neurips.cc/paper/2020/file/7288251b27c8f0e73f4d7f483b06a785-Paper.pdf}
\showURL{%
\tempurl}


\bibitem[\protect\citeauthoryear{Hu, Wei, Qian, Lin, Song, and Wen}{Hu
  et~al\mbox{.}}{2021}]%
        {hu2021class}
\bibfield{author}{\bibinfo{person}{Di Hu}, \bibinfo{person}{Yake Wei},
  \bibinfo{person}{Rui Qian}, \bibinfo{person}{Weiyao Lin},
  \bibinfo{person}{Ruihua Song}, {and} \bibinfo{person}{Ji-Rong Wen}.}
  \bibinfo{year}{2021}\natexlab{}.
\newblock \showarticletitle{Class-aware sounding objects localization via
  audiovisual correspondence}.
\newblock \bibinfo{journal}{\emph{IEEE Transactions on Pattern Analysis and
  Machine Intelligence}} (\bibinfo{year}{2021}).
\newblock


\bibitem[\protect\citeauthoryear{Jarrahi}{Jarrahi}{2018}]%
        {jarrahi2018artificial}
\bibfield{author}{\bibinfo{person}{Mohammad~Hossein Jarrahi}.}
  \bibinfo{year}{2018}\natexlab{}.
\newblock \showarticletitle{Artificial intelligence and the future of work:
  Human-AI symbiosis in organizational decision making}.
\newblock \bibinfo{journal}{\emph{Business horizons}} \bibinfo{volume}{61},
  \bibinfo{number}{4} (\bibinfo{year}{2018}), \bibinfo{pages}{577--586}.
\newblock


\bibitem[\protect\citeauthoryear{Kandel, Paepcke, Hellerstein, and Heer}{Kandel
  et~al\mbox{.}}{2012}]%
        {kandel_enterprise_2012}
\bibfield{author}{\bibinfo{person}{Sean Kandel}, \bibinfo{person}{Andreas
  Paepcke}, \bibinfo{person}{Joseph~M. Hellerstein}, {and}
  \bibinfo{person}{Jeffrey Heer}.} \bibinfo{year}{2012}\natexlab{}.
\newblock \showarticletitle{Enterprise Data Analysis and Visualization: An
  Interview Study}.
\newblock \bibinfo{journal}{\emph{IEEE Transactions on Visualization and
  Computer Graphics}} \bibinfo{volume}{18}, \bibinfo{number}{12}
  (\bibinfo{year}{2012}), \bibinfo{pages}{2917--2926}.
\newblock
\urldef\tempurl%
\url{https://doi.org/10.1109/TVCG.2012.219}
\showDOI{\tempurl}


\bibitem[\protect\citeauthoryear{Kaul, Xie, and Zisserman}{Kaul
  et~al\mbox{.}}{2022}]%
        {kaul2022label}
\bibfield{author}{\bibinfo{person}{Prannay Kaul}, \bibinfo{person}{Weidi Xie},
  {and} \bibinfo{person}{Andrew Zisserman}.} \bibinfo{year}{2022}\natexlab{}.
\newblock \showarticletitle{Label, verify, correct: A simple few shot object
  detection method}. In \bibinfo{booktitle}{\emph{Proceedings of the IEEE/CVF
  Conference on Computer Vision and Pattern Recognition}}.
  \bibinfo{pages}{14237--14247}.
\newblock


\bibitem[\protect\citeauthoryear{Kehl, Xu, Gusev, Bakouny, Choueiri, Riaz,
  Elmarakeby, Van~Allen, and Schrag}{Kehl et~al\mbox{.}}{2021}]%
        {kehl2021artificial}
\bibfield{author}{\bibinfo{person}{Kenneth~L Kehl}, \bibinfo{person}{Wenxin
  Xu}, \bibinfo{person}{Alexander Gusev}, \bibinfo{person}{Ziad Bakouny},
  \bibinfo{person}{Toni~K Choueiri}, \bibinfo{person}{Irbaz~Bin Riaz},
  \bibinfo{person}{Haitham Elmarakeby}, \bibinfo{person}{Eliezer~M Van~Allen},
  {and} \bibinfo{person}{Deborah Schrag}.} \bibinfo{year}{2021}\natexlab{}.
\newblock \showarticletitle{Artificial intelligence-aided clinical annotation
  of a large multi-cancer genomic dataset}.
\newblock \bibinfo{journal}{\emph{Nature communications}} \bibinfo{volume}{12},
  \bibinfo{number}{1} (\bibinfo{year}{2021}), \bibinfo{pages}{7304}.
\newblock


\bibitem[\protect\citeauthoryear{Kidron, Schechner, and Elad}{Kidron
  et~al\mbox{.}}{2005}]%
        {kidron2005pixels}
\bibfield{author}{\bibinfo{person}{Einat Kidron}, \bibinfo{person}{Yoav~Y
  Schechner}, {and} \bibinfo{person}{Michael Elad}.}
  \bibinfo{year}{2005}\natexlab{}.
\newblock \showarticletitle{Pixels that sound}. In
  \bibinfo{booktitle}{\emph{2005 IEEE Computer Society Conference on Computer
  Vision and Pattern Recognition (CVPR'05)}}, Vol.~\bibinfo{volume}{1}. IEEE,
  \bibinfo{pages}{88--95}.
\newblock


\bibitem[\protect\citeauthoryear{Kim, Heo, Choe, Chung, Kwon, Lee, Kwon, and
  Chung}{Kim et~al\mbox{.}}{2021}]%
        {kim21k_interspeech}
\bibfield{author}{\bibinfo{person}{You~Jin Kim}, \bibinfo{person}{Hee-Soo Heo},
  \bibinfo{person}{Soyeon Choe}, \bibinfo{person}{Soo-Whan Chung},
  \bibinfo{person}{Yoohwan Kwon}, \bibinfo{person}{Bong-Jin Lee},
  \bibinfo{person}{Youngki Kwon}, {and} \bibinfo{person}{Joon~Son Chung}.}
  \bibinfo{year}{2021}\natexlab{}.
\newblock \showarticletitle{{Look Who’s Talking: Active Speaker Detection in
  the Wild}}. In \bibinfo{booktitle}{\emph{Proc. Interspeech 2021}}.
  \bibinfo{pages}{3675--3679}.
\newblock
\urldef\tempurl%
\url{https://doi.org/10.21437/Interspeech.2021-2041}
\showDOI{\tempurl}


\bibitem[\protect\citeauthoryear{Kong, Cao, Iqbal, Wang, Wang, and
  Plumbley}{Kong et~al\mbox{.}}{2020}]%
        {Kong2020PANNsLP}
\bibfield{author}{\bibinfo{person}{Qiuqiang Kong}, \bibinfo{person}{Yin Cao},
  \bibinfo{person}{Turab Iqbal}, \bibinfo{person}{Yuxuan Wang},
  \bibinfo{person}{Wenwu Wang}, {and} \bibinfo{person}{Mark~D. Plumbley}.}
  \bibinfo{year}{2020}\natexlab{}.
\newblock \showarticletitle{PANNs: Large-Scale Pretrained Audio Neural Networks
  for Audio Pattern Recognition}.
\newblock \bibinfo{journal}{\emph{IEEE/ACM Transactions on Audio, Speech, and
  Language Processing}}  \bibinfo{volume}{28} (\bibinfo{year}{2020}),
  \bibinfo{pages}{2880--2894}.
\newblock


\bibitem[\protect\citeauthoryear{Langer}{Langer}{1989}]%
        {langer1989minding}
\bibfield{author}{\bibinfo{person}{Ellen~J Langer}.}
  \bibinfo{year}{1989}\natexlab{}.
\newblock \showarticletitle{Minding matters: The consequences of
  mindlessness--mindfulness}.
\newblock In \bibinfo{booktitle}{\emph{Advances in experimental social
  psychology}}. Vol.~\bibinfo{volume}{22}. \bibinfo{publisher}{Elsevier},
  \bibinfo{pages}{137--173}.
\newblock


\bibitem[\protect\citeauthoryear{Laput, Lasecki, Wiese, Xiao, Bigham, and
  Harrison}{Laput et~al\mbox{.}}{2015}]%
        {laput_2015_zensors}
\bibfield{author}{\bibinfo{person}{Gierad Laput}, \bibinfo{person}{Walter~S.
  Lasecki}, \bibinfo{person}{Jason Wiese}, \bibinfo{person}{Robert Xiao},
  \bibinfo{person}{Jeffrey~P. Bigham}, {and} \bibinfo{person}{Chris Harrison}.}
  \bibinfo{year}{2015}\natexlab{}.
\newblock \showarticletitle{Zensors: Adaptive, Rapidly Deployable,
  Human-Intelligent Sensor Feeds}. In \bibinfo{booktitle}{\emph{Proceedings of
  the 33rd Annual ACM Conference on Human Factors in Computing Systems}}
  (Seoul, Republic of Korea) \emph{(\bibinfo{series}{CHI '15})}.
  \bibinfo{publisher}{Association for Computing Machinery},
  \bibinfo{address}{New York, NY, USA}, \bibinfo{pages}{1935–1944}.
\newblock
\showISBNx{9781450331456}
\urldef\tempurl%
\url{https://doi.org/10.1145/2702123.2702416}
\showDOI{\tempurl}


\bibitem[\protect\citeauthoryear{Leavy, Meaney, Wade, and Greene}{Leavy
  et~al\mbox{.}}{2020}]%
        {Leavy2020MitigatingGB}
\bibfield{author}{\bibinfo{person}{Susan Leavy}, \bibinfo{person}{Gerardine
  Meaney}, \bibinfo{person}{Karen Wade}, {and} \bibinfo{person}{Derek Greene}.}
  \bibinfo{year}{2020}\natexlab{}.
\newblock \showarticletitle{Mitigating Gender Bias in Machine Learning Data
  Sets}.
\newblock \bibinfo{journal}{\emph{ArXiv}}  \bibinfo{volume}{abs/2005.06898}
  (\bibinfo{year}{2020}).
\newblock


\bibitem[\protect\citeauthoryear{Li, Azaria, and Myers}{Li
  et~al\mbox{.}}{2017}]%
        {li_sugilite:_2017}
\bibfield{author}{\bibinfo{person}{Toby Jia-Jun Li}, \bibinfo{person}{Amos
  Azaria}, {and} \bibinfo{person}{Brad~A. Myers}.}
  \bibinfo{year}{2017}\natexlab{}.
\newblock \showarticletitle{{SUGILITE}: {Creating} {Multimodal} {Smartphone}
  {Automation} by {Demonstration}}. In \bibinfo{booktitle}{\emph{Proceedings of
  the 2017 {CHI} {Conference} on {Human} {Factors} in {Computing} {Systems}}}
  \emph{(\bibinfo{series}{{CHI} '17})}. \bibinfo{publisher}{ACM},
  \bibinfo{address}{New York, NY, USA}, \bibinfo{pages}{6038--6049}.
\newblock
\showISBNx{978-1-4503-4655-9}
\urldef\tempurl%
\url{https://doi.org/10.1145/3025453.3025483}
\showDOI{\tempurl}


\bibitem[\protect\citeauthoryear{Li, Chen, Xia, Mitchell, and Myers}{Li
  et~al\mbox{.}}{2020}]%
        {li_sovite:_2020}
\bibfield{author}{\bibinfo{person}{Toby Jia-Jun Li}, \bibinfo{person}{Jingya
  Chen}, \bibinfo{person}{Haijun Xia}, \bibinfo{person}{Tom~M. Mitchell}, {and}
  \bibinfo{person}{Brad~A. Myers}.} \bibinfo{year}{2020}\natexlab{}.
\newblock \showarticletitle{{Multi-Modal} {Repairs} of {Conversational}
  {Breakdowns} in {Task-Oriented} {Dialogs}}. In
  \bibinfo{booktitle}{\emph{Proceedings of the 33rd {Annual} {ACM} {Symposium}
  on {User} {Interface} {Software} and {Technology}}}
  \emph{(\bibinfo{series}{{UIST} 2020})}. \bibinfo{publisher}{ACM}.
\newblock
\urldef\tempurl%
\url{https://doi.org/10.1145/3379337.3415820}
\showDOI{\tempurl}


\bibitem[\protect\citeauthoryear{Li, Radensky, Jia, Singarajah, Mitchell, and
  Myers}{Li et~al\mbox{.}}{2019}]%
        {li_pumice:_2019}
\bibfield{author}{\bibinfo{person}{Toby Jia-Jun Li}, \bibinfo{person}{Marissa
  Radensky}, \bibinfo{person}{Justin Jia}, \bibinfo{person}{Kirielle
  Singarajah}, \bibinfo{person}{Tom~M. Mitchell}, {and}
  \bibinfo{person}{Brad~A. Myers}.} \bibinfo{year}{2019}\natexlab{}.
\newblock \showarticletitle{{PUMICE}: {A} {Multi}-{Modal} {Agent} that {Learns}
  {Concepts} and {Conditionals} from {Natural} {Language} and
  {Demonstrations}}. In \bibinfo{booktitle}{\emph{Proceedings of the 32nd
  {Annual} {ACM} {Symposium} on {User} {Interface} {Software} and
  {Technology}}} \emph{(\bibinfo{series}{{UIST} 2019})}.
  \bibinfo{publisher}{ACM}.
\newblock
\urldef\tempurl%
\url{https://doi.org/10.1145/3332165.3347899}
\showDOI{\tempurl}


\bibitem[\protect\citeauthoryear{Licklider}{Licklider}{1960}]%
        {licklider1960man}
\bibfield{author}{\bibinfo{person}{Joseph~CR Licklider}.}
  \bibinfo{year}{1960}\natexlab{}.
\newblock \showarticletitle{Man-computer symbiosis}.
\newblock \bibinfo{journal}{\emph{IRE transactions on human factors in
  electronics}} \bibinfo{number}{1} (\bibinfo{year}{1960}),
  \bibinfo{pages}{4--11}.
\newblock


\bibitem[\protect\citeauthoryear{Lin, Doll{\'a}r, Girshick, He, Hariharan, and
  Belongie}{Lin et~al\mbox{.}}{2017}]%
        {lin2017feature}
\bibfield{author}{\bibinfo{person}{Tsung-Yi Lin}, \bibinfo{person}{Piotr
  Doll{\'a}r}, \bibinfo{person}{Ross Girshick}, \bibinfo{person}{Kaiming He},
  \bibinfo{person}{Bharath Hariharan}, {and} \bibinfo{person}{Serge Belongie}.}
  \bibinfo{year}{2017}\natexlab{}.
\newblock \showarticletitle{Feature pyramid networks for object detection}. In
  \bibinfo{booktitle}{\emph{Proceedings of the IEEE conference on computer
  vision and pattern recognition}}. \bibinfo{pages}{2117--2125}.
\newblock


\bibitem[\protect\citeauthoryear{Lin, Maire, Belongie, Hays, Perona, Ramanan,
  Doll{\'a}r, and Zitnick}{Lin et~al\mbox{.}}{2014}]%
        {lin2014microsoft}
\bibfield{author}{\bibinfo{person}{Tsung-Yi Lin}, \bibinfo{person}{Michael
  Maire}, \bibinfo{person}{Serge Belongie}, \bibinfo{person}{James Hays},
  \bibinfo{person}{Pietro Perona}, \bibinfo{person}{Deva Ramanan},
  \bibinfo{person}{Piotr Doll{\'a}r}, {and} \bibinfo{person}{C~Lawrence
  Zitnick}.} \bibinfo{year}{2014}\natexlab{}.
\newblock \showarticletitle{Microsoft coco: Common objects in context}. In
  \bibinfo{booktitle}{\emph{European conference on computer vision}}. Springer,
  \bibinfo{pages}{740--755}.
\newblock


\bibitem[\protect\citeauthoryear{Lin, Li, and Wang}{Lin et~al\mbox{.}}{2019}]%
        {lin2019dual}
\bibfield{author}{\bibinfo{person}{Yan-Bo Lin}, \bibinfo{person}{Yu-Jhe Li},
  {and} \bibinfo{person}{Yu-Chiang~Frank Wang}.}
  \bibinfo{year}{2019}\natexlab{}.
\newblock \showarticletitle{Dual-modality seq2seq network for audio-visual
  event localization}. In \bibinfo{booktitle}{\emph{ICASSP 2019-2019 IEEE
  International Conference on Acoustics, Speech and Signal Processing
  (ICASSP)}}. IEEE, \bibinfo{pages}{2002--2006}.
\newblock


\bibitem[\protect\citeauthoryear{Liu, Du, Du, Guo, and Chen}{Liu
  et~al\mbox{.}}{2020a}]%
        {liu2020faster}
\bibfield{author}{\bibinfo{person}{Minzhe Liu}, \bibinfo{person}{Li Du},
  \bibinfo{person}{Yuan Du}, \bibinfo{person}{Ruofan Guo}, {and}
  \bibinfo{person}{Xiaoliang Chen}.} \bibinfo{year}{2020}\natexlab{a}.
\newblock \showarticletitle{Faster Human-Machine Collaboration Bounding Box
  Annotation Framework Based on Active Learning}.
\newblock  (\bibinfo{year}{2020}).
\newblock


\bibitem[\protect\citeauthoryear{Liu, Ram, Vijaykeerthy, Bouneffouf, Bramble,
  Samulowitz, Wang, Conn, and Gray}{Liu et~al\mbox{.}}{2020b}]%
        {liu2020admm}
\bibfield{author}{\bibinfo{person}{Sijia Liu}, \bibinfo{person}{Parikshit Ram},
  \bibinfo{person}{Deepak Vijaykeerthy}, \bibinfo{person}{Djallel Bouneffouf},
  \bibinfo{person}{Gregory Bramble}, \bibinfo{person}{Horst Samulowitz},
  \bibinfo{person}{Dakuo Wang}, \bibinfo{person}{Andrew Conn}, {and}
  \bibinfo{person}{Alexander Gray}.} \bibinfo{year}{2020}\natexlab{b}.
\newblock \showarticletitle{An ADMM based framework for automl pipeline
  configuration}. In \bibinfo{booktitle}{\emph{Proceedings of the AAAI
  Conference on Artificial Intelligence}}, Vol.~\bibinfo{volume}{34}.
  \bibinfo{pages}{4892--4899}.
\newblock


\bibitem[\protect\citeauthoryear{Liu, Albanie, Nagrani, and Zisserman}{Liu
  et~al\mbox{.}}{2019}]%
        {liu2019use}
\bibfield{author}{\bibinfo{person}{Yang Liu}, \bibinfo{person}{Samuel Albanie},
  \bibinfo{person}{Arsha Nagrani}, {and} \bibinfo{person}{Andrew Zisserman}.}
  \bibinfo{year}{2019}\natexlab{}.
\newblock \showarticletitle{Use what you have: Video retrieval using
  representations from collaborative experts}.
\newblock \bibinfo{journal}{\emph{arXiv preprint arXiv:1907.13487}}
  (\bibinfo{year}{2019}).
\newblock


\bibitem[\protect\citeauthoryear{Louie, Coenen, Huang, Terry, and Cai}{Louie
  et~al\mbox{.}}{2020}]%
        {louie_noivce_2020}
\bibfield{author}{\bibinfo{person}{Ryan Louie}, \bibinfo{person}{Andy Coenen},
  \bibinfo{person}{Cheng~Zhi Huang}, \bibinfo{person}{Michael Terry}, {and}
  \bibinfo{person}{Carrie~J. Cai}.} \bibinfo{year}{2020}\natexlab{}.
\newblock \showarticletitle{Novice-AI Music Co-Creation via AI-Steering Tools
  for Deep Generative Models}. In \bibinfo{booktitle}{\emph{Proceedings of the
  2020 CHI Conference on Human Factors in Computing Systems}} (Honolulu, HI,
  USA) \emph{(\bibinfo{series}{CHI '20})}. \bibinfo{publisher}{Association for
  Computing Machinery}, \bibinfo{address}{New York, NY, USA},
  \bibinfo{pages}{1–13}.
\newblock
\showISBNx{9781450367080}
\urldef\tempurl%
\url{https://doi.org/10.1145/3313831.3376739}
\showDOI{\tempurl}


\bibitem[\protect\citeauthoryear{Ma, Hao, Chen, Chen, Lu, and Ko{\v{s}}ir}{Ma
  et~al\mbox{.}}{2019}]%
        {ma2019audio}
\bibfield{author}{\bibinfo{person}{Yaxiong Ma}, \bibinfo{person}{Yixue Hao},
  \bibinfo{person}{Min Chen}, \bibinfo{person}{Jincai Chen},
  \bibinfo{person}{Ping Lu}, {and} \bibinfo{person}{Andrej Ko{\v{s}}ir}.}
  \bibinfo{year}{2019}\natexlab{}.
\newblock \showarticletitle{Audio-visual emotion fusion (AVEF): A deep
  efficient weighted approach}.
\newblock \bibinfo{journal}{\emph{Information Fusion}}  \bibinfo{volume}{46}
  (\bibinfo{year}{2019}), \bibinfo{pages}{184--192}.
\newblock


\bibitem[\protect\citeauthoryear{Mitra, Hutto, and Gilbert}{Mitra
  et~al\mbox{.}}{2015}]%
        {mitra_comparing_2015}
\bibfield{author}{\bibinfo{person}{Tanushree Mitra}, \bibinfo{person}{C.J.
  Hutto}, {and} \bibinfo{person}{Eric Gilbert}.}
  \bibinfo{year}{2015}\natexlab{}.
\newblock \showarticletitle{Comparing Person- and Process-Centric Strategies
  for Obtaining Quality Data on Amazon Mechanical Turk}. In
  \bibinfo{booktitle}{\emph{Proceedings of the 33rd Annual ACM Conference on
  Human Factors in Computing Systems}} (Seoul, Republic of Korea)
  \emph{(\bibinfo{series}{CHI '15})}. \bibinfo{publisher}{Association for
  Computing Machinery}, \bibinfo{address}{New York, NY, USA},
  \bibinfo{pages}{1345–1354}.
\newblock
\showISBNx{9781450331456}
\urldef\tempurl%
\url{https://doi.org/10.1145/2702123.2702553}
\showDOI{\tempurl}


\bibitem[\protect\citeauthoryear{Mollas, Chrysopoulou, Karlos, and
  Tsoumakas}{Mollas et~al\mbox{.}}{2020}]%
        {Mollas2020ETHOSAM}
\bibfield{author}{\bibinfo{person}{Ioannis Mollas}, \bibinfo{person}{Zoe
  Chrysopoulou}, \bibinfo{person}{Stamatis Karlos}, {and}
  \bibinfo{person}{Grigorios Tsoumakas}.} \bibinfo{year}{2020}\natexlab{}.
\newblock \showarticletitle{ETHOS: a multi-label hate speech detection
  dataset}.
\newblock \bibinfo{journal}{\emph{Complex \& Intelligent Systems}}
  (\bibinfo{year}{2020}), \bibinfo{pages}{1--16}.
\newblock


\bibitem[\protect\citeauthoryear{Morgado, Vasconcelos, Langlois, and
  Wang}{Morgado et~al\mbox{.}}{2018}]%
        {morgado2018self}
\bibfield{author}{\bibinfo{person}{Pedro Morgado}, \bibinfo{person}{Nuno
  Vasconcelos}, \bibinfo{person}{Timothy Langlois}, {and}
  \bibinfo{person}{Oliver Wang}.} \bibinfo{year}{2018}\natexlab{}.
\newblock \showarticletitle{Self-supervised generation of spatial audio for 360
  video}.
\newblock \bibinfo{journal}{\emph{arXiv preprint arXiv:1809.02587}}
  (\bibinfo{year}{2018}).
\newblock


\bibitem[\protect\citeauthoryear{Muller, Lange, Wang, Piorkowski, Tsay, Liao,
  Dugan, and Erickson}{Muller et~al\mbox{.}}{2019}]%
        {muller_how_2019}
\bibfield{author}{\bibinfo{person}{Michael Muller}, \bibinfo{person}{Ingrid
  Lange}, \bibinfo{person}{Dakuo Wang}, \bibinfo{person}{David Piorkowski},
  \bibinfo{person}{Jason Tsay}, \bibinfo{person}{Q.~Vera Liao},
  \bibinfo{person}{Casey Dugan}, {and} \bibinfo{person}{Thomas Erickson}.}
  \bibinfo{year}{2019}\natexlab{}.
\newblock \showarticletitle{How Data Science Workers Work with Data: Discovery,
  Capture, Curation, Design, Creation}. In
  \bibinfo{booktitle}{\emph{Proceedings of the 2019 CHI Conference on Human
  Factors in Computing Systems}} (Glasgow, Scotland Uk)
  \emph{(\bibinfo{series}{CHI '19})}. \bibinfo{publisher}{Association for
  Computing Machinery}, \bibinfo{address}{New York, NY, USA},
  \bibinfo{pages}{1–15}.
\newblock
\showISBNx{9781450359702}
\urldef\tempurl%
\url{https://doi.org/10.1145/3290605.3300356}
\showDOI{\tempurl}


\bibitem[\protect\citeauthoryear{Muller, Wolf, Andres, Desmond, Joshi,
  Ashktorab, Sharma, Brimijoin, Pan, Duesterwald, and Dugan}{Muller
  et~al\mbox{.}}{2021}]%
        {muller_2021_ground-truth}
\bibfield{author}{\bibinfo{person}{Michael Muller},
  \bibinfo{person}{Christine~T. Wolf}, \bibinfo{person}{Josh Andres},
  \bibinfo{person}{Michael Desmond}, \bibinfo{person}{Narendra~Nath Joshi},
  \bibinfo{person}{Zahra Ashktorab}, \bibinfo{person}{Aabhas Sharma},
  \bibinfo{person}{Kristina Brimijoin}, \bibinfo{person}{Qian Pan},
  \bibinfo{person}{Evelyn Duesterwald}, {and} \bibinfo{person}{Casey Dugan}.}
  \bibinfo{year}{2021}\natexlab{}.
\newblock \showarticletitle{Designing Ground Truth and the Social Life of
  Labels}. In \bibinfo{booktitle}{\emph{Proceedings of the 2021 CHI Conference
  on Human Factors in Computing Systems}} (Yokohama, Japan)
  \emph{(\bibinfo{series}{CHI '21})}. \bibinfo{publisher}{Association for
  Computing Machinery}, \bibinfo{address}{New York, NY, USA}, Article
  \bibinfo{articleno}{94}, \bibinfo{numpages}{16}~pages.
\newblock
\showISBNx{9781450380966}
\urldef\tempurl%
\url{https://doi.org/10.1145/3411764.3445402}
\showDOI{\tempurl}


\bibitem[\protect\citeauthoryear{Murray and Wallace}{Murray and
  Wallace}{2011}]%
        {murray2011neural}
\bibfield{author}{\bibinfo{person}{Micah~M Murray} {and}
  \bibinfo{person}{Mark~T Wallace}.} \bibinfo{year}{2011}\natexlab{}.
\newblock \showarticletitle{The neural bases of multisensory processes}.
\newblock  (\bibinfo{year}{2011}).
\newblock


\bibitem[\protect\citeauthoryear{Neves and Leser}{Neves and Leser}{2014}]%
        {neves2014survey}
\bibfield{author}{\bibinfo{person}{Mariana Neves} {and} \bibinfo{person}{Ulf
  Leser}.} \bibinfo{year}{2014}\natexlab{}.
\newblock \showarticletitle{A survey on annotation tools for the biomedical
  literature}.
\newblock \bibinfo{journal}{\emph{Briefings in bioinformatics}}
  \bibinfo{volume}{15}, \bibinfo{number}{2} (\bibinfo{year}{2014}),
  \bibinfo{pages}{327--340}.
\newblock


\bibitem[\protect\citeauthoryear{Ning, Zhang, Sun, Tian, Zhang, and Li}{Ning
  et~al\mbox{.}}{2023}]%
        {ning2023empirical}
\bibfield{author}{\bibinfo{person}{Zheng Ning}, \bibinfo{person}{Zheng Zhang},
  \bibinfo{person}{Tianyi Sun}, \bibinfo{person}{Yuan Tian},
  \bibinfo{person}{Tianyi Zhang}, {and} \bibinfo{person}{Toby Jia-Jun Li}.}
  \bibinfo{year}{2023}\natexlab{}.
\newblock \showarticletitle{An empirical study of model errors and user error
  discovery and repair strategies in natural language database queries}. In
  \bibinfo{booktitle}{\emph{Proceedings of the 28th International Conference on
  Intelligent User Interfaces}}. \bibinfo{pages}{633--649}.
\newblock


\bibitem[\protect\citeauthoryear{Noroozi, Marjanovic, Njegus, Escalera, and
  Anbarjafari}{Noroozi et~al\mbox{.}}{2017}]%
        {noroozi2017audio}
\bibfield{author}{\bibinfo{person}{Fatemeh Noroozi}, \bibinfo{person}{Marina
  Marjanovic}, \bibinfo{person}{Angelina Njegus}, \bibinfo{person}{Sergio
  Escalera}, {and} \bibinfo{person}{Gholamreza Anbarjafari}.}
  \bibinfo{year}{2017}\natexlab{}.
\newblock \showarticletitle{Audio-visual emotion recognition in video clips}.
\newblock \bibinfo{journal}{\emph{IEEE Transactions on Affective Computing}}
  \bibinfo{volume}{10}, \bibinfo{number}{1} (\bibinfo{year}{2017}),
  \bibinfo{pages}{60--75}.
\newblock


\bibitem[\protect\citeauthoryear{Oviatt}{Oviatt}{1999a}]%
        {oviatt_mutual_1999}
\bibfield{author}{\bibinfo{person}{Sharon Oviatt}.}
  \bibinfo{year}{1999}\natexlab{a}.
\newblock \showarticletitle{Mutual disambiguation of recognition errors in a
  multimodal architecture}. In \bibinfo{booktitle}{\emph{Proceedings of the
  {SIGCHI} conference on {Human} {Factors} in {Computing} {Systems}}}.
  \bibinfo{publisher}{ACM}, \bibinfo{pages}{576--583}.
\newblock


\bibitem[\protect\citeauthoryear{Oviatt}{Oviatt}{1999b}]%
        {oviatt_ten_1999}
\bibfield{author}{\bibinfo{person}{Sharon Oviatt}.}
  \bibinfo{year}{1999}\natexlab{b}.
\newblock \showarticletitle{Ten {Myths} of {Multimodal} {Interaction}}.
\newblock \bibinfo{journal}{\emph{Commun. ACM}} \bibinfo{volume}{42},
  \bibinfo{number}{11} (\bibinfo{date}{Nov.} \bibinfo{year}{1999}),
  \bibinfo{pages}{74--81}.
\newblock
\showISSN{0001-0782}
\urldef\tempurl%
\url{https://doi.org/10.1145/319382.319398}
\showDOI{\tempurl}


\bibitem[\protect\citeauthoryear{Oviatt and Cohen}{Oviatt and Cohen}{2000}]%
        {oviatt_2000_perceptual}
\bibfield{author}{\bibinfo{person}{Sharon Oviatt} {and} \bibinfo{person}{Philip
  Cohen}.} \bibinfo{year}{2000}\natexlab{}.
\newblock \showarticletitle{Perceptual User Interfaces: Multimodal Interfaces
  That Process What Comes Naturally}.
\newblock \bibinfo{journal}{\emph{Commun. ACM}} \bibinfo{volume}{43},
  \bibinfo{number}{3} (\bibinfo{date}{March} \bibinfo{year}{2000}),
  \bibinfo{pages}{45–53}.
\newblock
\showISSN{0001-0782}
\urldef\tempurl%
\url{https://doi.org/10.1145/330534.330538}
\showDOI{\tempurl}


\bibitem[\protect\citeauthoryear{Owens and Efros}{Owens and Efros}{2018}]%
        {owens2018audio}
\bibfield{author}{\bibinfo{person}{Andrew Owens} {and}
  \bibinfo{person}{Alexei~A Efros}.} \bibinfo{year}{2018}\natexlab{}.
\newblock \showarticletitle{Audio-visual scene analysis with self-supervised
  multisensory features}. In \bibinfo{booktitle}{\emph{Proceedings of the
  European Conference on Computer Vision (ECCV)}}. \bibinfo{pages}{631--648}.
\newblock


\bibitem[\protect\citeauthoryear{Pavel, Reyes, and Bigham}{Pavel
  et~al\mbox{.}}{2020}]%
        {pavel2020rescribe}
\bibfield{author}{\bibinfo{person}{Amy Pavel}, \bibinfo{person}{Gabriel Reyes},
  {and} \bibinfo{person}{Jeffrey~P Bigham}.} \bibinfo{year}{2020}\natexlab{}.
\newblock \showarticletitle{Rescribe: Authoring and Automatically Editing Audio
  Descriptions}. In \bibinfo{booktitle}{\emph{Proceedings of the 33rd Annual
  ACM Symposium on User Interface Software and Technology}}.
  \bibinfo{pages}{747--759}.
\newblock


\bibitem[\protect\citeauthoryear{Qian, Hu, Dinkel, Wu, Xu, and Lin}{Qian
  et~al\mbox{.}}{2020}]%
        {qian2020multiple}
\bibfield{author}{\bibinfo{person}{Rui Qian}, \bibinfo{person}{Di Hu},
  \bibinfo{person}{Heinrich Dinkel}, \bibinfo{person}{Mengyue Wu},
  \bibinfo{person}{Ning Xu}, {and} \bibinfo{person}{Weiyao Lin}.}
  \bibinfo{year}{2020}\natexlab{}.
\newblock \showarticletitle{Multiple Sound Sources Localization from Coarse to
  Fine}. In \bibinfo{booktitle}{\emph{ECCV}}.
\newblock


\bibitem[\protect\citeauthoryear{Qiao, Sun, Liu, Xia, Luo, Zhang, and Kuo}{Qiao
  et~al\mbox{.}}{2022}]%
        {Qiao2022HumanintheLoopVS}
\bibfield{author}{\bibinfo{person}{Nan Qiao}, \bibinfo{person}{Yuyin Sun},
  \bibinfo{person}{Chongyu Liu}, \bibinfo{person}{Lu Xia},
  \bibinfo{person}{Jiajia Luo}, \bibinfo{person}{K. Zhang}, {and}
  \bibinfo{person}{Cheng-Hao Kuo}.} \bibinfo{year}{2022}\natexlab{}.
\newblock \showarticletitle{Human-in-the-Loop Video Semantic Segmentation
  Auto-Annotation}.
\newblock


\bibitem[\protect\citeauthoryear{Rahman, Xu, and Sigal}{Rahman
  et~al\mbox{.}}{2019}]%
        {rahman2019watch}
\bibfield{author}{\bibinfo{person}{Tanzila Rahman}, \bibinfo{person}{Bicheng
  Xu}, {and} \bibinfo{person}{Leonid Sigal}.} \bibinfo{year}{2019}\natexlab{}.
\newblock \showarticletitle{Watch, listen and tell: Multi-modal weakly
  supervised dense event captioning}. In \bibinfo{booktitle}{\emph{Proceedings
  of the IEEE/CVF International Conference on Computer Vision}}.
  \bibinfo{pages}{8908--8917}.
\newblock


\bibitem[\protect\citeauthoryear{Rashid, Albert, Cosley, Lam, McNee, Konstan,
  and Riedl}{Rashid et~al\mbox{.}}{2002}]%
        {rashid_getting_2002}
\bibfield{author}{\bibinfo{person}{Al~Mamunur Rashid}, \bibinfo{person}{Istvan
  Albert}, \bibinfo{person}{Dan Cosley}, \bibinfo{person}{Shyong~K. Lam},
  \bibinfo{person}{Sean~M. McNee}, \bibinfo{person}{Joseph~A. Konstan}, {and}
  \bibinfo{person}{John Riedl}.} \bibinfo{year}{2002}\natexlab{}.
\newblock \showarticletitle{Getting to Know You: Learning New User Preferences
  in Recommender Systems}. In \bibinfo{booktitle}{\emph{Proceedings of the 7th
  International Conference on Intelligent User Interfaces}} (San Francisco,
  California, USA) \emph{(\bibinfo{series}{IUI '02})}.
  \bibinfo{publisher}{Association for Computing Machinery},
  \bibinfo{address}{New York, NY, USA}, \bibinfo{pages}{127–134}.
\newblock
\showISBNx{1581134592}
\urldef\tempurl%
\url{https://doi.org/10.1145/502716.502737}
\showDOI{\tempurl}


\bibitem[\protect\citeauthoryear{Ratner, Bach, Ehrenberg, Fries, Wu, and
  R{\'e}}{Ratner et~al\mbox{.}}{2017}]%
        {ratner2017snorkel}
\bibfield{author}{\bibinfo{person}{Alexander Ratner},
  \bibinfo{person}{Stephen~H Bach}, \bibinfo{person}{Henry Ehrenberg},
  \bibinfo{person}{Jason Fries}, \bibinfo{person}{Sen Wu}, {and}
  \bibinfo{person}{Christopher R{\'e}}.} \bibinfo{year}{2017}\natexlab{}.
\newblock \showarticletitle{Snorkel: Rapid training data creation with weak
  supervision}. In \bibinfo{booktitle}{\emph{Proceedings of the VLDB Endowment.
  International Conference on Very Large Data Bases}},
  Vol.~\bibinfo{volume}{11}. \bibinfo{pages}{269}.
\newblock


\bibitem[\protect\citeauthoryear{Ren, He, Girshick, and Sun}{Ren
  et~al\mbox{.}}{2015}]%
        {ren2015faster}
\bibfield{author}{\bibinfo{person}{Shaoqing Ren}, \bibinfo{person}{Kaiming He},
  \bibinfo{person}{Ross Girshick}, {and} \bibinfo{person}{Jian Sun}.}
  \bibinfo{year}{2015}\natexlab{}.
\newblock \showarticletitle{Faster r-cnn: Towards real-time object detection
  with region proposal networks}.
\newblock \bibinfo{journal}{\emph{Advances in neural information processing
  systems}}  \bibinfo{volume}{28} (\bibinfo{year}{2015}),
  \bibinfo{pages}{91--99}.
\newblock


\bibitem[\protect\citeauthoryear{Richard, Lea, Ma, Gall, De~la Torre, and
  Sheikh}{Richard et~al\mbox{.}}{2021}]%
        {richard2021audio}
\bibfield{author}{\bibinfo{person}{Alexander Richard}, \bibinfo{person}{Colin
  Lea}, \bibinfo{person}{Shugao Ma}, \bibinfo{person}{Jurgen Gall},
  \bibinfo{person}{Fernando De~la Torre}, {and} \bibinfo{person}{Yaser
  Sheikh}.} \bibinfo{year}{2021}\natexlab{}.
\newblock \showarticletitle{Audio-and gaze-driven facial animation of codec
  avatars}. In \bibinfo{booktitle}{\emph{Proceedings of the IEEE/CVF Winter
  Conference on Applications of Computer Vision}}. \bibinfo{pages}{41--50}.
\newblock


\bibitem[\protect\citeauthoryear{Rietz and Maedche}{Rietz and Maedche}{2021}]%
        {rietz2021cody}
\bibfield{author}{\bibinfo{person}{Tim Rietz} {and} \bibinfo{person}{Alexander
  Maedche}.} \bibinfo{year}{2021}\natexlab{}.
\newblock \showarticletitle{Cody: An AI-based system to semi-automate coding
  for qualitative research}. In \bibinfo{booktitle}{\emph{Proceedings of the
  2021 CHI Conference on Human Factors in Computing Systems}}.
  \bibinfo{pages}{1--14}.
\newblock


\bibitem[\protect\citeauthoryear{Sambasivan, Kapania, Highfill, Akrong,
  Paritosh, and Aroyo}{Sambasivan et~al\mbox{.}}{2021}]%
        {sambasivan_2021_everyone}
\bibfield{author}{\bibinfo{person}{Nithya Sambasivan}, \bibinfo{person}{Shivani
  Kapania}, \bibinfo{person}{Hannah Highfill}, \bibinfo{person}{Diana Akrong},
  \bibinfo{person}{Praveen Paritosh}, {and} \bibinfo{person}{Lora~M Aroyo}.}
  \bibinfo{year}{2021}\natexlab{}.
\newblock \showarticletitle{“Everyone Wants to Do the Model Work, Not the
  Data Work”: Data Cascades in High-Stakes AI}. In
  \bibinfo{booktitle}{\emph{Proceedings of the 2021 CHI Conference on Human
  Factors in Computing Systems}} (Yokohama, Japan) \emph{(\bibinfo{series}{CHI
  '21})}. \bibinfo{publisher}{Association for Computing Machinery},
  \bibinfo{address}{New York, NY, USA}, Article \bibinfo{articleno}{39},
  \bibinfo{numpages}{15}~pages.
\newblock
\showISBNx{9781450380966}
\urldef\tempurl%
\url{https://doi.org/10.1145/3411764.3445518}
\showDOI{\tempurl}


\bibitem[\protect\citeauthoryear{Schein, Popescul, Ungar, and Pennock}{Schein
  et~al\mbox{.}}{2002}]%
        {schein2002methods}
\bibfield{author}{\bibinfo{person}{Andrew~I Schein},
  \bibinfo{person}{Alexandrin Popescul}, \bibinfo{person}{Lyle~H Ungar}, {and}
  \bibinfo{person}{David~M Pennock}.} \bibinfo{year}{2002}\natexlab{}.
\newblock \showarticletitle{Methods and metrics for cold-start
  recommendations}. In \bibinfo{booktitle}{\emph{Proceedings of the 25th annual
  international ACM SIGIR conference on Research and development in information
  retrieval}}. \bibinfo{pages}{253--260}.
\newblock


\bibitem[\protect\citeauthoryear{Schmitz, Soderland, Bart, Etzioni,
  et~al\mbox{.}}{Schmitz et~al\mbox{.}}{2012}]%
        {schmitz2012open}
\bibfield{author}{\bibinfo{person}{Michael Schmitz}, \bibinfo{person}{Stephen
  Soderland}, \bibinfo{person}{Robert Bart}, \bibinfo{person}{Oren Etzioni},
  {et~al\mbox{.}}} \bibinfo{year}{2012}\natexlab{}.
\newblock \showarticletitle{Open language learning for information extraction}.
  In \bibinfo{booktitle}{\emph{Proceedings of the 2012 joint conference on
  empirical methods in natural language processing and computational natural
  language learning}}. \bibinfo{pages}{523--534}.
\newblock


\bibitem[\protect\citeauthoryear{Senocak, Oh, Kim, Yang, and Kweon}{Senocak
  et~al\mbox{.}}{2018}]%
        {senocak2018learning}
\bibfield{author}{\bibinfo{person}{Arda Senocak}, \bibinfo{person}{Tae-Hyun
  Oh}, \bibinfo{person}{Junsik Kim}, \bibinfo{person}{Ming-Hsuan Yang}, {and}
  \bibinfo{person}{In~So Kweon}.} \bibinfo{year}{2018}\natexlab{}.
\newblock \showarticletitle{Learning to localize sound source in visual
  scenes}. In \bibinfo{booktitle}{\emph{Proceedings of the IEEE Conference on
  Computer Vision and Pattern Recognition}}. \bibinfo{pages}{4358--4366}.
\newblock


\bibitem[\protect\citeauthoryear{Settles}{Settles}{2012}]%
        {settles_activelearning_2012}
\bibfield{author}{\bibinfo{person}{Burr Settles}.}
  \bibinfo{year}{2012}\natexlab{}.
\newblock \showarticletitle{Active Learning}.
\newblock \bibinfo{journal}{\emph{Synthesis Lectures on Artificial Intelligence
  and Machine Learning}} \bibinfo{volume}{6}, \bibinfo{number}{1}
  (\bibinfo{year}{2012}), \bibinfo{pages}{1--114}.
\newblock
\urldef\tempurl%
\url{https://doi.org/10.2200/S00429ED1V01Y201207AIM018}
\showDOI{\tempurl}
\showeprint{https://doi.org/10.2200/S00429ED1V01Y201207AIM018}


\bibitem[\protect\citeauthoryear{Shnarch, Halfon, Gera, Danilevsky, Katsis,
  Choshen, Cooper, Epelboim, Zhang, Wang, et~al\mbox{.}}{Shnarch
  et~al\mbox{.}}{2022a}]%
        {shnarch2022label}
\bibfield{author}{\bibinfo{person}{Eyal Shnarch}, \bibinfo{person}{Alon
  Halfon}, \bibinfo{person}{Ariel Gera}, \bibinfo{person}{Marina Danilevsky},
  \bibinfo{person}{Yannis Katsis}, \bibinfo{person}{Leshem Choshen},
  \bibinfo{person}{Martin~Santillan Cooper}, \bibinfo{person}{Dina Epelboim},
  \bibinfo{person}{Zheng Zhang}, \bibinfo{person}{Dakuo Wang}, {et~al\mbox{.}}}
  \bibinfo{year}{2022}\natexlab{a}.
\newblock \showarticletitle{Label Sleuth: From Unlabeled Text to a Classifier
  in a Few Hours}.
\newblock \bibinfo{journal}{\emph{arXiv preprint arXiv:2208.01483}}
  (\bibinfo{year}{2022}).
\newblock


\bibitem[\protect\citeauthoryear{Shnarch, Halfon, Gera, Danilevsky, Katsis,
  Choshen, Cooper, Epelboim, Zhang, Wang, Yip, Ein-Dor, Dankin, Shnayderman,
  Aharonov, Li, Liberman, Slesarev, Newton, Ofek-Koifman, Slonim, and
  Katz}{Shnarch et~al\mbox{.}}{2022b}]%
        {labelsleuth2022}
\bibfield{author}{\bibinfo{person}{Eyal Shnarch}, \bibinfo{person}{Alon
  Halfon}, \bibinfo{person}{Ariel Gera}, \bibinfo{person}{Marina Danilevsky},
  \bibinfo{person}{Yannis Katsis}, \bibinfo{person}{Leshem Choshen},
  \bibinfo{person}{Martin~Santillan Cooper}, \bibinfo{person}{Dina Epelboim},
  \bibinfo{person}{Zheng Zhang}, \bibinfo{person}{Dakuo Wang},
  \bibinfo{person}{Lucy Yip}, \bibinfo{person}{Liat Ein-Dor},
  \bibinfo{person}{Lena Dankin}, \bibinfo{person}{Ilya Shnayderman},
  \bibinfo{person}{Ranit Aharonov}, \bibinfo{person}{Yunyao Li},
  \bibinfo{person}{Naftali Liberman}, \bibinfo{person}{Philip~Levin Slesarev},
  \bibinfo{person}{Gwilym Newton}, \bibinfo{person}{Shila Ofek-Koifman},
  \bibinfo{person}{Noam Slonim}, {and} \bibinfo{person}{Yoav Katz}.}
  \bibinfo{year}{2022}\natexlab{b}.
\newblock \showarticletitle{{Label} {Sleuth}: From Unlabeled Text to a
  Classifier in a Few Hours}. In \bibinfo{booktitle}{\emph{Proceedings of the
  2022 Conference on Empirical Methods in Natural Language Processing: System
  Demonstrations}}. \bibinfo{publisher}{Association for Computational
  Linguistics}.
\newblock
\urldef\tempurl%
\url{https://arxiv.org/abs/2208.01483}
\showURL{%
\tempurl}


\bibitem[\protect\citeauthoryear{Song, Zhang, Wang, and Gildea}{Song
  et~al\mbox{.}}{2018}]%
        {song2018graph}
\bibfield{author}{\bibinfo{person}{Linfeng Song}, \bibinfo{person}{Yue Zhang},
  \bibinfo{person}{Zhiguo Wang}, {and} \bibinfo{person}{Daniel Gildea}.}
  \bibinfo{year}{2018}\natexlab{}.
\newblock \showarticletitle{A graph-to-sequence model for AMR-to-text
  generation}.
\newblock \bibinfo{journal}{\emph{arXiv preprint arXiv:1805.02473}}
  (\bibinfo{year}{2018}).
\newblock


\bibitem[\protect\citeauthoryear{Tian, Guan, Goodman, Moore, and Xu}{Tian
  et~al\mbox{.}}{2018a}]%
        {tian2018attempt}
\bibfield{author}{\bibinfo{person}{Yapeng Tian}, \bibinfo{person}{Chenxiao
  Guan}, \bibinfo{person}{Justin Goodman}, \bibinfo{person}{Marc Moore}, {and}
  \bibinfo{person}{Chenliang Xu}.} \bibinfo{year}{2018}\natexlab{a}.
\newblock \showarticletitle{An attempt towards interpretable audio-visual video
  captioning}.
\newblock \bibinfo{journal}{\emph{arXiv preprint arXiv:1812.02872}}
  (\bibinfo{year}{2018}).
\newblock


\bibitem[\protect\citeauthoryear{Tian, Hu, and Xu}{Tian et~al\mbox{.}}{2021}]%
        {Tian_2021_CVPR}
\bibfield{author}{\bibinfo{person}{Yapeng Tian}, \bibinfo{person}{Di Hu}, {and}
  \bibinfo{person}{Chenliang Xu}.} \bibinfo{year}{2021}\natexlab{}.
\newblock \showarticletitle{Cyclic Co-Learning of Sounding Object Visual
  Grounding and Sound Separation}. In \bibinfo{booktitle}{\emph{Proceedings of
  the IEEE/CVF Conference on Computer Vision and Pattern Recognition (CVPR)}}.
  \bibinfo{pages}{2745--2754}.
\newblock


\bibitem[\protect\citeauthoryear{Tian, Li, and Xu}{Tian et~al\mbox{.}}{2020}]%
        {tian2020unified}
\bibfield{author}{\bibinfo{person}{Yapeng Tian}, \bibinfo{person}{Dingzeyu Li},
  {and} \bibinfo{person}{Chenliang Xu}.} \bibinfo{year}{2020}\natexlab{}.
\newblock \showarticletitle{Unified multisensory perception: Weakly-supervised
  audio-visual video parsing}. In \bibinfo{booktitle}{\emph{Computer
  Vision--ECCV 2020: 16th European Conference, Glasgow, UK, August 23--28,
  2020, Proceedings, Part III 16}}. Springer, \bibinfo{pages}{436--454}.
\newblock


\bibitem[\protect\citeauthoryear{Tian, Shi, Li, Duan, and Xu}{Tian
  et~al\mbox{.}}{2018b}]%
        {tian2018audio}
\bibfield{author}{\bibinfo{person}{Yapeng Tian}, \bibinfo{person}{Jing Shi},
  \bibinfo{person}{Bochen Li}, \bibinfo{person}{Zhiyao Duan}, {and}
  \bibinfo{person}{Chenliang Xu}.} \bibinfo{year}{2018}\natexlab{b}.
\newblock \showarticletitle{Audio-visual event localization in unconstrained
  videos}. In \bibinfo{booktitle}{\emph{Proceedings of the European Conference
  on Computer Vision (ECCV)}}. \bibinfo{pages}{247--263}.
\newblock


\bibitem[\protect\citeauthoryear{Vajda, Rangoni, and Cecotti}{Vajda
  et~al\mbox{.}}{2015}]%
        {vajda_2015_semiautomatic}
\bibfield{author}{\bibinfo{person}{Szil\'{a}rd Vajda}, \bibinfo{person}{Yves
  Rangoni}, {and} \bibinfo{person}{Hubert Cecotti}.}
  \bibinfo{year}{2015}\natexlab{}.
\newblock \showarticletitle{Semi-Automatic Ground Truth Generation Using
  Unsupervised Clustering and Limited Manual Labeling}.
\newblock \bibinfo{journal}{\emph{Pattern Recogn. Lett.}} \bibinfo{volume}{58},
  \bibinfo{number}{C} (\bibinfo{date}{June} \bibinfo{year}{2015}),
  \bibinfo{pages}{23–28}.
\newblock
\showISSN{0167-8655}
\urldef\tempurl%
\url{https://doi.org/10.1016/j.patrec.2015.02.001}
\showDOI{\tempurl}


\bibitem[\protect\citeauthoryear{Von~Ahn, Maurer, McMillen, Abraham, and
  Blum}{Von~Ahn et~al\mbox{.}}{2008}]%
        {von2008recaptcha}
\bibfield{author}{\bibinfo{person}{Luis Von~Ahn}, \bibinfo{person}{Benjamin
  Maurer}, \bibinfo{person}{Colin McMillen}, \bibinfo{person}{David Abraham},
  {and} \bibinfo{person}{Manuel Blum}.} \bibinfo{year}{2008}\natexlab{}.
\newblock \showarticletitle{recaptcha: Human-based character recognition via
  web security measures}.
\newblock \bibinfo{journal}{\emph{Science}} \bibinfo{volume}{321},
  \bibinfo{number}{5895} (\bibinfo{year}{2008}), \bibinfo{pages}{1465--1468}.
\newblock


\bibitem[\protect\citeauthoryear{Wada}{Wada}{2016}]%
        {labelme2016}
\bibfield{author}{\bibinfo{person}{Kentaro Wada}.}
  \bibinfo{year}{2016}\natexlab{}.
\newblock \bibinfo{title}{{labelme: Image Polygonal Annotation with Python}}.
\newblock \bibinfo{howpublished}{\url{https://github.com/wkentaro/labelme}}.
\newblock


\bibitem[\protect\citeauthoryear{Wang, Andres, Weisz, Oduor, and Dugan}{Wang
  et~al\mbox{.}}{2021a}]%
        {dakuo_autods_2021}
\bibfield{author}{\bibinfo{person}{Dakuo Wang}, \bibinfo{person}{Josh Andres},
  \bibinfo{person}{Justin~D. Weisz}, \bibinfo{person}{Erick Oduor}, {and}
  \bibinfo{person}{Casey Dugan}.} \bibinfo{year}{2021}\natexlab{a}.
\newblock \showarticletitle{AutoDS: Towards Human-Centered Automation of Data
  Science}. In \bibinfo{booktitle}{\emph{Proceedings of the 2021 CHI Conference
  on Human Factors in Computing Systems}} (Yokohama, Japan)
  \emph{(\bibinfo{series}{CHI '21})}. \bibinfo{publisher}{Association for
  Computing Machinery}, \bibinfo{address}{New York, NY, USA}, Article
  \bibinfo{articleno}{79}, \bibinfo{numpages}{12}~pages.
\newblock
\showISBNx{9781450380966}
\urldef\tempurl%
\url{https://doi.org/10.1145/3411764.3445526}
\showDOI{\tempurl}


\bibitem[\protect\citeauthoryear{Wang, Maes, Ren, Shneiderman, Shi, and
  Wang}{Wang et~al\mbox{.}}{2021c}]%
        {wang_designing_2021}
\bibfield{author}{\bibinfo{person}{Dakuo Wang}, \bibinfo{person}{Pattie Maes},
  \bibinfo{person}{Xiangshi Ren}, \bibinfo{person}{Ben Shneiderman},
  \bibinfo{person}{Yuanchun Shi}, {and} \bibinfo{person}{Qianying Wang}.}
  \bibinfo{year}{2021}\natexlab{c}.
\newblock \bibinfo{booktitle}{\emph{Designing AI to Work WITH or FOR People?}}
\newblock \bibinfo{publisher}{ACM}, \bibinfo{address}{New York, NY, USA}.
\newblock
\showISBNx{9781450380959}
\urldef\tempurl%
\url{https://doi.org/10.1145/3411763.3450394}
\showURL{%
\tempurl}


\bibitem[\protect\citeauthoryear{Wang, Weisz, Muller, Ram, Geyer, Dugan,
  Tausczik, Samulowitz, and Gray}{Wang et~al\mbox{.}}{2019}]%
        {wang_human-ai_2019}
\bibfield{author}{\bibinfo{person}{Dakuo Wang}, \bibinfo{person}{Justin~D.
  Weisz}, \bibinfo{person}{Michael Muller}, \bibinfo{person}{Parikshit Ram},
  \bibinfo{person}{Werner Geyer}, \bibinfo{person}{Casey Dugan},
  \bibinfo{person}{Yla Tausczik}, \bibinfo{person}{Horst Samulowitz}, {and}
  \bibinfo{person}{Alexander Gray}.} \bibinfo{year}{2019}\natexlab{}.
\newblock \showarticletitle{Human-AI Collaboration in Data Science: Exploring
  Data Scientists' Perceptions of Automated AI}.
\newblock \bibinfo{journal}{\emph{Proc. ACM Hum.-Comput. Interact.}}
  \bibinfo{volume}{3}, \bibinfo{number}{CSCW}, Article \bibinfo{articleno}{211}
  (\bibinfo{date}{Nov.} \bibinfo{year}{2019}), \bibinfo{numpages}{24}~pages.
\newblock
\urldef\tempurl%
\url{https://doi.org/10.1145/3359313}
\showDOI{\tempurl}


\bibitem[\protect\citeauthoryear{Wang, Liang, Huang, Zhang, Li, and Yu}{Wang
  et~al\mbox{.}}{2021b}]%
        {wang2021toward}
\bibfield{author}{\bibinfo{person}{Yujia Wang}, \bibinfo{person}{Wei Liang},
  \bibinfo{person}{Haikun Huang}, \bibinfo{person}{Yongqi Zhang},
  \bibinfo{person}{Dingzeyu Li}, {and} \bibinfo{person}{Lap-Fai Yu}.}
  \bibinfo{year}{2021}\natexlab{b}.
\newblock \showarticletitle{Toward Automatic Audio Description Generation for
  Accessible Videos}. In \bibinfo{booktitle}{\emph{Proceedings of the 2021 CHI
  Conference on Human Factors in Computing Systems}}. \bibinfo{pages}{1--12}.
\newblock


\bibitem[\protect\citeauthoryear{Wongsuphasawat, Qu, Moritz, Chang, Ouk, Anand,
  Mackinlay, Howe, and Heer}{Wongsuphasawat et~al\mbox{.}}{2017}]%
        {wongsuphasawat_voyager_2017}
\bibfield{author}{\bibinfo{person}{Kanit Wongsuphasawat},
  \bibinfo{person}{Zening Qu}, \bibinfo{person}{Dominik Moritz},
  \bibinfo{person}{Riley Chang}, \bibinfo{person}{Felix Ouk},
  \bibinfo{person}{Anushka Anand}, \bibinfo{person}{Jock Mackinlay},
  \bibinfo{person}{Bill Howe}, {and} \bibinfo{person}{Jeffrey Heer}.}
  \bibinfo{year}{2017}\natexlab{}.
\newblock \showarticletitle{Voyager 2: Augmenting Visual Analysis with Partial
  View Specifications}. In \bibinfo{booktitle}{\emph{Proceedings of the 2017
  CHI Conference on Human Factors in Computing Systems}} (Denver, Colorado,
  USA) \emph{(\bibinfo{series}{CHI '17})}. \bibinfo{publisher}{Association for
  Computing Machinery}, \bibinfo{address}{New York, NY, USA},
  \bibinfo{pages}{2648–2659}.
\newblock
\showISBNx{9781450346559}
\urldef\tempurl%
\url{https://doi.org/10.1145/3025453.3025768}
\showDOI{\tempurl}


\bibitem[\protect\citeauthoryear{Wu, Ribeiro, Heer, and Weld}{Wu
  et~al\mbox{.}}{2021}]%
        {wu2021polyjuice}
\bibfield{author}{\bibinfo{person}{Tongshuang Wu}, \bibinfo{person}{Marco~Tulio
  Ribeiro}, \bibinfo{person}{Jeffrey Heer}, {and} \bibinfo{person}{Daniel~S
  Weld}.} \bibinfo{year}{2021}\natexlab{}.
\newblock \showarticletitle{Polyjuice: Generating Counterfactuals for
  Explaining, Evaluating, and Improving Models}. In
  \bibinfo{booktitle}{\emph{Proceedings of the 59th Annual Meeting of the
  Association for Computational Linguistics}}.
\newblock


\bibitem[\protect\citeauthoryear{Wu, Kirillov, Massa, Lo, and Girshick}{Wu
  et~al\mbox{.}}{2019a}]%
        {wu2019detectron2}
\bibfield{author}{\bibinfo{person}{Yuxin Wu}, \bibinfo{person}{Alexander
  Kirillov}, \bibinfo{person}{Francisco Massa}, \bibinfo{person}{Wan-Yen Lo},
  {and} \bibinfo{person}{Ross Girshick}.} \bibinfo{year}{2019}\natexlab{a}.
\newblock \bibinfo{title}{Detectron2}.
\newblock
  \bibinfo{howpublished}{\url{https://github.com/facebookresearch/detectron2}}.
\newblock


\bibitem[\protect\citeauthoryear{Wu and Yang}{Wu and Yang}{2021}]%
        {wu2021exploring}
\bibfield{author}{\bibinfo{person}{Yu Wu} {and} \bibinfo{person}{Yi Yang}.}
  \bibinfo{year}{2021}\natexlab{}.
\newblock \showarticletitle{Exploring Heterogeneous Clues for Weakly-Supervised
  Audio-Visual Video Parsing}. In \bibinfo{booktitle}{\emph{Proceedings of the
  IEEE/CVF Conference on Computer Vision and Pattern Recognition}}.
  \bibinfo{pages}{1326--1335}.
\newblock


\bibitem[\protect\citeauthoryear{Wu, Zhu, Yan, and Yang}{Wu
  et~al\mbox{.}}{2019b}]%
        {wu2019DAM}
\bibfield{author}{\bibinfo{person}{Yu Wu}, \bibinfo{person}{Linchao Zhu},
  \bibinfo{person}{Yan Yan}, {and} \bibinfo{person}{Yi Yang}.}
  \bibinfo{year}{2019}\natexlab{b}.
\newblock \showarticletitle{Dual Attention Matching for Audio-Visual Event
  Localization}. In \bibinfo{booktitle}{\emph{Proceedings of the IEEE
  International Conference on Computer Vision (ICCV)}}.
\newblock


\bibitem[\protect\citeauthoryear{Wyatte}{Wyatte}{2019}]%
        {wyatte2019biasing}
\bibfield{author}{\bibinfo{person}{Dean Wyatte}.}
  \bibinfo{year}{2019}\natexlab{}.
\newblock \showarticletitle{De-biasing Weakly Supervised Learning by
  Regularizing Prediction Entropy}.
\newblock  (\bibinfo{year}{2019}).
\newblock


\bibitem[\protect\citeauthoryear{Xiao, Zhu, Chen, Zhao, Jiang, and Zheng}{Xiao
  et~al\mbox{.}}{2018}]%
        {Xiao2018AddressingTB}
\bibfield{author}{\bibinfo{person}{Zhujun Xiao}, \bibinfo{person}{Yanzi Zhu},
  \bibinfo{person}{Yuxin Chen}, \bibinfo{person}{Ben~Y. Zhao},
  \bibinfo{person}{Junchen Jiang}, {and} \bibinfo{person}{Haitao Zheng}.}
  \bibinfo{year}{2018}\natexlab{}.
\newblock \showarticletitle{Addressing Training Bias via Automated Image
  Annotation}.
\newblock \bibinfo{journal}{\emph{arXiv: Computer Vision and Pattern
  Recognition}} (\bibinfo{year}{2018}).
\newblock


\bibitem[\protect\citeauthoryear{Xtract.io}{Xtract.io}{2020}]%
        {xtract}
\bibfield{author}{\bibinfo{person}{Xtract.io}.}
  \bibinfo{year}{2020}\natexlab{}.
\newblock \bibinfo{booktitle}{\emph{Xtract.io video annotation tool}}.
\newblock
\urldef\tempurl%
\url{https://www.xtract.io/lp/image-annotation-tool}
\showURL{%
\tempurl}


\bibitem[\protect\citeauthoryear{Zeng, Yu, and Oyama}{Zeng
  et~al\mbox{.}}{2018}]%
        {zeng2018audio}
\bibfield{author}{\bibinfo{person}{Donghuo Zeng}, \bibinfo{person}{Yi Yu},
  {and} \bibinfo{person}{Keizo Oyama}.} \bibinfo{year}{2018}\natexlab{}.
\newblock \showarticletitle{Audio-visual embedding for cross-modal music video
  retrieval through supervised deep CCA}. In \bibinfo{booktitle}{\emph{2018
  IEEE International Symposium on Multimedia (ISM)}}. IEEE,
  \bibinfo{pages}{143--150}.
\newblock


\bibitem[\protect\citeauthoryear{Zhang, Wang, Zhang, Zhu, Chen, and
  Zhang}{Zhang et~al\mbox{.}}{2022}]%
        {zhang2022onelabeler}
\bibfield{author}{\bibinfo{person}{Yu Zhang}, \bibinfo{person}{Yun Wang},
  \bibinfo{person}{Haidong Zhang}, \bibinfo{person}{Bin Zhu},
  \bibinfo{person}{Siming Chen}, {and} \bibinfo{person}{Dongmei Zhang}.}
  \bibinfo{year}{2022}\natexlab{}.
\newblock \showarticletitle{OneLabeler: A Flexible System for Building Data
  Labeling Tools}. In \bibinfo{booktitle}{\emph{CHI Conference on Human Factors
  in Computing Systems}}. \bibinfo{pages}{1--22}.
\newblock


\bibitem[\protect\citeauthoryear{Zhang, Gao, Dhaliwal, and Li}{Zhang
  et~al\mbox{.}}{2023}]%
        {zhang2023visar}
\bibfield{author}{\bibinfo{person}{Zheng Zhang}, \bibinfo{person}{Jie Gao},
  \bibinfo{person}{Ranjodh~Singh Dhaliwal}, {and} \bibinfo{person}{Toby Jia-Jun
  Li}.} \bibinfo{year}{2023}\natexlab{}.
\newblock \showarticletitle{VISAR: A Human-AI Argumentative Writing Assistant
  with Visual Programming and Rapid Draft Prototyping}.
\newblock \bibinfo{journal}{\emph{arXiv preprint arXiv:2304.07810}}
  (\bibinfo{year}{2023}).
\newblock


\bibitem[\protect\citeauthoryear{Zhao, Gan, Rouditchenko, Vondrick, McDermott,
  and Torralba}{Zhao et~al\mbox{.}}{2018}]%
        {zhao2018sound}
\bibfield{author}{\bibinfo{person}{Hang Zhao}, \bibinfo{person}{Chuang Gan},
  \bibinfo{person}{Andrew Rouditchenko}, \bibinfo{person}{Carl Vondrick},
  \bibinfo{person}{Josh McDermott}, {and} \bibinfo{person}{Antonio Torralba}.}
  \bibinfo{year}{2018}\natexlab{}.
\newblock \showarticletitle{The sound of pixels}. In
  \bibinfo{booktitle}{\emph{ECCV}}.
\newblock


\bibitem[\protect\citeauthoryear{Zhou}{Zhou}{2018}]%
        {zhou2018brief}
\bibfield{author}{\bibinfo{person}{Zhi-Hua Zhou}.}
  \bibinfo{year}{2018}\natexlab{}.
\newblock \showarticletitle{A brief introduction to weakly supervised
  learning}.
\newblock \bibinfo{journal}{\emph{National science review}}
  \bibinfo{volume}{5}, \bibinfo{number}{1} (\bibinfo{year}{2018}),
  \bibinfo{pages}{44--53}.
\newblock


\end{thebibliography}
